\documentclass[reprint,
twocolumn,
floatfix,
superscriptaddress,
nofootinbib,
amsmath,amssymb,
aps,
]{revtex4-2}

\usepackage{booktabs}\usepackage[x11names, table]{xcolor}
\usepackage{graphicx}\usepackage{dcolumn}\usepackage{bm}\usepackage{hyperref}\usepackage{cleveref}\usepackage[mathlines]{lineno}\usepackage{todonotes}
\usepackage[justification=justified, singlelinecheck=false]{caption}

\usepackage{multirow}
\usepackage[normalem]{ulem}
\usepackage{custom}
\usepackage{orcidlink}
\usepackage{tabularray}
\usepackage{enumitem}
\usepackage{comment}

\definecolor{amethyst}{rgb}{0.6, 0.4, 0.8}
\definecolor{jblue}{rgb}{0 0.4470 0.7410}

\newcommand{\gray}[1]{\textcolor{black!70}{#1}}

\definecolor{og}{RGB}{34, 139, 34}

\begin{document}
\title{
Studying the QCD Matter produced in Heavy-Ion Collisions \\
using the MUSES Calculation Engine
}

\author{Johannes~Jahan\,\orcidlink{0009-0002-4557-4652}}
\affiliation{
 Department of Physics, University of Houston, Houston, TX 77204, USA
}
\author{Kevin~P.~Pala\,\orcidlink{0000-0002-1677-3476}}
\affiliation{
 Instituto de F\'{i}sica, Universidade de S\~ao Paulo, Rua do Mat\~ao 1371, 05508-090 S\~ao Paulo-SP, Brazil
}
\author{Yumu~Yang\,\orcidlink{0009-0001-8979-9343}}
\affiliation{
 The Grainger College of Engineering, Illinois Center for Advanced Studies of the Universe, Department of Physics, University of Illinois at Urbana-Champaign, Urbana, IL 61801, USA
}
\author{Isabella~Danhoni\,\orcidlink{0000-0003-0126-393X}}
\affiliation{
 The Grainger College of Engineering, Illinois Center for Advanced Studies of the Universe, Department of Physics, University of Illinois at Urbana-Champaign, Urbana, IL 61801, USA
}
\author{Prachi~Garella\,\orcidlink{0000-0002-9980-4855}}
\affiliation{
 Department of Physics, University of Houston, Houston, TX 77204, USA
}
\author{Jonathan~Gonzales}
\affiliation{
 Department of Physics, University of Houston, Houston, TX 77204, USA
}
\author{Joaquin~Grefa\,\orcidlink{0000-0001-7590-9364}}
\affiliation{
 Center for Nuclear Research, Department of Physics, Kent State University, Kent, OH 44243, USA
}
\affiliation{
 Department of Physics, University of Houston, Houston, TX 77204, USA
}
\author{Mauricio~Hippert\,\orcidlink{0000-0001-5802-3908}}
\affiliation{
Centro Brasileiro de Pesquisas F\'isicas, Rua Dr. Xavier Sigaud 150, Rio de Janeiro, RJ, 22290-180, Brazil
}
\author{Surkhab~Kaur~Virk}
\affiliation{
 The Grainger College of Engineering, Illinois Center for Advanced Studies of the Universe, Department of Physics, University of Illinois at Urbana-Champaign, Urbana, IL 61801, USA
}
\author{Micheal~Kahangirwe\,\orcidlink{0000-0001-9144-6240}}
\affiliation{
 Center for Nuclear Research, Department of Physics, Kent State University, Kent, OH 44243, USA
}
\author{Musa~R.~Khan\,\orcidlink{0009-0002-4737-9065}}
\affiliation{
 Department of Physics, University of Houston, Houston, TX 77204, USA
}
\author{Feyisola~Nana\,\orcidlink{0009-0005-4862-1580}}
\affiliation{
 The Grainger College of Engineering, Illinois Center for Advanced Studies of the Universe, Department of Physics, University of Illinois at Urbana-Champaign, Urbana, IL 61801, USA
}
\author{Mateus~Reinke~Pelicer\,\orcidlink{0000-0002-2189-706X}}
\affiliation{
 The Grainger College of Engineering, Illinois Center for Advanced Studies of the Universe, Department of Physics, University of Illinois at Urbana-Champaign, Urbana, IL 61801, USA
}
\author{Tulio~E.~Restrepo\,\orcidlink{0000-0003-3579-5520}}
\affiliation{
 Department of Physics, University of Houston, Houston, TX 77204, USA
}
\author{Hitansh~Shah\,\orcidlink{0009-0008-1870-3157}}
\affiliation{
 Department of Physics, University of Houston, Houston, TX 77204, USA
}
\author{T.~Andrew~Manning\,\orcidlink{0000-0003-2545-9195}}
\affiliation{
 National Center for Supercomputing Applications, University of Illinois Urbana-Champaign, Urbana, IL 61801, USA
}
\author{Mark~Alford\,\orcidlink{ }}
\affiliation{
 Department of Physics, Washington University in St. Louis, St. Louis, MO 63130, USA
}
\author{Dekrayat~Almaalol\,\orcidlink{ }}
\affiliation{
 Center for Nuclear Theory, Department of Physics and Astronomy, Stony Brook University, Stony Brook, NY 11794, USA
}
\author{Ahmed~Abuali\,\orcidlink{0000-0003-1198-3201}}
\affiliation{
 Department of Physics, University of Houston, Houston, TX 77204, USA
}
\author{Alexander~Clevinger\,\orcidlink{0000-0001-6478-7066}}
\affiliation{
 Center for Nuclear Research, Department of Physics, Kent State University, Kent, OH 44243, USA
}
\author{Nikolas~Cruz-Camacho\,\orcidlink{0009-0004-7870-0039}}
\affiliation{
 The Grainger College of Engineering, Illinois Center for Advanced Studies of the Universe, Department of Physics, University of Illinois at Urbana-Champaign, Urbana, IL 61801, USA
}
\author{Carlos~Conde-Ocazionez\,\orcidlink{0000-0002-5393-0565}}
\affiliation{
 The Grainger College of Engineering, Illinois Center for Advanced Studies of the Universe, Department of Physics, University of Illinois at Urbana-Champaign, Urbana, IL 61801, USA
}
\author{Francesco~Di~Clemente\,\orcidlink{0000-0002-8257-3819}}
\affiliation{
 Department of Physics, University of Houston, Houston, TX 77204, USA
}
\author{David~Friedenberg\,\orcidlink{0009-0008-6766-5169}}
\affiliation{
 Cyclotron Institute, Texas A\&M University, College Station, TX 77843, USA
}
\affiliation{
 Department of Physics and Astronomy, Texas A\&M University, College Station, TX 77843, USA
}
\author{Hosein~Gholami\,\orcidlink{0009-0003-3194-926X}}
\affiliation{Technische Universit{\"a}t Darmstadt, Fachbereich Physik, Institut f{\"u}r Kernphysik,
Theoriezentrum, Schlossgartenstr. 2, D-64289 Darmstadt, Germany
}
\author{Marco~Hofmann\,\orcidlink{0000-0002-4947-1693}}
\affiliation{Technische Universit{\"a}t Darmstadt, Fachbereich Physik, Institut f{\"u}r Kernphysik,
Theoriezentrum, Schlossgartenstr. 2, D-64289 Darmstadt, Germany
}
\author{Jeremy~W.~Holt\,\orcidlink{0000-0003-4373-3856}}
\affiliation{
 Cyclotron Institute, Texas A\&M University, College Station, TX 77843, USA
}
\affiliation{
 Department of Physics and Astronomy, Texas A\&M University, College Station, TX 77843, USA
}
\author{Isaac~Legred\,\orcidlink{0000-0002-9523-9617}}
\affiliation{
 The Grainger College of Engineering, Illinois Center for Advanced Studies of the Universe, Department of Physics, University of Illinois at Urbana-Champaign, Urbana, IL 61801, USA
}
\author{Jamie~M.~Karthein\,\orcidlink{0000-0003-2041-5206}}
\affiliation{
 Cyclotron Institute, Texas A\&M University, College Station, TX 77843, USA
}
\affiliation{
 Department of Physics and Astronomy, Texas A\&M University, College Station, TX 77843, USA
}
\author{Toru~Kojo\,\orcidlink{0000-0001-5656-3652}}
\affiliation{
 Theory Center, IPNS, High Energy Accelerator Research Organization (KEK), 1-1 Oho, Tsukuba, Ibaraki, 305-0801, Japan
}
\affiliation{
 Graduate Institute for Advanced Studies, SOKENDAI, 1-1 Oho, Tsukuba, Ibaraki, 305-0801, Japan
}
\affiliation{
 Department of Physics, Tohoku University, Aoba-yama, Sendai, Miyagi, 980-8578, Japan
}
\author{Konstantin~Maslov\,\orcidlink{0000-0002-4334-1548}}
\affiliation{
 Department of Physics, University of Houston, Houston, TX 77204, USA
}
\author{Paolo~Parotto\,\orcidlink{0000-0002-4686-941X}}\affiliation{Dipartimento di Fisica, Universit\`a di Torino and INFN Torino, Via P. Giuria 1, I-10125 Torino, Italy}
\author{Leonardo~Pena\,\orcidlink{0009-0005-5631-8557}}
\affiliation{
 The Grainger College of Engineering, Illinois Center for Advanced Studies of the Universe, Department of Physics, University of Illinois at Urbana-Champaign, Urbana, IL 61801, USA
}
\author{Gr\'egoire~Pihan\,\orcidlink{0000-0002-7072-4480}}
\affiliation{
 Department of Physics, University of Houston, Houston, TX 77204, USA
}
\author{Christopher~Plumberg\,\orcidlink{0000-0001-6678-3966}}
\affiliation{
 Natural Science Division, Pepperdine University, Malibu, CA 90263, USA
}
\author{Roman~Poberezhniuk\,\orcidlink{0000-0002-5559-3718}}
\affiliation{
 Department of Physics, University of Houston, Houston, TX 77204, USA
}
\author{Romulo~Rougemont\,\orcidlink{0000-0002-1558-1624}}
\affiliation{
 Instituto de F\'{i}sica, Universidade Federal de Goi\'{a}s, Avenida Esperan\c{c}a - Campus Samambaia, CEP 74690-900, Goi\^{a}nia, Goi\'{a}s, Brazil
}
\author{Jordi~Salinas~San~Mart\'in\,\orcidlink{0000-0001-6203-4458}}
\affiliation{
 The Grainger College of Engineering, Illinois Center for Advanced Studies of the Universe, Department of Physics, University of Illinois at Urbana-Champaign, Urbana, IL 61801, USA
}
\author{Rajesh~Kumar\,\orcidlink{0000-0003-2746-3956}}
\affiliation{
 Center for Nuclear Research, Department of Physics, Kent State University, Kent, OH 44243, USA
}
\affiliation{
 Department of Physics, MRPD Government College Talwara, Punjab, 144216, India
}
\author{Volodymyr~Vovchenko\,\orcidlink{0000-0002-2189-4766}}
\affiliation{
 Department of Physics, University of Houston, Houston, TX 77204, USA
}
\author{Veronica~Dexheimer\,\orcidlink{0000-0001-5578-2626}}
\affiliation{
 Center for Nuclear Research, Department of Physics, Kent State University, Kent, OH 44243, USA
}
\author{Jorge~Noronha\, \orcidlink{0000-0002-9817-0272}}
\affiliation{
 The Grainger College of Engineering, Illinois Center for Advanced Studies of the Universe, Department of Physics, University of Illinois at Urbana-Champaign, Urbana, IL 61801, USA
}
\author{Jacquelyn Noronha-Hostler\,\orcidlink{0000-0003-3229-4958}}
\affiliation{
 The Grainger College of Engineering, Illinois Center for Advanced Studies of the Universe, Department of Physics, University of Illinois at Urbana-Champaign, Urbana, IL 61801, USA
}
\author{Claudia Ratti\,\orcidlink{0000-0002-8335-567X}}
\affiliation{
 Department of Physics, University of Houston, Houston, TX 77204, USA
}
\author{Nicol\'as Yunes\,\orcidlink{0000-0001-6147-1736}}
\affiliation{
 The Grainger College of Engineering, Illinois Center for Advanced Studies of the Universe, Department of Physics, University of Illinois at Urbana-Champaign, Urbana, IL 61801, USA
}

\date{\today}

\begin{abstract}
The equation of state of hot and dense matter is essential for describing heavy-ion collisions at all collision energies. Here, we explore the capabilities of the latest version of the MUSES Calculation Engine, \textit{Calliope}, focusing on software modules and workflows that compute the equation of state and observable properties of the matter produced in heavy-ion collisions. These include several equations of state, ranging from first-principles lattice QCD to phenomenological approaches, with or without a critical point, and with phase-space dimensionality ranging from two dimensions defined by temperature $T$ and baryon chemical potential $\mu_B$, to four dimensions after the addition of strangeness and electric-charge chemical potentials $\mu_S$ and $\mu_Q$. We also discuss modules that provide additional thermodynamic quantities and observables relevant for heavy-ion modeling, including elements of the pressure Hessian matrix and transport coefficients. Workflow examples are constructed that merge two equations of state thermodynamically consistently to extend phase-diagram coverage, and feed the results into an equation of state inverter to produce inputs suitable for hydrodynamic simulations. Finally, we apply this framework to perform a relativistic viscous hydrodynamic simulation with  equations of state with an extended $T$ and $\mu_B$ coverage and a movable critical point, including effects from transport coefficients that phenomenologically encode critical scaling, at collision energies $\sqrt{s_{NN}}=7.7, 19.6$, and $39$ GeV. 
\end{abstract}

\maketitle

\tableofcontents{}

\pagebreak

\section{Introduction}
\label{sec:intro}

Quantum chromodynamics (QCD) \cite{Achenbach:2023pba} is the non-Abelian $SU(3)$ gauge theory of the Standard Model that describes the strong interactions of quarks and gluons. Exploring the QCD phase diagram requires temperatures ($T$) and net baryon densities ($n_B$) far beyond those naturally realized on Earth \cite{Sorensen:2023zkk}. Ultra-relativistic heavy-ion collisions at the Relativistic Heavy-Ion Collider (RHIC) and the Large Hadron Collider (LHC) briefly produce hot QCD matter under conditions akin to those of the early Universe, while neutron stars provide access to cold (warm in the case of their mergers~\cite{Most:2022wgo}) matter at extreme densities \cite{Dexheimer:2020zzs}. Ideally, one would like to perform joint analyses that connect constraints from the laboratory to those from the cosmos \cite{MUSES:2023hyz}. Achieving this goal, however, requires careful accounting of the very different physical regimes and constraints realized in each system, as well as open-source tools that (i) compute microphysical thermodynamics across the relevant parts of the multidimensional phase diagram and (ii) interface consistently with dynamical simulation frameworks used for direct comparisons to experimental and astrophysical data.

\subsection{Heavy-ion collisions}

Heavy-ion collisions generate a unique state of matter known as the quark-gluon plasma (QGP) \cite{Gyulassy:2004zy}.  The transition from hadronic matter to the QGP occurs as a smooth crossover at high center-of-mass energies $\sqrt{s_{NN}}$, when matter and antimatter are produced in comparable amounts~\cite{Aoki:2006we}, which corresponds to a baryon chemical potential close to zero on average over the system, i.e., \emph{global} $\mu_B \simeq 0$.
At lower $\sqrt{s_{NN}}$, increased baryon stopping leads to net-baryon densities $n_B>0$ and correspondingly finite 
\emph{global} $\mu_B$. At finite $\mu_B$, the deconfinement transition may become first-order (see, e.g., \cite{Bzdak:2019pkr,Dexheimer:2020zzs,Lovato:2022vgq,Sorensen:2023zkk,Du:2024wjm} for recent reviews), which motivates the possibility of a critical point (CP) on the QCD phase diagram separating crossover and first-order regimes \cite{Stephanov:1998dy}.
In the coming years, heavy-ion experiments may provide decisive constraints on whether the QCD critical point exists and where it lies on the phase diagram.
The STAR experiment at RHIC conducted its second beam energy scan (BES-II) \cite{Tlusty:2018rif}, where net-proton fluctuations have been precisely measured \cite{STAR:2025zdq}. Additionally, STAR ran a fixed-target program to extend the search to even denser conditions, where data is currently being analyzed. Future experiments at FAIR (GSI Darmstadt) \cite{CBM:2016kpk,Durante:2019hzd} and J-Parc \cite{Nagamiya:2022xfb} will continue the study of high-density QCD matter beyond RHIC.

In heavy-ion collisions, the characteristic timescales are too short for weak-interaction processes to play any role in the dynamical evolution, so the relevant conserved charges are those associated with QCD and quantum electrodynamics (QED). In particular, net strangeness is conserved throughout the simulated evolution. Since the initial state has zero net-strangeness content (only nuclei, no hypernuclei are collided), the system remains globally strangeness neutral, despite producing abundant strange (anti)hadrons. Concretely, heavy-ion collisions measure large yields of strange particles 
($K$, $\Lambda$, $\Xi$, and $\Omega$), 
but always in combinations that maintain a vanishing net-strangeness. At large $\sqrt{s_{NN}}$, matter and antimatter are nearly balanced among both mesons and baryons, e.g., $N_{\Lambda}\sim N_{\bar{\Lambda}}$ and $N_{K^+}\sim N_{K^-}$. In contrast, at low $\sqrt{s_{NN}}$, baryons are energetically favored, so hyperons are more abundant than anti-hyperons, e.g., $N_{\Lambda}>N_{\bar{\Lambda}}$. Strangeness neutrality still applies, favoring mesons that carry anti-strangeness, e.g., $N_{K^+}>N_{K^-}$. This leads to a finite \emph{global} strangeness chemical potential $\mu_S$, associated with the heavy-ion system \cite{Bazavov:2012vg,Bazavov:2014xya,Bellwied:2015rza,Fu:2018qsk,Fu:2018swz,Borsanyi:2022qlh}.

Similarly, electric charge is conserved in heavy-ion collisions, and the charge fraction of the colliding nuclei, $Y_Q=Z/A$ (where $Z$ is proton number and $A$ is baryon number), fixes the global charge-to-baryon ratio. Isospin-symmetric nuclear matter corresponds to $Z=A/2$ and implies a vanishing electric charge chemical potential when no net strangeness is present. Since heavy nuclei often have $Y_Q\sim 0.4$, they generically imply a small, negative \emph{global} electric charge chemical potential $\mu_Q$, in order to satisfy electric charge conservation \cite{Bazavov:2012vg,Nana:2024okk,Grefa:2026meq}. 

In the present work, we distinguish local quantities from single-event-averaged and event-ensemble-averaged ones:
\begin{itemize}
    \item we reserve $\mu_X$ (where $X=B,S,Q$ for baryon number, strangeness, or electric charge) for \emph{local} chemical potentials   that may vary strongly across the evolving fireball. In other words, these depend on the local space-time position of the fluid cells;
    \item we use $\langle ~ \rangle$ to denote an average across a \emph{single heavy-ion event} (with a specified averaging prescription) for any quantity;
    \item we use $\langle\!\langle ~ \rangle\!\rangle$ to denote an ensemble average over multiple events in a given centrality class.
\end{itemize}
With this notation, the constraints that enforce $SQ$ (strangeness and electric charge) global charge conservation across a single event are:
\begin{equation}\label{eqn:StrNeu}
    \langle n_S\rangle=0,\quad \frac{\langle n_Q\rangle}{\langle n_B\rangle} =Y_Q,
\end{equation}
where the net-number densities are $\langle n_X\rangle=\sum_i X_i n_i$ and $i$ is summed over particles in the system. 
In other words, Eq.~\eqref{eqn:StrNeu} must be exactly correct, when averaged over the entire event. However, we note that these equations make no statement about how the $S$ and $Q$ charges are distributed locally across momentum or coordinate space. 

Using experimental data of particle yield ratios and thermal fits, one can access information about the event-ensemble averaged point of freeze-out
\footnote{Freeze-out is the stage in a rapidly expanding heavy-ion collision when the interaction rate between particles falls below the expansion rate, causing the system to decouple at an effective thermodynamic point, characterized by a temperature and chemical potentials, whose particle yields and spectra can then be compared to detector measurements. 
} 
for a given $\sqrt{s_{NN}}$ and $Y_Q$ \cite{Cleymans:1998fq,Cleymans:2005xv}. Then, averaging over a single event, we can define a point of freeze-out $\langle T^{FO}\rangle $ and $\langle\mu_B^{FO}\rangle$, where the constraints from  Eq.~\eqref{eqn:StrNeu} also lead to constraints in $\langle\mu_S^{FO}\rangle,\langle\mu_Q^{FO}\rangle$. However, $\langle T^{FO}\rangle,\langle\mu_B^{FO}\rangle$ can and do fluctuate event-to-event so the information extracted from thermal fits provides an event ensemble averaged $\left\{\langle\langle T^{FO}\rangle\rangle,\langle\langle\mu_B^{FO}\rangle\rangle,\langle\langle\mu_S^{FO}\rangle\rangle,\langle\langle\mu_Q^{FO}\rangle\rangle\right\}$. 
A point of caution that we will return to in Sec.~\ref{sec:observables} is that heavy-ion collisions feature significant \emph{local} fluctuations, so deviations from $\langle T\rangle,\langle \mu_B\rangle,\langle\mu_S\rangle,\langle\mu_Q\rangle$ can be important, even in a single event.

Two core challenges arise when determining the equation of state  (EoS) relevant for heavy-ion collisions. First, the EoS can only be computed from first principles in a limited regime of the QCD phase diagram due to the fermion sign problem \cite{Troyer:2004ge,Philipsen:2012nu}. At finite baryon chemical potential, the fermion determinant in the QCD path integral acquires a nonzero imaginary part, making standard Monte Carlo techniques inapplicable. At low $\mu_B$  and high $T$, lattice QCD provides reliable predictions through extrapolations from $\mu_B=0$ or analytical continuation from imaginary chemical potential \cite{deForcrand:2002hgr,DElia:2002tig,Bonati:2015bha,DElia:2016jqh,Guenther:2017hnx,Bonati:2018nut,Borsanyi:2018grb,Bellwied:2015rza,Bollweg:2020yum,Borsanyi:2020fev,Bollweg:2022rps,Ding:2024sux}. This way, the EoS can be reconstructed up to $\mu_B/T\sim 3.5$  and for $T\gtrsim 130$ MeV \cite{Borsanyi:2021sxv} from first principles. Extensions to all three chemical potentials are also becoming available, with similar limitations in the values of $\mu_X/T$ \cite{Noronha-Hostler:2019ayj,Monnai:2019hkn,Monnai:2024pvy,Abuali:2025tbd}.
Beyond that range, phenomenological models are typically used to compute the EoS, and one must ensure that these models are matched to lattice-QCD constraints in a thermodynamically consistent manner---a problem we tackle in this work.

Second, heavy-ion collisions are inherently dynamical and out-of-equilibrium systems, requiring dynamical simulation frameworks for direct comparisons to data. At high-to-medium $\sqrt{s_{NN}}$ (i.e., RHIC to LHC collision energy range), state-of-the-art modeling relies on relativistic viscous hydrodynamics with multiple conserved charges, followed by a hadronic afterburner \cite{Campbell:2022qmc}, sometimes resorting to a core-corona approach where both run concurrently to model the simultaneous evolution of deconfined bulk matter and the hadronic medium  \cite{Werner:2007bf,Becattini:2008ya,Kanakubo:2021qcw}. 
The standard approach includes shear and bulk viscosity effects, often with relatively simple parameterizations that are not directly tied to micro-physical calculations \cite{Auvinen:2017fjw,Jahan:2025cbp,Jahan:2024wpj,Gotz:2025wnv}, e.g., by taking the temperature times shear viscosity over enthalpy ratio $\eta T/(\varepsilon+P)$ as constant (here $\varepsilon$ is the energy density and $P$ is the pressure).
Viscous effects enter both the hydrodynamical evolution and the conversion to hadrons at freeze-out through non-equilibrium ``$\delta f$'' corrections, with distinct contributions for shear viscosity \cite{Teaney:2003kp,Bozek:2009dw,Luzum:2010ad}, bulk viscosity \cite{Monnai:2009ad,Pratt:2010jt,Dusling:2011fd,Noronha-Hostler:2013gga}, and diffusion \cite{Greif:2017byw}. To explore the BES-II program, one requires at minimum baryon number conservation in fluid-dynamical simulations \cite{Denicol:2018wdp}, and ideally also the conservation of strangeness and electric charge. A number of hydrodynamic frameworks include ideal conservation of either $B$, $BQ$, or $BSQ$ charges \cite{Karpenko:2013wva,Pihan:2024lxw,Monnai:2024pvy,Plumberg:2024leb}. However, out-of-equilibrium charge transport requires diffusion, and since the $B$, $S$, and $Q$ currents are correlated, a full diffusion matrix is required \cite{Greif:2017byw,Fotakis:2019nbq,Fotakis:2022usk}. The $BSQ$ diffusion matrix has been explored in reduced-dimensional setups, and is now possible in $(3+1)$D, using the new \texttt{CCAKE}~2.0 framework \cite{Pala:2025qoa} (previous work in \cite{Denicol:2018wdp} studied the effects of a single baryon current).

The vast majority of dynamical simulations to date have assumed baseline equations of state (EoSs)  
without a CP. Exceptions exist in lower-dimensional studies, e.g., $(0+1)$D or $(1+1)$D (spacial+time), which found that out-of-equilibrium effects can smear critical signatures and that ``dynamical lensing'' may occur, depending on the orientation of a critical point on the QCD phase diagram \cite{Dore:2022qyz}. Other work has focused on deriving hydrodynamic equations of motion with additional non-hydrodynamic modes that can account for criticality (see reviews \cite{Bzdak:2019pkr,An:2021wof,Basar:2024srd} and recent works \cite{An:2022jgc,Karthein:2025hvl}). Only in \cite{Monnai:2021kgu} are we aware of work that evolved a lattice QCD-based  EoS containing a critical point within hydrodynamics, but without implementing critical scaling either in the equations of motion or in the transport coefficients. 
However, transport coefficients are also sensitive to critical scaling and this effect could potentially enhance critical features \cite{Monnai:2016kud,Dore:2022qyz}. 
In summary, dynamical frameworks require both an EoS and transport coefficients as inputs, and ideally one would like to compute them within a consistent framework, especially in the regime relevant for a critical point and first-order transition.

\subsection{Connecting heavy-ion collisions \\ and neutron stars}

A major goal of the dense matter community is to connect constraints from low-energy heavy-ion collisions (such as BES-II) to those from neutron stars and their mergers \cite{Lovato:2022vgq, Sorensen:2023zkk,Dexheimer:2020zzs,MUSES:2023hyz}. While the net baryon densities reached in both environments can be comparable (both can reach up to $n_B^{\max}\sim 6n_{\rm{sat}}$ \cite{Oliinychenko:2022uvy, Mroczek:2023zxo,Most:2018eaw}, with $n_{\rm{sat}}$ being the nuclear saturation density), they generally do not correspond to the same point on the QCD phase diagram. The relevant QCD phase diagram is four-dimensional (4D), usually parameterized by $(T,\mu_B,\mu_S,\mu_Q)$, and all thermodynamic variables (including number densities and density fractions) vary on this full 4D space. Thus, even if the baryon density range probed in neutron stars and heavy-ion collisions were similar, the systems could still explore qualitatively different physics due to differences in $T$, $\mu_B$, $\mu_S$, and $\mu_Q$.

Heavy-ion collisions 
probe finite temperatures $T\sim [50,600]$ MeV, and vary within extensive ranges of chemical potentials, $\mu_B\sim [-400,800]$ MeV, $\mu_S\sim [-400,400]$ MeV (constrained by strangeness neutrality), and $\mu_Q\sim [-100,100]$ MeV (set by the colliding nuclei, which are nearly isospin symmetric such that $\mu_Q\sim 0$ globally,  but have a significantly larger range \emph{locally})  -- see, e.g., ~\cite{Plumberg:2024leb,Monnai:2024pvy,Pala:2025qoa,JETSCAPE:2025wac}.
In contrast, neutron stars are at approximately $T\sim 0$ MeV, with post-merger matter typically reaching on average $T\sim 30-60$ MeV (although some simulations reach up to $T=100$ MeV) \cite{Most:2018eaw, Radice:2020ddv, Most:2021ktk}, and feature significantly different ranges for $\mu_Q$ and $\mu_S$~\cite{Most:2019onn}. In other words, even when the $n_B$ range and parts of the $T$ range appear similar, qualitatively different phases of matter may be accessed due to differences in $\mu_S$ and $\mu_Q$ -- see, e.g., \cite{Cruz-Camacho:2024odu,Most:2019onn}.

These differences arise primarily due to the much longer timescales at play for neutron stars and their mergers, for which both QED and electroweak interactions shape the equilibrium composition. A neutron star must be electrically neutral, so positively charged particles are needed to balance the negatively charged leptons (potentially also contributions from negative hadrons or quarks).
Furthermore, neutron stars live long enough to reach electroweak ($\beta$) equilibrium, which together with charge neutrality requires a negative $\mu_Q$, whose magnitude becomes increasingly large in dense hadronic matter \cite{Akmal:1998cf,Heiselberg:1999mq}. This results in a density-dependent QCD\footnote{Here, we are careful to denote ``$QCD$" for these quantities because the entire system is electrically neutral due to the presence of leptons, but the QCD sector (nuclei, hadrons, and quarks) can thus have a finite electric charge.} charge fraction 
$Y_Q(n_B)= n_Q^{QCD} / n_B = -n_Q^{lep} / n_B$ 
and potentially a non-zero strangeness fraction $Y_S(n_B)=n_S^{QCD}/n_B$ depending on the degrees of freedom present. 
This is a fundamentally different constraint structure than in heavy-ion collisions, where strangeness neutrality is enforced and the system has a finite total electric charge.

Despite these caveats, progress has been made toward building tools that connect these limits of the phase diagram. Joint comparisons between neutron stars and heavy-ion collisions have begun in recent years \cite{Huth:2021bsp, Most:2022wgo,Tsang:2023vhh, Yao:2023yda, Steinheimer:2025hsr,Frohaug:2025okz}. Symmetry energy expansions, to go from isospin-symmetric matter relevant for heavy-ion collisions into asymmetric matter relevant for neutron stars and vice-versa, 
have started to include strangeness \cite{Yang:2025wop, Danhoni:2025qpn} and temperature effects \cite{Mroczek:2024sfp}. 
Initial comparisons have also explored how the entropy-per-baryon (or entropy density per baryon density, $s/n_B$) ratio can actually connect regions of the QCD phase diagram between neutron stars and heavy-ion collisions \cite{Most:2022wgo}. Yet, other work has considered a critical point at $\beta$-equilibrium, followed by a first-order phase transition in neutron stars \cite{Ecker:2025vnb} (see also \cite{Karthein:2026rok}); however, this assumes no other phase transition occurs as $T\rightarrow 0$, and it remains unclear how such extrapolations map onto the different $\mu_S$ and $\mu_Q$ constraints relevant for heavy-ion collisions.

\subsection{This work}

The MUSES (Modular Unified Solver of the Equation of State) Calculation Engine  represents a step forward in such endeavor, by providing a composable workflow management system that orchestrates calculation and data processing for a collection of software modules. These modules calculate both the EoS and observables of dense and hot matter. Recently, we have publicly released a set of MUSES modules, and discussed in detail how one can match different zero- and low-temperature charge-neutral, equilibrated EoSs using different thermodynamic variables and different methods to produce neutron-star observables \cite{ReinkePelicer:2025vuh}. 

With this paper, we are releasing a new set of modules that complement the ones available so far. 
The focus here is on modules that are primarily useful to model heavy-ion collisions. Examples include a first-principles, continuum estimated EoS in which $T$, $\mu_B$, $\mu_S$ and $\mu_Q$ can be varied independently over the broadest range available in the literature, EoSs that contain a critical point whose location and features can be changed by the users, and the hadron resonance gas (HRG) EoS in which hadronic interactions can be tuned to implement the ideal, excluded volume, or quantum van der Waals scenarios. Each EoS has its own well-defined range of validity, but several of them can be combined with the \textit{Synthesis} module to achieve a broad coverage of the phase diagram. Taken together, these developments establish the infrastructure needed to perform systematic studies of dense QCD matter and to enable future joint analyses that combine constraints from heavy-ion collisions and neutron stars within a unified, open-source framework \cite{Agarwal:2025ezo}.

Because the natural dynamical variables in the hydrodynamic evolution are densities rather than ($T, \mu_B,\mu_S,\mu_Q$), we use our \textit{EoS Inverter} module to perform the needed inversion and discuss the consequences of this process and its range of validity.
A representative subset of these EoSs are then tested in dynamical simulations using the
\texttt{AMPT}+\texttt{CCAKE}+\texttt{SMASH} framework from the \texttt{NuclearConfectionary} \cite{Pala:2025qoa} at
$\sqrt{s_{NN}}=7.7,\,19.6,$ and $39$ GeV. 
By varying the assumed location of the critical point in the EoS, these simulations
help to demonstrate how the extended coverage of the MUSES EoS tables can be used to track
which regions of the QCD phase diagram are sampled during the fireball evolution,
assess the best locations for identifying critical behavior,
and quantify how the finite domain of validity of expansion-based EoSs may affect
predictions for experimental observables.

This manuscript is organized as follows. In Section \ref{state} we offer a brief description of all modules available in the current MUSES \textit{Calliope} release. Section \ref{sec:HI_EOS} contains a more detailed explanation of all EoS modules that are suitable in their current form to describe heavy-ion collisions at RHIC and the LHC. Section \ref{sec:observables} presents the currently available modules that calculate observables that can be useful for heavy-ion collision simulations. In Section \ref{sec:workflow} we show how to merge different equations of state in their respective ranges of validity, in order to build an EoS with broader temperature and chemical potential coverage. Examples of hydrodynamic simulations of the fireball created in heavy-ion collisions, using some of the MUSES equations of state presented here, are provided in Section \ref{sec:hydro}. Our summary and outlook are presented in Section \ref{summary}. 

\emph{Notation:} We use natural units $\hbar=k_B=c=1$. 

\section{State of the art of MUSES products}
\label{state}

In this section, we list all modules that will be available in the new version of the MUSES Calculation Engine: \textbf{\textit{\mbox{Calliope}}} (to be released publicly upon acceptance of this manuscript for publication), used to generate all results presented in this work. 
We expand our original 10 MUSES modules (7 for EoSs, 2 for Observables, and 1 \textit{Synthesis}) \cite{ReinkePelicer:2025vuh} into 17 modules (12 for EoSs, 3 for Observables, 1 \textit{Synthesis}, and 1 \textit{EoS Inverter}) with significant upgrades. 
In addition, the external solver \texttt{CCAKE}~\cite{ccakesite} has adopted the MUSES data standards, enabling interoperability with Calculation Engine output and extending the calculation pipeline beyond native Calculation Engine workflows.
While the remainder of this manuscript focuses on the subset of modules and workflows directly relevant to heavy-ion phenomenology, the list below includes \textit{all} modules registered with the latest Calculation Engine release, for the sake of completeness.
In a future publication, we will discuss the usage of the remainder of these modules, as well as their connections to heavy-ion collision modules.

We note that a detailed introduction to the different elements of the MUSES cyberinfrastructure was already provided in Ref.~\cite{ReinkePelicer:2025vuh}, including definitions of the technical terminology and explanations about MUSES modules, the MUSES Calculation Engine \cite{calculation_engine_all_versions}, and their functionalities.
Although most of the software features have been described in our previous publication, more recently we added an interactive web form that allows users to construct and launch simple linear workflows without needing any knowledge of YAML format or specific workflow definition syntax. 
Users can thus easily build workflows comprising one or more processes using this graphical user interface (GUI) by selecting modules from a drop-down menu, filling in a user friendly form to configure each process, and defining connections between process outputs and inputs. While this new feature cannot construct complex workflow topologies, it satisfies the calculation needs of many researchers while making the Calculation Engine much more accessible to those with less computing expertise.

\begin{itemize}[leftmargin=*]
    \item \textbf{Heavy-ion modules:}
    \begin{itemize}[leftmargin=*]
        \item \textbf{HRG:}
        the hadron resonance gas  module of MUSES employs \textsc{Thermal-FIST} v1.6 via a C++ wrapper~\cite{Vovchenko:2019pjl,ThermalFISTGitHub,vovchenko_2026_18666424} to compute thermodynamic quantities and charge fluctuations in the full $(T,\mu_B,\mu_S,\mu_Q)$ space, using PDG-based particle lists and decay tables. 
        The module supports 3 modes: the ideal HRG (IHRG), excluded volume, and the interacting Quantum van der Waals HRG (QvdW-HRG). The excluded volume mode has repulsive interactions only, and QvdW-HRG has both
       repulsive and attractive interactions included via an excluded-volume and attractive mean-field~\cite{Vovchenko:2016rkn,Vovchenko:2017zpj}.
        In the present release, attraction and repulsion act only among baryons and among antibaryons with universal parameters $a$ and $b$, respectively.
        A common meson eigenvolume $b_M$ can also be specified to suppress hadronic degrees of freedom at high $T$, and pions can also be treated as interacting particles through an effective mass model matched to chiral perturbation theory at $T=0$.
        Given a user-supplied grid in $\{T,\mu_B,\mu_S,\mu_Q\}$ and interaction parameters, the module returns standard EoS quantities and the full Hessian of pressure with respect to $(T,\mu_{B,Q,S})$. The supported input ranges are $T \in[0, 300]~\mathrm{MeV}$, $\mu_B\in[-1,1]~\mathrm{GeV}$, $\mu_Q\in[-2m_\pi, 2m_\pi]$ and $\mu_S\in[-m_K, m_K]$, covering the regime relevant for heavy-ion applications and matching to other MUSES components.

        \item \textbf{4D-Taylor (\textit{BQS}):} 
        the 4D Taylor-expanded lattice (referred to as \textit{4D-Taylor} or \textit{BQS}) module calculates the lattice QCD-based EoS by Taylor-expanding the pressure at zero chemical potential
        in powers of $\hat{\mu}_X=\mu_X/T$  \cite{Noronha-Hostler:2019ayj}.   
        To compute the EoS, the default input includes a parameterization based on the recently developed lattice QCD (LQCD) EoS at zero $\mu_X$ with continuum-estimated susceptibilities \cite{jahan_2025_16749145}.
        An original parametrization based on older lattice results at $N_\tau=12$ lattice divisions in the temporal direction \cite{Borsanyi:2018grb} can also still be used.
        The \textit{BQS} module can compute an EoS table for $T\in[30,600]$ MeV, and for  $\mu_X\in[-450,450]$ MeV for $X=B,Q,S$. However, notice that there is no treatment of pion/kaon condensation in this expansion.
        See Ref.~\cite{Noronha-Hostler:2019ayj} for details about the construction, and Ref.~\cite{jahan_2025_14639786} for the open-source code.

        \item \textbf{4D-TExS:} the 4D $T^\prime$-Expansion Scheme (\textit{4D-TExS}) module calculates a LQCD-based EoS, employing a generalization of the $T^\prime$-Expansion Scheme (TExS) \cite{Borsanyi:2021sxv, Borsanyi:2022qlh} at finite $X$ for $X=B,Q,S$.
        Using a new set of continuum-estimated second- and fourth-order  susceptibilities computed at $\mu_X=0$ on the lattice \cite{jahan_2025_14639786}, the thermodynamics is derived through the calculation of the density at finite chemical potentials, from an expanded temperature $T^\prime(T, \hat{\mu})$ \cite{Abuali:2025tbd}.
        While the original 2D formulation \cite{Borsanyi:2021sxv} was estimated to be reliable for $\hat{\mu}_B$ up to 3.5, the current limits of validity of the 4D formulation can only be based on thermodynamic consideration of stability and causality.
        The 4D-TExS module allows to compute an EoS with $T\in[30,800]$ MeV and $\mu_X \in [-800,800]$ MeV for all three conserved charges, although results cannot be obtained in some regions at large chemical potentials due to numerical limitations. Besides, notice that there is no treatment of pion/kaon condensation in this expansion.
        See Ref.~\cite{jahan_2025_16749257} for the open-source code.

        \item \textbf{Ising 2D-TExS:} 
        the Ising 2D–$T^\prime$-Expansion Scheme (\textit{Ising 2D-TExS}) module combines the recently proposed 2D $(T,\mu_B)$ lattice QCD extrapolation method \cite{Borsanyi:2021sxv, Borsanyi:2022qlh} with the universal critical behavior of the 3D Ising-model universality class to yield an EoS that incorporates the phase transition with a first-order line and extends the $\mu_B$ coverage beyond Taylor-based approaches \cite{Parotto:2018pwx}. The framework provides tunable parameters that let users specify the location and properties of the critical point, including the size and shape of the critical region. It agrees with lattice QCD results near $\mu_B=0$ and remains valid up to $\mu_B =700~\text{MeV}$ over the temperature range $T\in[30,\text{800}]~\text{MeV}$ \cite{Kahangirwe:2024cny}.
        Using the accompanying C++ code (see Ref.~\cite{kahangirwe_2025_14637802} for details), users can also customize the resolution and grid of the EoS table as a function of $(T,\mu_B)$.

        \item \textbf{Holography (\textit{NumRelHolo}):} 
        the Holographic (\textit{NumRelHolo}) module generates a 2D $(T, \mu_B)$ EoS from a bottom-up, non-conformal Einstein–Maxwell–Dilaton (EMD) model in asymptotically AdS spacetime, based on the gauge/gravity correspondence~\cite{Maldacena:1997re}. 
        The EMD potentials are determined by fitting to lattice QCD data at $\mu_B=0$ for entropy density $s/T^3$ and second-order baryon susceptibility $\chi_2^B$. Within this framework, the crossover region evolves into a line of first-order phase transition ending at a critical point~\cite{DeWolfe:2010he,Critelli:2017oub,Grefa:2021qvt,Rougemont:2023gfz}. Current Bayesian constraints on the holographic parameter space and the critical point location are summarized in~\cite{Hippert:2023bel}. 
        The MUSES holographic module provides EoSs within a range of $T\in [20, 800]$ MeV and  $\mu_B \in [-1000, 1000]$ MeV.
        See Ref.~\cite{yang_2026_19154307} for the open-source C++ code. 
    \end{itemize}

\item \textbf{Neutron star modules:}
\begin{itemize}[leftmargin=*]
        \item \textbf{Crust-DFT:} the low $n_B$ region of a neutron star, usually referred to as the crust, contains nuclei. Within MUSES, the  Density Functional Theory (\textit{Crust-DFT}) module is the only one so far that can describe nuclei, and therefore is the only one that can be applied to low $n_B$, in addition to intermediate/high  $n_B$. It calculates the EoS of neutrons and protons in equilibrium with an ensemble of nuclei 
        for $n_B \in [2.0\times10^{-12}, 2.0]~\text{fm}^{-3}$, proton fraction $Y_Q \in [0.01, 0.7]$, and $T\in[0,127]$ MeV  (although it contains no hyperons, resonances, thermal mesons, or deconfined quarks, which limits its applicability at high $T$ and low or high $\mu_B$) \cite{Du:2018vyp,Du:2021rhq}. 
        See Ref.~\cite{steiner_2025_14714273} for the open-source C++ code.

        \item \textbf{$\chi$EFT:} the Chiral Effective Field Theory (\textit{$\chi$EFT}) module delivers an \textit{ab-initio} treatment of bulk hadronic matter, computed within many-body perturbation theory \cite{Wellenhofer:2014hya, Wellenhofer:2015qba}. Its implementation in MUSES includes nucleons with several realistic nucleon-nucleon and three-nucleon interactions fit to low-energy nuclear scattering and bound states \cite{Machleidt:2011zz}. Including chiral interactions at varying energy resolution scales ($\Lambda \in \{414, 450, 500\}$ MeV) enables uncertainty quantification of the nuclear EoS in the applicable regime. The module does not incorporate nuclear clustering, so its applicability is limited to the density regime around nuclear saturation,
        for $n_B \in [0.02, 0.36]~\text{fm}^{-3}$ and proton fraction $Y_Q \in [0, 0.5]$ (or, conversely, isospin asymmetry $\delta_Q = (n_n - n_p) / (n_n + n_p) \in [0,1]$) at $T=0$, corresponding to the outer core of neutron stars \cite{Machleidt:2011zz,Drischler:2021kxf}.
        See Ref.~\cite{zenodo_cheft} for the open-source C++ code.

        \item \textbf{CMF++:} the Chiral Mean Field (\textit{CMF++}) module provides an effective relativistic description of strongly interacting matter in terms of hadronic and quark degrees of freedom, including the baryon octet and decuplet, and light plus strange quarks. It reproduces key features of the QCD phase diagram, such as the nuclear liquid–gas phase transition, chiral symmetry restoration, and the transition to deconfined quark matter~\cite{Dexheimer:2008ax,Dexheimer:2009hi,Hempel:2013tfa,Kumar:2024owe}, and has been successfully applied to neutron stars and heavy-ion phenomenology~\cite{Negreiros:2010hk,Dexheimer:2018dhb,Steinheimer:2009nn,Grefa:2026meq,Most:2018eaw,Steinheimer:2026xeg}. 
        In the MUSES Calculation Engine, \textbf{CMF++} is a \texttt{C++} implementation of the CMF model tailored to neutron-star applications, which provides in its current form the EoS at $T=0$ for cold and dense matter. It does not include meson condensation or nuclear cluster degrees of freedom and is therefore not applicable at $n_B$ below the nuclear liquid–gas transition or in regions of large $\mu_Q,\mu_S$. Therefore, the current applicability of the module is estimated to be $T=0$, $\mu_B \in [920, 2000]$ MeV,  $|\mu_Q| \lesssim m_\pi$ and $|\mu_S|\lesssim m_K$. Extensions to finite $T$ and more general $\mu_X$ conditions are being pursued but not yet implemented.
        See Ref.~\cite{Cruz-Camacho:2024odu} for details of the \texttt{C++} implementation and Ref.~\cite{zenodo_cmf} for the open-source code.

        \item \textbf{NJL:} the new Nambu--Jona-Lasinio (\textit{NJL}) module solves a three-flavor NJL model \cite{Rehberg:1995kh,Gastineau:2001zke,Klahn:2006iw} including effects of chiral symmetry breaking, color superconductivity, and a repulsive vector interaction self-consistently in the mean-field approximation. This module can be used to obtain EoSs for dense quark matter at zero and nonzero temperature and arbitrary fractions of electric charge and strangeness. 
        It can, for example, reproduce the quark-matter part of the EoSs in the quark-hadron continuity model \cite{Baym:2017whm,Baym:2019iky,Kojo:2021wax}, as well as the quark-matter part employed in the Maxwell-constructed equations of state of Refs.\ \cite{Gholami:2024ety,Christian:2025dhe}.
        Furthermore, the model can be calculated in a renormalization group consistent way \cite{Gholami:2024diy, Braun:2018svj}, such that cutoff artifacts are removed \cite{Gholami:2025guq}, providing more reliability at large values of temperature and chemical potentials.
        In principle, one can calculate the NJL model across any range of $(T,\mu_X)$; The module's applicability is, $T\in [0, 600]$ MeV, $\mu_B \in[ 0, 3000]$ MeV, $\mu_Q\in[ -400, 400]$ MeV and $\mu_S\in [ -400, 400]$ MeV. 
        Note, however, that our NJL model does not describe the confined phase, and its range of applicability is limited for temperatures at and above the hadron-quark phase transition and/or large $\mu_B$. Furthermore, at the current stage, the module does not account for the condensation of pseudoscalar meson or diquark condensates, which is expected to be preferred in some part of this $(T,\mu_B,\mu_S,\mu_Q)$ range. The inclusion of pseudoscalar condensates in an RG-consistent manner is planned to be included following Ref.\ \cite{Gholami:2026qpi}.
        See Ref.~\cite{zenodo_njl} for the open-source code.
        
        \item \textbf{pQCD:}
        at vanishing $T$ and high enough $n_B$, QCD matter becomes weakly coupled and perturbative QCD (pQCD) calculations are applicable. In the new \textit{pQCD} MUSES module, 2+1 flavor pQCD calculations at next-to-leading-order \cite{Graf:2015tda} with realistic quark masses at $T=0$ are included, allowing calculations for general ($\mu_B, \mu_S, \mu_Q$) \cite{Danhoni:2025qpn}. Although the code can be evaluated at broad ranges of chemical potentials, the pQCD module is only valid in the weak-coupling regime, which varies with the renormalization scale $\Lambda(\mu_B, \mu_S, \mu_Q)$ whose functional form is determined by the user. Note that when the strange quark mass $m_s\neq 0$ and/or depending on the choice of $\Lambda$, the pQCD EoS will have a flavor dependence.  Therefore, the module is mostly applicable at extremely large densities of $n_B\in[40n_{sat},\infty]$, with $|\mu_{Q, S}| < \infty$. 
        See Ref.~\cite{reinke_pelicer_2026_19265257} for the open-source code.
        
        \item \textbf{Meta-model:} 
        the \textit{Meta-model} module provides a flexible nucleonic parametrization of the EoS in terms of $(n_B, \delta_Q)$, based on the empirical bulk nuclear matter properties, such as $\{n_{\mathrm{sat}}, E/A, K_0, E_\mathrm{sym}, L_\mathrm{sym}, K_\mathrm{sym},...\} $, following the formalism of~\cite{Margueron:2017eqc,Margueron:2017lup}. In this approach, the total nucleon energy per particle, $E/A$, is written as the sum of an analytically computed relativistic kinetic term plus a potential given by a power series in $n_B$ and isospin asymmetry $\delta_Q$ around $n_{\mathrm sat}$ and $\delta_Q = 0$. The coefficients of the potential are fixed by matching them to the empirical nuclear matter parameters. This makes the module particularly useful for systematically exploring how uncertainties in bulk nuclear matter parameters propagate to the EoS and neutron-star observables~\cite{Tews:2018iwm,Carreau:2019zdy,Tsang:2023vhh,Yang:2025yoo,HegadeKR:2026iou}. The current implementation does not include nuclei or crustal clustering, so it should be used only beyond the liquid-gas transition. Additionally, it does not allow for strangeness, nor quark degrees of freedom. While the meta-model is only a controlled expansion around saturation, its range of applicability can be estimated to be, roughly, $n_B \in [n_{\mathrm LG}, 4 n_{\mathrm{sat}}]$, for isospin asymmetries in $\delta_Q \in [0, 1]$. 
        
        \item \textbf{Lepton module:} leptons are essential for the description of stellar matter, in particular the EoS for neutron stars and white dwarfs. The \textit{Lepton} module 
        can i) produce the EoS for a free Fermi relativistic gas of leptons, ii) read a table with thermodynamic quantities associated with hadronic or quark matter and add leptons to enforce charge neutrality, or iii) read a table with thermodynamic quantities associated with hadronic or quark matter and add leptons to enforce charge neutrality and $\beta$-equilibrium. In the latter case, regardless of the input EoS table being gridded in  ($T, \mu_B, \mu_Q$) or ($T, n_B, Y_Q$), the module can compute the $\beta$-equilibrium trajectory $\mu_Q(T, n_B)$. The leptons in the module include electrons, muons, taus, and neutrinos (as well as their antiparticles) at zero and finite temperature. If trapped neutrinos are included, as in proto-neutron stars, the lepton fraction can also be fixed. For cases ii) and iii), the input table may be gridded in up to 4-dimension\footnote{This only works if strangeness is gridded in terms of $\mu_S$, but we still need to add the capability of finding the $\mu_S$ trajectory in a table gridded in $Y_S$. } ($T, \mu_B, \mu_S, \mu_Q)$, and the applicability of the resulting EoS is then determined by the validity of the input table. See Ref.~\cite{pelicer_2025_14654137} for the open-source code. 

\end{itemize}

\begin{table*}[!ht]
        \centering
\renewcommand{\arraystretch}{1.2}
        \begin{tabular}{|c|c|c|c|c|c|c|c|c|}
        \hline
        \textbf{Module} & 
        \textbf{$T$ (MeV)} & 
        \textbf{$\mu_B$ (MeV)} & 
        \textbf{$\mu_S$ (MeV)} & 
        \textbf{$\mu_Q$ (MeV)} & 
        \textbf{$n_B$ (fm$^{-3}$)} & 
        \textbf{$Y_Q$} & 
        \textbf{CP} & 
        \textbf{Link} 
        \\ \hline
\textbf{HRG} 
        & 0 to 300 & -1000 to 1000 & -$m_K$ to $m_K$ & -$2m_\pi$ to $2m_\pi$ & --- & --- & Yes\footnotemark[1] & TBA
        \\ \hline
\textbf{4D-Taylor (\textit{BQS})}
        & 30 to 600 & -450 to 450 & -450 to 450 & -450 to 450 & --- & --- & -- & \cite{jahan_2025_14639786}
        \\ \hline
\textbf{4D-TExS} 
        & 0 to 800 & -800 to 800 & -800 to 800 & -800 to 800 & --- & --- & -- & \cite{jahan_2025_16749257}
        \\ \hline
\textbf{Ising-2DTExS} 
        & 30 to 800 & -700 to 700 & --- & --- & --- & --- & Yes & \cite{kahangirwe_2025_14637802}
        \\ \hline
\textbf{NumRelHolo}
        & 20 to 800 & 0 to 1000  & --- & --- & --- & --- & Yes & \cite{yang_2026_19154307}
        \\ \hline
\textbf{Crust-DFT} 
& 0
        & --- & --- & --- & $2.0\times10^{-12}$ to $2.0$ & 0.01 to 0.7 & -- & \cite{steiner_2025_14714273}
        \\  \hline
\textbf{$\chi$-EFT} 
        & 0 & --- & --- & --- & $1.6\times10^{-4}$ to 0.36 & 0 to 0.5 & -- & \cite{zenodo_cheft}
        \\ \hline
\textbf{CMF++} 
        & 0 & 0 to $+\infty$ & $-\infty$ to $+\infty$ & $-\infty$ to $+\infty$ & --- & --- & Yes\footnotemark[1]\footnotemark[2] & \cite{zenodo_cmf}
        \\ \hline
\textbf{NJL}
        & 0 to 600 &0 to 3000 &$-400$ to $400$ & $-400$ to $400$ & --- & --- & Yes & \cite{zenodo_njl}
        \\ \hline
\textbf{pQCD} 
        & 0 & $1000$ to $10^4$ & $-\infty$ to $+\infty$ & $-\infty$ to $+\infty$ & --- & --- & -- & TBA
        \\ \hline
\textbf{Meta-model} 
        & 0 & --- & --- & --- & 0 to $+\infty$ & 0 to 0.5 & -- & TBA
        \\ \hline
\textbf{Lepton} 
        & 0 to $+\infty$ & --- & --- & $-\infty$ to $+\infty$ & --- & ---  & --- & \cite{pelicer_2025_14654137}
        \\ \hline
        \end{tabular}
\footnotetext[1]{\textit{This CP corresponds to the nuclear liquid-gas phase transition.}}
\footnotetext[2]{\textit{The CMF model presents a critical point for deconfinement at finite temperatures, a regime not yet available in \texttt{CMF++}}}
\caption{
        Comparison of the main features of the different EoS modules to be available in MUSES \textit{Calliope}: 
        covered ranges (for input parameters), presence of a critical point (CP), and links to code repositories 
        (for modules already published).
        Indicated ranges mostly overshoot the respective physical regimes of applicability, and are simply indicative of the code capabilities. For more details on the limits of applicability of the underlying models, we refer the reader to the respective module documentations.
}
        \label{tab:HI_modules}
    \end{table*}
    
\begin{figure*}[!ht]
        \centering
        \includegraphics[width=.49\linewidth]{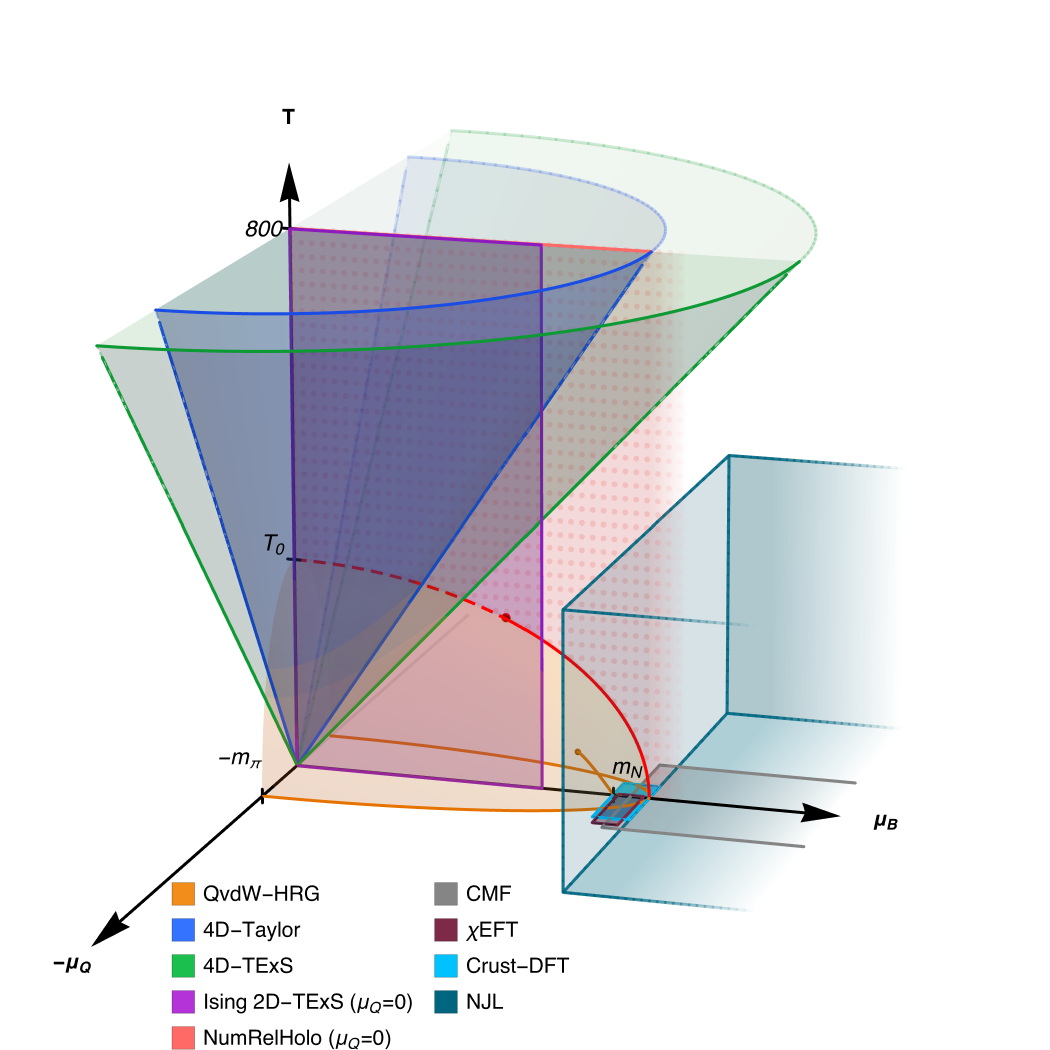}
        \includegraphics[width=.49\linewidth]{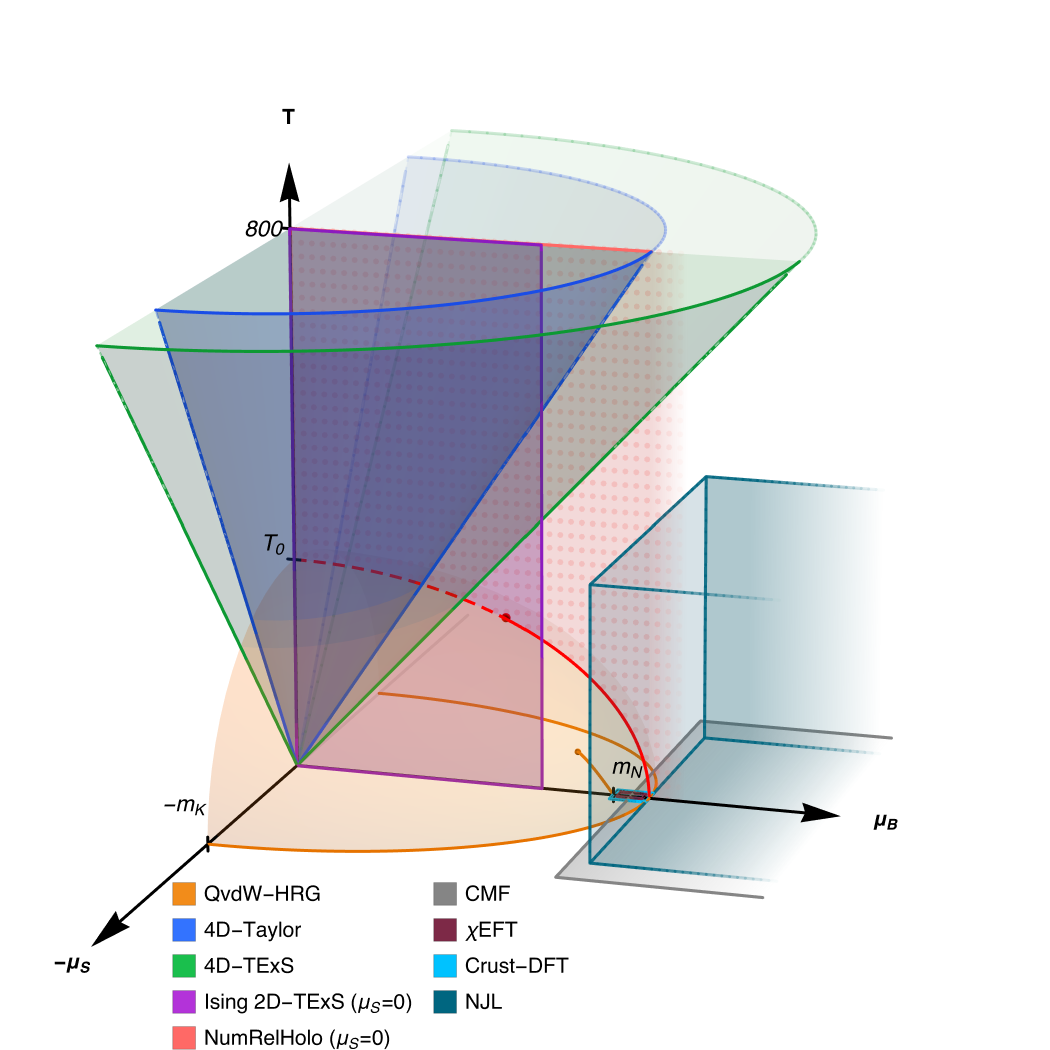}
        \caption{Schematics of the regions of validity 
        for heavy-ion and some new EoS modules of MUSES \textit{Calliope} discussed in this work, and $T\sim0$ neutron star modules introduced in Ref.~\cite{ReinkePelicer:2025vuh} as a comparison. The left panel shows finite $(T, \mu_B, \mu_Q)$, while the right panel shows finite $(T, \mu_B, \mu_S)$ subspace.
        The \textit{pQCD} module is not shown because it is valid at much larger chemical potentials (at $T=0$). 
        }
        \label{fig:modules_coverage}
    \end{figure*}
    
\item \textbf{Observables \& Tools:}
    \begin{itemize}[leftmargin=*]
        \item \textbf{Flavor Eq.:} given enough time (in the time scale of the weak force), the particle flavor in neutron stars equilibrates. The Flavor Equilibration (\textit{Flavor Eq.}) module calculates the weak equilibrium charge fraction, the flavor (isospin) relaxation rate, and the frequency-dependent bulk viscosity of nucleonic matter \cite{Alford:2023gxq,Alford:2024xfb}. It goes beyond the Fermi surface approximation to compute the direct Urca rates. At the moment, it cannot yet be coupled to EoSs that include hyperonic or quark degrees of freedom or merged EoSs (as the module requires microscopic information). The module works for input tables gridded in $(T, n_B, Y_Q)$, and imposes no restrictions on the ranges of the input table, and validity of the output inherits the validity of the input. See Ref.~\cite{zenodo_flavor_equil} for the open-source code. 
    
        \item \textbf{QLIMR:} 
        the Quadrupole, Love number, Inertia, Mass, Radius (\textit{QLIMR}) module computes key macroscopic observables of non-rotating or slowly rotating neutron stars, including mass–radius relations, tidal deformability, moments of inertia, and quadrupole moments. It is based on the TOV equations~\cite{Tolman:1939jz,Oppenheimer:1939ne} and the Hartle–Thorne slow-rotation expansion of the Einstein equations. It employs established quasi-universal I-Love-Q relations among neutron star properties \cite{Hartle:1968si, Flanagan:2007ix, Yagi:2013awa} to efficiently derive consistent stellar observables from a given EoS.
        See Ref.~\cite{conde_ocazionez_2025_14525356} for the open-source code.

        \item \textbf{Partial Pressures:}
        in the hadronic phase, the total pressure of the system can be represented as the sum of partial pressures based on hadrons, grouped through their baryon, strangeness, and electric charge content. Assuming an ideal-HRG scenario to describe $P(T,\mu_X)$, we obtain a relationship between linear combinations of diagonal and off-diagonal susceptibilities up to fourth-order, and the corresponding hadronic partial pressures at $\mu_X=0$ \cite{Gonzales:2026mfx}.  
        The \textit{Partial Pressures} module computes these linear combinations, based on state-of-the-art, continuum estimated susceptibilities from lattice QCD \cite{Abuali:2025tbd}, or from the HRG model. Users have the possibility to use their own set of susceptibilities as inputs.

        \item \textbf{Holographic transport coefficients:}
        the holographic EMD model provides a unified framework for investigating both the equilibrium and dynamical properties of the strongly coupled quark–gluon plasma. In addition to thermodynamic observables at finite $(T,\mu_B)$, it enables the computation of transport and energy-loss observables, which are otherwise directly inaccessible to first-principles lattice QCD calculations.
        Within MUSES, the \textit{NumRelHolo} module has been extended \cite{Khan:2026uqc} to include the calculation of transport coefficients in a fully self-consistent way using the same holographic background fields that determine the EoS. The module provides key quantities such as the shear and bulk viscosities, baryon and thermal conductivities, baryon diffusion coefficient, jet quenching parameter, heavy-quark drag force, and Langevin diffusion coefficients as functions of $(T,\mu_B)$, within the same ranges available for the EoS calculation \cite{Rougemont:2015ona,Rougemont:2015wca,Rougemont:2023gfz,Grefa:2022sav}.

        \item \textbf{Synthesis:}
        the \textit{Synthesis} module allows one to merge two different EoSs into a single one that covers a broader range of $(T,\mu_X)$. 
        The Synthesis module was originally developed to merge EoSs in the context of neutron star physics, which means $T\sim 0$ and finite $\mu_B\gtrsim930~\rm{MeV}$. The first version of the module allows the user to construct a first-order phase transition using either Maxwell or Gibbs phase constructions, or to smoothly match the EoSs with different matching functions~\cite{ReinkePelicer:2025vuh,ReinkePelicer:2025ado}.
        We are here extending the Synthesis module with a new merging procedure based on stability criteria and statistical mixture \cite{Yang:2026brr}, which works in any $(T,\mu_X)$ range. This more general framework can be tuned to produce a crossover, a first-order transition, or even a critical point separating the crossover from the first-order transition. Further details of the procedure are provided in Section \ref{sec:ext_of_syn}.  The validity of the resulting EoS is determined by the  union of the domain of validity of the input EoSs, given that the matching is well-behaved.

        \item \textbf{EoS Inverter:}
        the \textit{EoS Inverter} module allows inverting a table regularly gridded in $(T,\mu_X)$ into a table in terms of entropy regularly gridded in $(s,n_X)$ or $(\varepsilon,n_X)$ by solving a coupled non-linear system of equations. This is particularly relevant for hydrodynamical evolution, where each fluid cell evolves in terms of the densities of the conserved charges, and $s$ or $\varepsilon$. The module allows the user to choose different grid structures for the inverted table (linear, log, decade). The module falls back to a set of simplified analytic EoSs if no solution is found in the original table, ensuring the inverted table returns a causal and thermodynamically stable result\footnote{Causality implies that the speed of sound squared at constant entropy per baryon, keeping the other charge fractions fixed, is $\in [0,1]$ \cite{Gavassino:2023xkt}. Thermodynamical stability is the standard criterion that the entropy, in the presence of the relevant constraints, is maximum in thermodynamical equilibrium \cite{Landau:1980mil}.}. We do not yet support first-order phase transitions in the \textit{EoS Inverter} module but will explore this in future work. 
        For a more thorough discussion, see Ref.~\cite{Plumberg:2024leb}. 
    \end{itemize}    
\end{itemize}

Table~\ref{tab:HI_modules} compiles all the EoS modules which will be available in the MUSES Calculation Engine, \textit{\mbox{Calliope}} (to be released publicly upon journal acceptance), 
with their respective ranges of applicability based on input variables that the user can select. 
While some modules run as functions of $\mu_X$, others run in terms of $n_B$ and $Y_Q$. 
The existence of a critical point in the corresponding EoS is also indicated, and links to the corresponding Zenodo repositories are provided for modules already available on the MUSES Calculation Engine (links to new modules will be provided together with the upcoming public release of \textit{Calliope}, upon journal acceptance).  
Note that the indicated ranges are numerical limits set by the developers to access most of the regime of applicability of the underlying models, but do not reflect the actual physical limitations discussed above in most cases. 

In Fig.~\ref{fig:modules_coverage}, we show a schematic picture of the regions where our heavy-ion modules (the focus of this work) and current neutron star modules can be applied.
The two 3D plots represent finite $(T, \mu_B, \mu_Q)$ and $(T, \mu_B, \mu_S)$ subspaces, respectively.
They provide physics-motivated guidance on the regions of applicability, as a complement to the information provided in Table~\ref{tab:HI_modules}.
The deconfinement transition line in the $(T, \mu_B)$ plane is shown in red, crossing the crossover  established by lattice QCD calculations at $T_0 = 158.0$~MeV \cite{Borsanyi:2020fev}, transitioning at large finite $\mu_B$ into a first-order phase transition via a hypothetical critical point. The nuclear liquid-gas phase transition is also indicated in the $(T, \mu_B)$ plane, with the nucleon mass $m_N$ indicated as a visual reference. Similarly, we indicate the values of the pion mass $m_\pi$ and kaon mass $m_K$ to mark the start of meson condensation regions when $|\mu_Q| > m_\pi$ and $|\mu_S| > \mu_K$, respectively. 
More details on the regime of validity corresponding to each individual heavy-ion module are given in Sec.~\ref{sec:HI_EOS}.

 \section{Heavy-ion EoS modules}
\label{sec:HI_EOS}

In this section, we discuss in detail the different MUSES modules computing EoSs suited for heavy-ion collision modeling.
For each of them, we briefly introduce the underlying theory or effective models employed to compute the EoS of QCD matter and discuss the module capabilities and features. 

While each of these EoS modules produces an EoS table in \texttt{CSV} format that can be downloaded directly after its execution, the MUSES Calculation Engine now integrates multi-module workflows involving the heavy-ion EoS modules. Several of these workflows are described in detail in Sec.~\ref{sec:workflow}.
To facilitate the coupling between different modules, we employ the same default format for the ordering of the EoS tables produced by heavy-ion modules as the one defined in Ref.~\cite{ReinkePelicer:2025vuh} for neutron star EoS modules, storing all quantities as dimension-full, yet with some differences in the units used.

This default format fixes the ordering of thermodynamics variables $\varepsilon$, $P$, and first-order quantities, namely the first derivatives of $P$ with respect to $T$ and $\mu_X$. The column ordering is described in Table~\ref{tab:HI_eos_format}, together with the associated units. 
Note that, since some modules offer the possibility to store only the quantities of interest in the output EoS table through the use of \textit{switchers} (mostly to save disk memory), this ordering will change as some columns might be skipped. 
Notably, modules computing only 2D EoSs at finite $(T,\mu_B)$ do not have columns for $\mu_Q$ and $\mu_S$ by default, for instance. 
Finally, the ordering for quantities marked in gray (if computed by the module), i.e., second-order derivatives of $P$, isentropic speed of sound and heat capacity, is not systematically similar for all modules, but common to most of them (when computable). 
The composition and ordering of the data contained in the produced EoS table is always described in the associated output metadata file, generated by every MUSES module.

\begin{table}[!h]
\centering
\begin{tabular}{cll}
\toprule
\textbf{Col.} & \textbf{Quantity} & \textbf{Units (HI $\|$ NS)} \\ 
\midrule
    1 & Temperature ($T$) & MeV \\
    2 & Baryon chemical potential ($\mu_B$) & MeV \\
    3 & Strange chemical potential ($\mu_S$) & MeV \\
    4 & Charge chemical potential ($\mu_Q$) & MeV \\
    5 & Baryon density ($n_B$) & MeV$^{3}$ $\|$ fm$^{-3}$ \\
    6 & Strangeness density ($n_S$) & MeV$^{3}$ $\|$ fm$^{-3}$\\
    7 & Charge density ($n_Q$) & MeV$^{3}$ $\|$ fm$^{-3}$\\
    8 & Energy density ($\varepsilon$) & MeV$^{4}$ $\|$ MeV$\cdot$fm$^{-3}$\\
    9 & Pressure ($P$) & MeV$^{4}$ $\|$ MeV$\cdot$fm$^{-3}$\\
    10 & Entropy density ($s$) & MeV$^{3}$ $\|$ fm$^{-3}$\\
\arrayrulecolor{gray}
\midrule
    \gray{11} & \gray{$\chi_2^B$} & \gray{MeV$^{2}$} \\
    \gray{12} & \gray{$\chi_2^S$} & \gray{MeV$^{2}$} \\
    \gray{13} & \gray{$\chi_2^Q$} & \gray{MeV$^{2}$} \\
    \gray{14} & \gray{$\chi_{11}^{BS}$} & \gray{MeV$^{2}$} \\
    \gray{15} & \gray{$\chi_{11}^{BQ}$} & \gray{MeV$^{2}$} \\
    \gray{16} & \gray{$\chi_{11}^{SQ}$} & \gray{MeV$^{2}$} \\
    \gray{17} & \gray{$dn_B / dT$} & \gray{MeV$^{2}$} \\
    \gray{18} & \gray{$dn_S / dT$} & \gray{MeV$^{2}$} \\
    \gray{19} & \gray{$dn_Q / dT$} & \gray{MeV$^{2}$} \\
    \gray{20} & \gray{$ds / dT$} & \gray{MeV$^{2}$} \\
    \gray{21} & \gray{Isentropic speed of sound ($c_s^2$)} & \gray{\;\;---} \\
    \gray{22} & \gray{Heat capacity ($C_V$)} & \gray{MeV$^{3}$} \\
\arrayrulecolor{black}
\bottomrule
\end{tabular}
\caption{Default format of the output files for MUSES EoS modules. Different units used by heavy-ion (HI) and neutron star (NS) modules are specified.
Note that the ordering of quantities in gray is not set by default across all MUSES modules, but at least common to the majority of them, assuming they are computed. Mind as well that while the EoS tables store dimension-full susceptibilities, we do employ dimensionless susceptibilities throughout this paper.
}
\label{tab:HI_eos_format}
\end{table}

\subsection{Module for hadronic phase: HRG } 
\label{subsect:HRG}

The hadron resonance gas model describes QCD matter in the confined hadronic phase as a mixture of all established hadrons and resonances in thermal and chemical equilibrium. 
The HRG model reproduces lattice QCD thermodynamics up to $T\lesssim 150$–$160$ MeV \cite{Borsanyi:2011sw,Borsanyi:2013bia,HotQCD:2014kol,Alba:2017mqu,SanMartin:2023zhv}, and is a common model for describing the chemical freeze-out stage of heavy-ion collisions \cite{Yen:1997rv,Braun-Munzinger:1999hun,Becattini:2000jw,Braun-Munzinger:2001hwo,Becattini:2005xt,Andronic:2008gu,Alba:2014eba,Andronic:2017pug,Vovchenko:2016ebv,Vovchenko:2019pjl,Alba:2020jir}.

The MUSES Calculation Engine will incorporate \textsc{Thermal-FIST} (Thermal, Fast and Interactive Statistical Toolkit)~\cite{Vovchenko:2019pjl} version 1.6 as its \textit{HRG} module via a wrapper; the \texttt{C++} source (GPLv3) is available in Ref.~\cite{ThermalFISTGitHub}.
Thermal-FIST computes thermodynamics, hadron yields, and charge fluctuations in the full 4D space $(T,\mu_B,\mu_S,\mu_Q)$.
The baseline input consists of a particle list from PDG tables, with hadron masses, widths, spins, and quantum numbers. 
By default, the \textit{HRG} module uses the PDG2025 particles list \cite{ParticleDataGroup:2024cfk}, but the user can specify the path to another particle list of their choice instead, through the \texttt{particle\_list} option of \texttt{HRG\_parameters} in the module's YAML configuration file.

In the ideal HRG, the pressure is a sum over hadronic species
\begin{align}
P_{\rm id}(T,\{\mu\}) &= \sum_i p^{\rm id}_i(T,\mu_i), 
\end{align}
where $\tilde{\mu}_i$ is the effective chemical potential
\begin{align}
\label{mu-i}
\tilde{\mu}_i &= B_i\mu_B + Q_i\mu_Q + S_i\mu_S + C_i\mu_C,
\end{align}
with a contribution to the pressure from a single particle type $i$ given by
\begin{align}
\nonumber
p^{\rm id}_i(T,\mu_i) &= \frac{g_i}{6\pi^2}
\int_0^\infty \frac{k^4\,dk}{\sqrt{k^2+m_i^2}} \\[1ex]
&\times 
\left[\exp\!\left(\frac{\sqrt{k^2+m_i^2}-\tilde{\mu}_i}{T}\right)+\eta_i\right]^{-1},
\end{align}
where $g_i$ is the spin-isospin degeneracy of species $i$, $m_i$ is its mass, and $\eta_i=+1$ for fermions, $\eta_i=-1$ for bosons, $k$ is the momentum, and $\eta_i=0$ leads to the Boltzmann approximation.

Beyond the ideal gas, an excluded volume EoS is possible or both repulsive and attractive mean-field interactions are included within a 
multi-component QvdW-HRG approach~\cite{Vovchenko:2016rkn,Vovchenko:2017zpj}. 
Short-range repulsion is encoded by an excluded-volume matrix $\tilde{b}_{ij}$, while attraction enters via coefficients $a_{ij}$, characterizing the interactions for each pair of species $i$ and $j$. 
The grand canonical pressure reads
\begin{align}
\label{eq:vdw-p}
P(T,\{\mu\}) &= \sum_i p^{\!*}_i \;-\; \sum_{i,j} a_{ij}\,n_i n_j, 
\end{align}
where $p^{\!*}_i\equiv p^{\rm id}_i\big(T,\mu_i^{\!*}\big)$ denotes the ideal quantum-gas pressure (Fermi/Bose) of species $i$ and $\mu_i^{\!*}$ are the shifted chemical potentials due to both repulsion and attraction.
Calculations proceed by solving a transcendental system of equations for $\mu_i^{\!*}$:
\begin{align}
\label{eq:vdw-mu}
\mu_i^{\!*} &= \mu_i \;-\; \sum_j \tilde b_{ij}\,p^{\!*}_j + \sum_j (a_{ij}+a_{ji})\,n_j,
\end{align}
using a multi-dimensional Broyden method.
For a case when multiple solutions may exist (for instance, in a first-order phase transition region), multiple initial conditions are employed and the physical branch is selected by the Gibbs criterion of maximal pressure $P$.

The number densities are found from the linear system of equations 
\begin{align}
\sum_j\!\left(\delta_{ij} + \tilde b_{ji}\,n^{\!*}_i\right) n_j \;=\; n^{\!*}_i,
\qquad
n^{\!*}_i \equiv n^{\rm id}_i\!\big(T,\mu_i^{\!*}\big)
\label{eq:vdw-n}
\end{align}
where $\delta_{ij}$ is the Kroenecker delta. 
A common prescription defines $\tilde b_{ij}$ from hard-sphere radii $r_i$ via the geometric pair excluded volume,
\begin{equation}
\tilde b_{ij}=\frac{2\,b_{ii}\,b_{ij}}{b_{ii}+b_{jj}},\qquad 
b_{ij}=\frac{2\pi}{3}(r_i+r_j)^3,
\end{equation}
which reproduces the correct low-density virial limit for multi-component hard spheres~\cite{Gorenstein:1999ce}.
The \textit{diagonal EV} special case assigns each species an eigenvolume $v_i=\tfrac{16\pi}{3}r_i^3$ independently and sets $\tilde b_{ij}\equiv v_i$~\cite{Vovchenko:2016ebv}.
The attractive sector is specified by $a_{ij}$; in practice one restricts to (anti)baryon-(anti)baryon terms ($a_{BB'},a_{\bar B\bar B'}\neq0$), while setting baryon-antibaryon, meson-baryon, and meson-meson attractions to zero to avoid double counting of resonant channels already included explicitly.
The QvdW-HRG reduces to IHRG if both $a_{ij}$ and $b_{ij}$ vanish; and it reduces to  an excluded volume EoS if $a_{ij}=0$ but $b_{ij}\neq 0$.

The QvdW-HRG model reproduces key features of nuclear matter such as the liquid-gas phase transition and  its associated critical point. 
A widely used minimalistic implementation assumes that both repulsion and attraction act only between baryons (and, by symmetry, between antibaryons), with universal interaction parameters $a_{B B'}=a_{\bar B\bar B'}=a$ and $\tilde b_{B B'}=\tilde b_{\bar B\bar B'}=b$ for all (anti-)baryon pairs.
This two-parameter realization of the QvdW-HRG is sufficient to reproduce the empirical properties of symmetric nuclear matter at $T=0$, such as the binding energy per nucleon and saturation density, with typical choices $a = 0.329\,\text{GeV}.\text{fm}^3$ and $b = 3.42\,\text{fm}^3$~\cite{Vovchenko:2015vxa,Vovchenko:2015pya}.
The model yields a nuclear liquid-gas phase transition, with a critical point at $T_c \approx 19.7\,\text{MeV}$ 
and $n_c \approx 0.07\,\text{fm}^{-3}$, in fair agreement with phenomenological constraints from nuclear multi-fragmentation~\cite{Elliott:2013pna}. 
The resulting EoS also provides a fair description of lattice QCD thermodynamics at $\mu_B = 0$, 
including susceptibilities of conserved charges~\cite{Vovchenko:2016rkn}.

The MUSES framework utilizes the QvdW-HRG model from Thermal-FIST and allows the user to specify the (anti-)baryon interaction parameters for attraction and repulsion, $a$ and $b$, as well as a common (anti-) meson excluded-volume parameter $b_M$.
By construction, $a_{\bar{B}\bar{B}} = a_{BB}=a$ and $b_{\bar{B}\bar{B}} = b_{BB}=b$, and $b_M$ is applied identically to mesons and anti-mesons.
The motivation for a non-zero $b_M$ is to suppress hadronic degrees of freedom at high temperatures, where the QGP is expected.
This option is useful for a smooth merging of the hadronic EoS with other MUSES modules in the crossover region, 
like \textit{NumRelHolo} (see Sec.~\ref{sec:holography}). Assigning small but nonzero values, typically 
$b_{M}\!\sim\!0.1$-$0.3~\mathrm{fm}^3$ (corresponding to 
$r_M\!\sim\!0.15$-$0.25~\mathrm{fm}$) reduces the total $P$ and $s$ at high $T$ without spoiling the existing agreement of the HRG model with lattice data at lower $T$ (see, e.g., \cite{Vovchenko:2014pka,McLaughlin:2021dph}).
The van der Waals interactions (both attractive and repulsive), in all other sectors, specifically baryon-antibaryon, meson-baryon, and meson-antimeson, are set to zero.

In addition to standard thermodynamic properties, such as the $P$, $\varepsilon$, and $s$, the module outputs the Hessian matrix of the pressure in the grand canonical ensemble:
\begin{equation}
\label{eq:hessian}
H
=
\begin{pmatrix}
\dfrac{\partial^2 P}{\partial T^2} & 
\dfrac{\partial^2 P}{\partial T\,\partial \mu_Y}\\[4pt]
\dfrac{\partial^2 P}{\partial \mu_X\,\partial T} & 
\dfrac{\partial^2 P}{\partial \mu_X\,\partial \mu_Y}
\end{pmatrix}
\equiv
\begin{pmatrix}
\partial s/\partial T &
\alpha_Y\\[4pt]
\alpha_X & 
\chi^{XY}
\end{pmatrix}.
\end{equation}
Here 
$\alpha_X=(\partial n_X/\partial T)_{\mu}$ are thermal expansion coefficients, evaluated at constant chemical potentials. 
The bottom right corner of the Hessian matrix corresponds to
second-order net-charge susceptibilities for the three conserved charges $\chi_2^B$, $\chi_2^Q$, $\chi_2^S$ including cross terms, namely $\chi_{11}^{BQ}$, $\chi_{11}^{BS}$, and $\chi_{11}^{QS}$.
In the general case of arbitrary order, the net-charge susceptibilities are defined as:
\begin{align}
    \chi^{BQS}_{ijk}(T) =  \frac{\partial^{i+j+k} (P/T^4)}{\partial \hat{\mu}_B^i \, \partial \hat{\mu}_Q^j \, \partial \hat{\mu}_S^k}. \label{eq:susceptibilities}
\end{align}
For the ideal HRG, the derivatives in Eq.~\eqref{eq:hessian} are obtained by sums over species, since 
$P=\sum_i p_i^{\rm id}(T,\mu_i)$ with $\mu_i$ given by Eq.~\eqref{mu-i}.
In the QvdW-HRG, the Hessian requires differentiation of the implicit interacting system (\ref{eq:vdw-p}-\ref{eq:vdw-mu}) with respect to $T$ and 
$\mu_X$
and solving the resulting linear response equations.

The MUSES \textit{HRG} module can be used to compute the hadronic EoS for $T\in[0, 300]$ MeV, and for \mbox{$|\mu_B| \leq 1000$} MeV,  \mbox{$|\mu_Q| \leq 2 m_\pi (\sim 300$} MeV) and \mbox{$|\mu_S| \leq m_K (\sim 493$} MeV).
It takes as input a grid of ($T$, $\mu_B$, $\mu_S$,  $\mu_Q$) values, 
specified in \texttt{parameters} of the YAML configuration file through a triplet \texttt{x\_min}, \texttt{x\_max} and \texttt{dx} for each of the four variables $\texttt{x}\in[\texttt{T},\,\texttt{mu\_B},\,\texttt{mu\_S},\, \texttt{mu\_Q}]$.    
In \texttt{HRG\_parameters}, the values of the van der Waals parameters are set through \texttt{a}, \texttt{b} and \texttt{bm} respectively. Additionally, one can decide to treat pions as interacting particles, using an effective mass model matched to chiral perturbation theory for pions at $T=0$ by setting \texttt{interacting-pions} to \texttt{true}. 
This allows for a description of the pion condensation region reasonably well up to a value of $\mu_Q$ around twice the pion mass \cite{Vovchenko:2020crk}.
Finally, \texttt{switchers} allow the user to select which thermodynamic quantities they want to be computed and stored in the output EoS table, by setting any of the \texttt{get\_y} to \texttt{true} of \texttt{false}, with 
$\texttt{y} \in [\texttt{P}, \texttt{E\_dens}, \texttt{s\_dens}, \texttt{B\_dens}, \texttt{Q\_dens}, \texttt{S\_dens}, \allowbreak \texttt{PressHessianMatrix}]$. Note that the latter will also trigger the calculation and storage of isentropic $c_s^2$ and $C_V$.

Additional \textsc{Thermal-FIST} capabilities, which may be incorporated into future MUSES updates, include: species- and flavor-dependent interaction parameters $\{\tilde b_{ij},a_{ij}\}$, such as dedicated isospin~\cite{Poberezhnyuk:2018mwt}, strangeness~\cite{Vovchenko:2016rkn}, or charm~\cite{Goswami:2025pnr} sectors enabling different strengths of repulsion or attraction by hadron family; light nuclei (d, t, ${}^3$He, $\alpha$, \dots) to be included as additional HRG species with independent interaction parameters, permitting a consistent description of clusterized matter; finite resonance widths via Breit--Wigner mass distributions which affect thermodynamics.
In addition, dense-matter generalizations of QvdW through extensions to density-dependent mean fields and excluded volume can improve the high-density behavior of the HRG EoS, and potentially allow for a smoother merging with other modules. 
Finally, \textsc{Thermal-FIST} can compute fluctuations and correlations of conserved charges up to fourth order, including charm, and natively includes cumulants of non-conserved quantities (in particular particle yields) up to second order, which will be added in future updates.

\subsection{Lattice QCD modules}
\label{subsec:lattice-EOSs}

\subsubsection{Expansion schemes at finite $\mu_X$}
\label{subsubsec:lattice-exp}

Lattice QCD calculations are presently restricted to vanishing real chemical potentials due to the sign problem (with the exception of finite isospin chemical potential, which does not suffer from it \cite{Kogut:2002zg, Kogut:2002tm, Brandt:2017oyy}). 
To extend the EoS to finite conserved-charge chemical potentials $\mu_X$, we employ expansion schemes built on $2{+}1$-flavor lattice QCD results. 

The expansion schemes take as input the conserved-charge susceptibilities defined in Eq.~\eqref{eq:susceptibilities}. We combine and fit these susceptibilities across datasets by matching to HRG at low $T$, and to lattice QCD results together with their Stefan--Boltzmann limits at high $T$, using both an older $N_\tau=12$ set \cite{Borsanyi:2018grb} and more recent continuum-estimated results for the second- and fourth-order terms ($\chi^{BSQ}_{ijk}$ with $i+j+k=2,4$) \cite{jahan_2025_16749145}. These fitted susceptibilities provide the foundation for the three schemes described below.
\begin{itemize}
    \item {\it Taylor expansion.} 
 The pressure is given as an expansion around $\hat{\mu}_{B,Q,S} = 0$:
 \begingroup\small
\begin{equation}
    \frac{P(T, \hat{\mu}_B, \hat{\mu}_S, \hat{\mu}_Q)}{T^4} = \sum_{i,j,k \geq 0} \frac{\chi^{BSQ}_{ijk}(T)}{i! \, j! \, k!} \,  \, (\hat{\mu}_B)^i (\hat{\mu}_S)^j (\hat{\mu}_Q)^k.
\label{eq:Taylor}
\end{equation}
\endgroup

In practice, the series is truncated at a finite total order in $\hat{\mu}_X$, which is equal to four in our case.

\item {\it $T^\prime$-Expansion Scheme.} Informed by simulations at imaginary chemical potential and explorations along lines of constant baryon density over $\mu_B$, this method \cite{Borsanyi:2021sxv} shifts $(T, \hat{\mu}_B)$ to an effective temperature $T'(T, \hat{\mu}_B)$ such that
\begin{align}
    \frac{n_B(T, \mu_B)}{T^3} = \hat{\mu}_B \, \chi_2^B \big( T'(T, \hat{\mu}_B), 0 \big)
    \label{eq:Tprime_baryon_density}
\end{align}
with an expansion in $\hat{\mu}_B^2$:
\begingroup\small
\begin{equation}
    T'(T, \hat{\mu}_B) = T \Big[ 1 + \kappa_2^B(T) \, \hat{\mu}_B^2 + \kappa_4^B(T) \, \hat{\mu}_B^4 + \cdots \Big]. 
    \hspace{-.6cm}
    \label{eq:Tprime}
\end{equation}
\endgroup
The coefficients $\kappa_{2,4}^B(T)$ can be determined directly on the lattice (e.g., from imaginary-$\mu_B$ data) or matched order-by-order to the Taylor susceptibilities.

\item {\it $T^\prime$ expansion scheme with Stefan-Boltzmann (SB) normalization.} 
To improve the asymptotic behavior at high $T$ and beyond the purely baryonic sector, this variant \cite{Borsanyi:2022qlh} enforces the correct SB limit by rescaling with free-gas susceptibilities:
\begin{align}
    \frac{n_B(T, \mu_B)}{T^3} = \frac{\bar{\chi}_1^B(\hat{\mu}_B)}{\bar{\chi}_2^B(0)} \, \chi_2^B \big( T'(T, \hat{\mu}_B), 0 \big),
  \label{Eq:TprimenB_SB}
\end{align}
where overbars denote SB-limit quantities, and the scaled temperature is given by
\begingroup\small
\begin{align}
    T'(T, \hat{\mu}_B) = T \Big[ 1 + \lambda_2^B(T) \, \hat{\mu}_B^2 + \lambda_4^B(T) \, \hat{\mu}_B^4 + \cdots \Big].
    \hspace{-.6cm}
    \label{eq:TprimeSB}
\end{align}
\endgroup
The coefficients $\lambda_{2,4}^B(T)$  can be determined the same way as $\kappa_{2,4}^B$. 
\end{itemize}
The convergence properties of these extrapolation schemes across different effective models are detailed in Ref.\ \cite{Kahangirwe:2024xyl}.

These expansion schemes are subsequently used in Sections \ref{subsubsec:BQS}, \ref{subsubsec:4D-TExS}, and \ref{subsubsec:Ising2D-TExS} to construct the EoS at finite density.

\subsubsection{4D-Taylor expansion}
\label{subsubsec:BQS}

The \textit{4D-Taylor (BQS)} lattice expansion module constructs a 4D EoS for QCD at finite $(T,~\mu_B,~\mu_S,~\mu_Q)$, using lattice QCD results at vanishing chemical potential. 
This is achieved by expanding the pressure as a Taylor series in powers of ($\hat{\mu}_B,~\hat{\mu}_S,~\hat{\mu}_Q$), as described in Eq. \eqref{eq:Taylor}.

The first term in the expansion corresponds to $i=j=k = 0 $ which is $P$ calculated on the lattice at $\mu_X=0$, while the other finite-order Taylor expansion coefficients are the conserved charge susceptibilities defined in Eq.\eqref{eq:susceptibilities}. In the MUSES \textit{BQS} module, we only utilize the susceptibilities with $i+j+k \leq 4$. 
Since lattice QCD-based susceptibilities do not cover the low $T$ limit required for the hydrodynamical evolution of the system, they are smoothly merged to HRG model results (See \ref{subsect:HRG}). At high $T$, a smooth approach to the Stefan-Boltzmann limit is also imposed. Once this is done, we parameterize these coefficients in a range of $30\leq T \leq 600$ MeV as follows:
\begin{align}
\chi_{ijk}^{BSQ}(T)
&=
\frac{
\sum_{n=0}^{9} a_n^i / x^n
}{
\sum_{n=0}^{9} b_n^i / x^n
}
+ c_0 ,
\label{eq:chis_bqs_fit}
\end{align}
This parametrization works well for almost all coefficients $a_n^i, b_n^i$, and $c_0$. However, the optimal functional form of the parameterization for $\chi_2^B$ was taken from Ref. \cite{Parotto:2018pwx}:
\begin{equation}
\chi_{2}^{B}(T) = 
e^{-h_{1}/x' - h_{2}/x'^{2}}
\cdot f_{3} \cdot 
\left(1 + \tanh(f_{4}x' + f_{5})\right).
\label{eq:chi2B_fit}
\end{equation}
In Eqs.\ (\ref{eq:chis_bqs_fit}-\ref{eq:chi2B_fit}), $x = T/154$ MeV and $x'= T/200$ MeV, respectively, while the best fit for the parameters is obtained by performing a $\chi^2$ minimization process. 

The \textit{BQS} module provides two different input parameter choices for the parameterization, depending on the lattice-based susceptibility results that the user wants to choose, which can be selected in the YAML configuration file with \texttt{coef\_file}.
The default choice is the parameters for the continuum estimates of the susceptibilities obtained with the 4stout action with physical quark masses \cite{jahan_2025_16749145}, recently published in \cite{Abuali:2025tbd}, which can be selected by setting \texttt{coef\_file} to \texttt{2025}.
The second choice for the parameters corresponds to the susceptibilities results from lattice QCD simulations at finite lattice spacing, with spatial dimension $N_S = 48$ and temporal dimension $N_{\tau} = 12$ in the temperature range of $135\leq T \leq 220$, from Ref. \cite{Borsanyi:2018grb}, usable by setting \texttt{coef\_file} to \texttt{2019}.
The reason for this last option is to give the user the possibility to reproduce the original 4D Taylor expansion parametrization from \cite{Noronha-Hostler:2019ayj}. 
We discuss the results of the 4D-Taylor expansion, showed in Fig.~\ref{fig:4D-lattice-EOSs}, in 
Sec.~\ref{subsubsec:4D-TExS}.

In the \textit{BQS} module, the user can specify the range and step size for $(T,\mu_B,\mu_S,\mu_Q)$ in their input, 
using the exact same keys in \texttt{parameters} as for the \textit{HRG} module (see Sec. \ref{subsect:HRG}).
The allowed range for obtaining the EoS in this module is $T\in[30,600]$ MeV, and  $\mu_X\in[-450,450]$ MeV for all chemical potentials. 
Note however that the expansion is estimated to be valid up to $\hat{\mu}_X \sim 2.5$ for all three chemical potentials. 
Similarly to the \textit{HRG} module, the \texttt{switchers} allow one to select which observables to store in the produced EoS table. 
The list of \texttt{get\_y} quantities comprises the usual $\texttt{y} \in [\texttt{P}, \texttt{E\_dens}, \texttt{s\_dens}, \texttt{B\_dens}, \texttt{Q\_dens}, \texttt{S\_dens}]$, but also \texttt{get\_c\_s} for the isentropic speed of sound, and \texttt{get\_P\_derivatives\_finite\_mu} to store all terms from the Hessian matrix of pressure at finite $\mu_X$, as defined in Eq.~\eqref{eq:hessian}.
In addition, \texttt{get\_Chis\_zero\_mu}, \texttt{get\_dChisdT\_zero\_mu} and \texttt{get\_d2ChisdT2\_zero\_mu} indicate if the susceptibilities, their first- and second-order $T$-derivatives calculated at $\mu_X=0$, respectively, should be stored (in a separate file than for the EoS results, if applicable). 

\subsubsection{4D-TExS}
\label{subsubsec:4D-TExS}

The \textit{4D-TExS} module constructs another 4D lattice QCD-based EoS at finite \((T,\mu_B,\mu_S,\mu_Q)\), by generalizing the 2D \((T,\mu_B)\) $T^\prime$-Expansion Scheme of Eq.~\eqref{Eq:TprimenB_SB} to arbitrary directions in the three chemical potential space. The construction maps the cartesian coordinates \((\hat{\mu}_B,\hat{\mu}_Q,\hat{\mu}_S)\) to the corresponding spherical coordinates \((\hat{\mu},\theta,\varphi)\) via
\begin{align} \nonumber     \label{eq:mustoangles}
    \hat{\mu}_B &= \hat{\mu}\, c_{\theta}\, ,\\
    \hat{\mu}_Q &= \hat{\mu}\, s_{\theta} c_{\varphi}\, , \\ \nonumber
    \hat{\mu}_S &= \hat{\mu}\, s_{\theta} s_{\varphi}\, ,
\end{align}
where \(s_{\gamma} \equiv \sin\gamma\) and \(c_{\gamma} \equiv \cos\gamma\). The inverse relations read:
\begin{align} \nonumber     \label{eq:anglestomus}
    \hat{\mu} &= \sqrt{\hat{\mu}_B^2 + \hat{\mu}_Q^2 + \hat{\mu}_S^2}\, , \\
    \theta  &= \arccos\!\left(\hat{\mu}_B/\hat{\mu}\right)\, , \\ 
    \varphi &= \operatorname{sgn}(\hat{\mu}_S)\,\arccos\!\left(\hat{\mu}_Q/\sqrt{\hat{\mu}_Q^2 + {\hat{\mu}_S}^2}\right)\, . \nonumber
\end{align}

With this prescription, any given triplet \((\hat{\mu}_B,\hat{\mu}_Q,\hat{\mu}_S)\) fixes the angles \((\theta,\varphi)\), so the problem reduces to a 1D extrapolation in the generalized chemical potential \(\hat{\mu}\) along that direction. In analogy with Eq.~\eqref{Eq:TprimenB_SB}, the generalized expansion reads \cite{Abuali:2025tbd}:
\begin{align}
    X_1^{\theta,\varphi}(T, \hat{\mu}) =
    \frac{\overline{X}_1^{\theta,\varphi}(\hat{\mu})}{\overline{X}_2^{\theta,\varphi}(0)}\,
    X_2^{\theta,\varphi}\!\left(T^{\, \prime \, \theta,\varphi}(T,\hat{\mu}), 0\right),
    \label{eq:4D-TExS}
\end{align}
where \(X_1^{\theta,\varphi}\) and \(X_2^{\theta,\varphi}\) denote, respectively, the generalized density and the second-order generalized susceptibility along the direction \((\theta,\varphi)\), with bars indicating Stefan-Boltzmann limits and their argument specifying the generalized chemical potential at which they are calculated. The effective temperature is defined, in analogy with Eq.~\eqref{eq:TprimeSB}, as:
\begin{equation}
    T^{\, \prime \, \theta,\varphi}(T,\hat{\mu}) = T\!\left[ 1 + \lambda_{2}^{\theta,\varphi}(T)\,\hat{\mu}^2 + \ldots \right] \, ,
    \label{eq:4D-Tprime}
\end{equation}
with the expansion coefficient
\begin{align}
    \lambda_{2}^{\theta,\varphi}(T)
    &= \frac{1}{6 T \left(X_2^{\theta,\varphi}(T)\right)^{\prime}} \!
       \left[
         X_4^{\theta,\varphi}(T)
         - \frac{\overline{X}_4^{\theta,\varphi}(0)}{\overline{X}_2^{\theta,\varphi}(0)}\,X_2^{\theta,\varphi}(T)
       \right] \! .
    \label{eq:generalizedlambda}
\end{align}
Here, \(X_4^{\theta,\varphi}(T)\) is the generalized fourth-order susceptibility along \((\theta,\varphi)\). Whenever \(\hat{\mu}=0\), we suppress the explicit \(\hat{\mu}\)-dependence for brevity.

The quantities \(X_n^{\theta,\varphi}(T)\) generalize the usual susceptibilities \(\chi_n(T)\) as directional derivatives of $P$ with respect to the direction-dependent chemical potential \(\hat{\mu}\):
\begin{equation}
    X_n^{\theta,\varphi}(T) = \left.\frac{\partial^n (P/T^4)}{\partial \hat{\mu}^n}\right|_{\hat{\mu}=0}^{\theta,\varphi},
\end{equation}
i.e., derivatives taken along the direction specified by \((\theta,\varphi)\). At \(\hat{\mu}=0\), the \(X_n^{\theta,\varphi}\) relate to the standard susceptibilities \(\chi_{ijk}^{BSQ}(T)\) by straightforward application of the chain rule as detailed in \cite{Abuali:2025tbd}.

The  Stefan-Boltzmann limit of the generalized density at finite \(\hat{\mu}\), \(\overline{X}_1^{\theta,\varphi}(\hat{\mu})\), follows from the Taylor-expanded $P$ in Eq.~\eqref{eq:Taylor}:
\begin{equation}
\label{eq:X1_SBL}
    \overline{X}_1^{\theta,\varphi}(\hat{\mu}) 
    = \hat{\mu}\,\overline{X}_2^{\theta,\varphi}
    + \frac{\hat{\mu}^3}{6}\,\overline{X}_4^{\theta,\varphi}.
\end{equation}

Here, the expansion is carried up to \(\lambda_2^{\theta,\varphi}(T)\), i.e., at leading order in the $T'$-expansion, which corresponds to next-to-leading order (NLO) in the Taylor expansion (i.e., including susceptibilities up to fourth order). The procedure generalizes straightforwardly to higher orders, provided higher-order susceptibilities are available.

Given fixed values of \((\hat{\mu}_B,\hat{\mu}_Q,\hat{\mu}_S)\) at any $T$, we proceed as follows:
\begin{enumerate}
    \item Determine \(\hat{\mu},\theta,\varphi\).
    \item Construct \(\lambda_2^{\theta,\varphi}(T)\) from \(X_2^{\theta,\varphi}(T)\) and \(X_4^{\theta,\varphi}(T)\) using Eq.~\eqref{eq:generalizedlambda}.
    \item Evaluate \(X_1^{\theta,\varphi}(T,\hat{\mu})\) using Eq.~\eqref{eq:4D-TExS}.
\end{enumerate}

Starting from the generalized density \(X_1(T,\hat{\mu})\), the pressure is obtained as
\begin{align} 
    \label{eq:pressure}
    \frac{P(T,\hat{\mu})}{T^4} &= \frac{P(T,0)}{T^4} + \int_0^{\hat{\mu}}  d\hat{\mu}^{\prime}\, X_1(T,\hat{\mu}^{\prime}) \, .
\end{align}
In a similar manner, all first-order quantities derived from pressure are computed from analytical expressions based on $X_1(T, \hat{\mu})$ \cite{Abuali:2025tbd}.

The \textit{4D-TExS} module can compute an EoS table for $T\in[0, 800]$ MeV and {$\mu_X\in[-800, 800]$} MeV. 
However, these ranges are not uniform in the whole 4D space: while the aforementioned $T$ range is limited by the provided $P(T,0)$ and $s(T,0)$ from lattice QCD calculations, computing the EoS at finite $\hat{\mu}$ requires to evaluate $X_2$ at $T^\prime(T, \hat{\mu})$, as shown in Eq.~\eqref{eq:4D-TExS}. 
Hence, in the regions where the value of $T^\prime(T, \hat{\mu})$ is out of the $T$ range available for the provided second-order susceptibilities (presently $0 \leq T \leq 2000$ MeV), calculations are skipped and all quantities are set to \texttt{nan}.
This mostly happens for large values of $\hat{\mu}$ and large absolute values of $\lambda_{2}(T)$, as can be seen from Eq.~\eqref{eq:4D-Tprime}.

\begin{figure}
    \centering
    \includegraphics[trim={0 0 0 .6cm},clip,width=.865\linewidth]{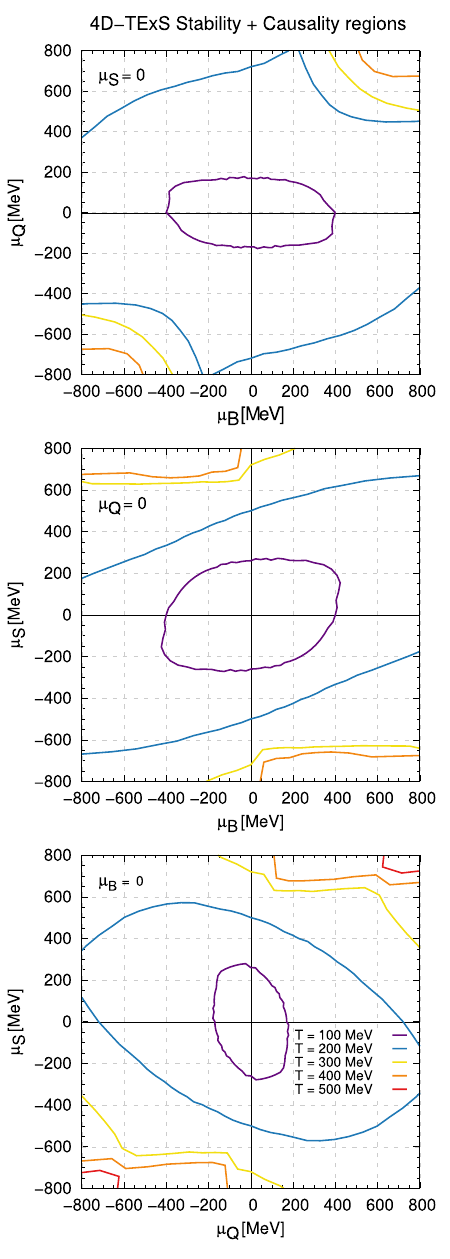}
    \caption{Contours of the onset of instability or acausality of the 4D-TExS EoS \cite{Abuali:2025tbd} in the $(\mu_B, \mu_Q)$ plane at $\mu_S = 0$ (top), $(\mu_B, \mu_S)$ plane at $\mu_Q = 0$ (middle) and $(\mu_Q, \mu_S)$ plane at $\mu_B = 0$ (bottom). 
    Different colors represent different constant temperatures, ranging from 100 MeV to 500 MeV. }
    \label{fig:4DTExS-stability}
\end{figure}

Note that, beyond the previous limitations, no criterion of validity is set to limit the calculation of the EoS with the \textit{4D-TExS} module. The $T^\prime$-Expansion Scheme is estimated to be valid up to $\hat{\mu} \sim 3.5$ at finite $\hat{\mu}_B$ \cite{Borsanyi:2021sxv}, but the user should verify where the obtained EoS is actually stable and causal.
The regions of stability and causality of the 4D-TExS EoS are shown in Fig.~\ref{fig:4DTExS-stability}, as functions of the three permutations of finite $(\mu_X, \mu_Y)$ at $\mu_Z=0$ (with $X,Y,Z \in B,Q,S$) and displayed within the ranges of computation available in the \mbox{\textit{4D-TExS}} module. 
The different contours represent different values of constant $T$, and surround the regions where the EoS is both causal and thermodynamically stable: to verify these conditions, we ensure respectively that $c_s^2 \in [0,1]$ (at constant particle fractions), and that the Hessian matrix of pressure is positive definite, which can be reduced to a 2D problem in the formalism of the 4D-TExS through the condition:
\begin{equation}
    \label{eq:4D-TExS_stability}
    X_2^{\theta,\varphi}(T,\hat{\mu}) \frac{ds(T,\hat{\mu})}{dT} - \left( \frac{dX_1^{\theta,\varphi}(T,\hat{\mu})}{dT}\right)^2 > 0
    \; .
\end{equation}
The region of stability and causality spans across larger values of $(\mu_X, \mu_Y)$ for all three planes when $T$ increases, as expected from the fact that the expansion is defined in terms of $\mu_X/T$. 
One observes that the 4D-TExS EoS seems to be valid across larger values of $\mu_B$ than $\mu_S$ and than $\mu_Q$, successively. Moreover, while not visible through the different panels of Fig.~\ref{fig:4DTExS-stability}, these regions of validity shrink when increasing the value of $\mu_Z$.
Estimating the region of validity of the $T^\prime$-Expansion Scheme truncated to $\mathcal{O}(\hat{\mu}_X^2)$ in 4D is currently limited by the large statistical errors on fourth- and sixth-order susceptibilities of $B,Q,S$ (and $T$-derivatives) needed to compute the next expansion coefficient $\lambda_4$ in Eq.~\ref{eq:4D-Tprime}. 
Hence, Fig.~\ref{fig:4DTExS-stability} is currently the only reference to estimate the regions of validity of the 4D-TExS EoS, although its description of the thermodynamics of nuclear matter might become less accurate when getting close to the limits of stability and causality.

To select the computation grid of the desired EoS, the \textit{4D-TExS} module relies on the exact same parameters as the \textit{BQS} module (see Sec.~\ref{subsubsec:BQS}) for its YAML configuration file.
In the same fashion than the two previous modules, a list of \texttt{switchers} of the form \texttt{get\_y} is used to select the observables to store in the produced EoS table, with $\texttt{y} \in [\texttt{P}, \texttt{E\_dens}, \texttt{s\_dens}, \texttt{B\_dens}, \texttt{Q\_dens}, \texttt{S\_dens}, \allowbreak \texttt{PressHessianMatrix}, \texttt{cs2}]$, the latter being used for  $c_s^2$.
The only addition with respect to the \textit{BQS} module is the \texttt{errors\_thermo} key, which can be set to \texttt{true} or \texttt{false} to indicate whether the user wants to store the errors on the computed thermodynamics quantities in the EoS table.
If not specified, this parameter is set to \texttt{false} by default to minimize the size of the output, but requesting the storage of the errors will not affect the computation time of the module. 
Note also that elements of the Hessian matrix of pressure are currently calculated from numerical derivatives: interpolated-spline derivatives for $T$-derivatives of entropy and charge densities, and finite difference of charge densities for susceptibilities. 
Future updates are already planned to introduce these calculations based on the actual analytical formulae for these quantities, to improve both the precision and speed of these calculations.
\\

\begin{figure*}
    \centering
    \includegraphics[width=1\linewidth]{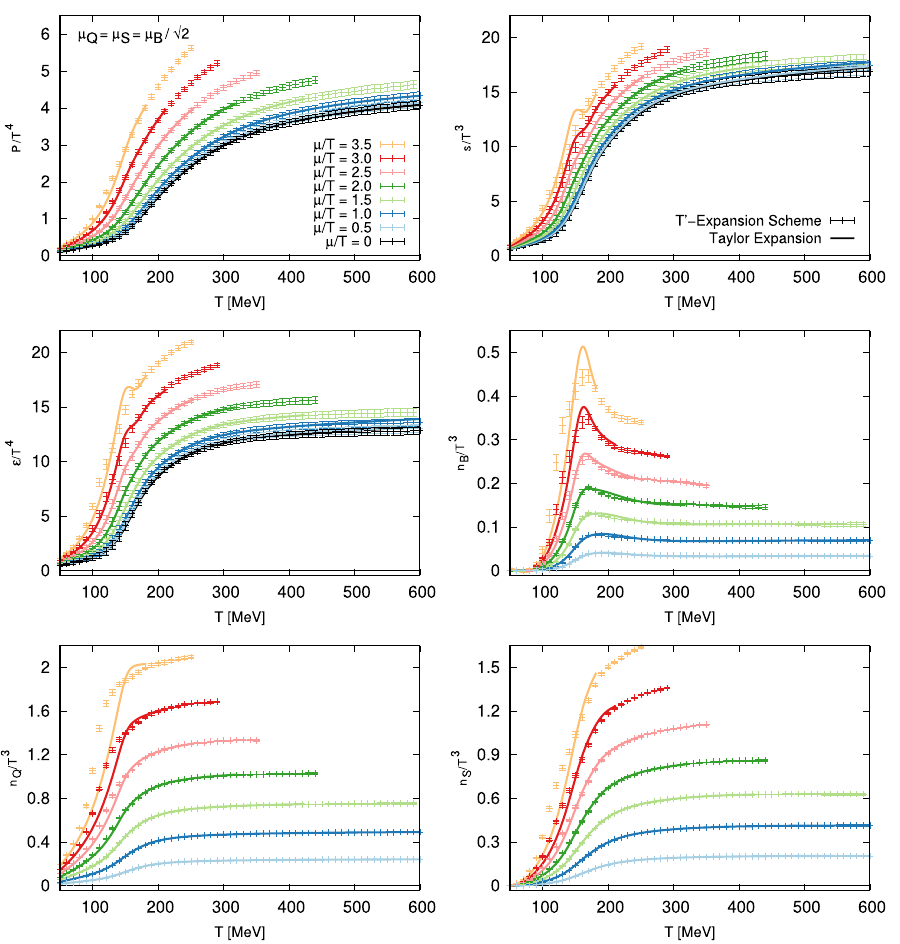}
    \caption{Normalized thermodynamic quantities: pressure, entropy, energy, baryon, electric charge, and strangeness densities as functions of the temperature, for increasing values of $\hat{\mu}$, in a direction where $\mu_Q = \mu_S = \mu_B/\sqrt{2}$.
    Results from the 4D-TExS EoS, displayed with points, are compared with the 4D Taylor expansion EoS, represented by continuous lines.
    }
    \label{fig:4D-lattice-EOSs}
\end{figure*}

Fig.~\ref{fig:4D-lattice-EOSs} shows a comparison of thermodynamic quantities as functions of $T$, namely the dimensionless  pressure, entropy density, energy density, and $n_X$, computed from both \textit{4D-TExS} and \textit{BQS} modules. 
Both sets of results are shown for values of generalized $\hat{\mu}$ ranging from 0 to 3.5, along a trajectory where $\mu_Q = \mu_S = \mu_B/\sqrt{2}$, and up to the maximum chemical potentials computable by each corresponding module.
Results from the \textit{4D-TExS} modules are displayed using thick points with error-bars, which are derived using jackknife statistics from the set of sampled continuum-estimated susceptibilities described in Sec.~\ref{subsubsec:lattice-exp} \cite{jahan_2025_16749145}.
Results from the \textit{BQS} module, obtained with the same set of susceptibilities (using the default option \texttt{coef\_file=2025}), are represented by continuous lines.
Both expansions are consistent with each other within error bars for all thermodynamic quantities displayed, and only start to show discrepancies for $\hat{\mu} \geq 3.0$. 
These differences are of course expected, since the Taylor expansion truncated at fourth-order is only reliable up to $\hat{\mu} \sim 2.5$, as explained in Sec.~\ref{subsubsec:BQS}. 
Beyond this value, one can clearly observe its breakdown as oscillations start to appear in the entropy around the transition region, which might eventually lead to non-monotonicity as a function of $T$, signaling an unstable EoS.

\subsection{Phenomenological modules for the \\ transition region}

\subsubsection{Ising 2D-TExS}
\label{subsubsec:Ising2D-TExS}
The \textit{Ising 2D-TExS} module extends the finite-$\mu_B$ coverage in the $(T,\mu_B)$ plane and models the deconfinement transition by combining the 2D lattice $T^\prime$-expansion with 3D-Ising critical scaling mapped to QCD, thereby improving upon the Taylor-based framework~\cite{Parotto:2018pwx}. In the $T^\prime$-Expansion Scheme, $n_B$ is written as Eq.~\eqref{eq:Tprime_baryon_density} with finite-$n_B$ effects encapsulated in the expansion of $T'(T,\hat{\mu}_B)$ in powers of $\hat{\mu}_B$ given in Eq. \eqref{eq:Tprime}. The critical point is incorporated by embedding its singular contribution directly into $T'(T,\hat{\mu}_B)$, as detailed below.

Lattice QCD provides $\chi_2^B(T)$, $\kappa_2^B(T)$, and $P(T,0)$ at discrete values of $T$. To obtain a smooth EoS over the range
$25~\text{MeV}\le T\le 800~\text{MeV}$,
we splice continuum-extrapolated Wuppertal–Budapest results~\cite{Borsanyi:2010cj,Borsanyi:2013bia,Bellwied:2015lba,Borsanyi:2018grb} to an HRG model at low $T$~\cite{Vovchenko:2019pjl}, and use a physically motivated parametrization that (i) approaches the Stefan–Boltzmann limit at high $T$ and (ii) reproduces thermodynamics with physical hadron masses at low $T$ (see~\cite{Kahangirwe:2024cny} for details).

\begin{figure*}[!htbp]
  \centering
 \includegraphics[width=\textwidth]{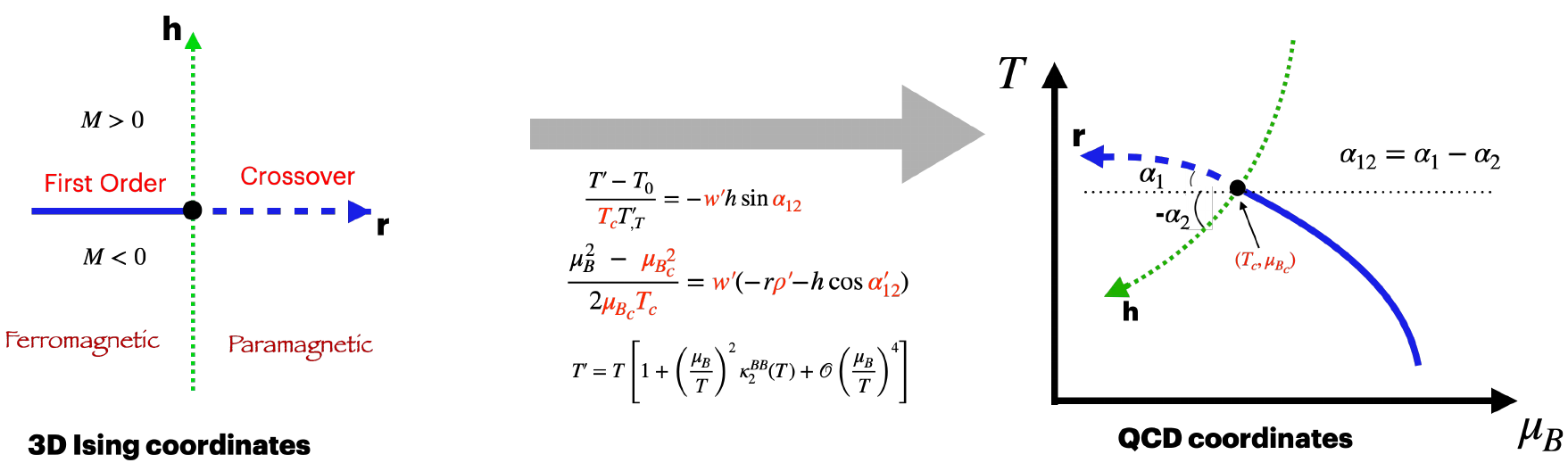}
\caption{Mapping of the critical point at $(r=0, h=0)$ from the 3D Ising model (left) to QCD coordinates $(T_c, {\mu_B}_c)$ in the Ising 2D-TExS module (right). The parameters in red ${\mu_B}_c$, \(w'\), \(\rho'\), and \(\alpha'_{12}\) are the free mapping parameters.   Figure taken from Ref.~\cite{Kahangirwe:2024cny}.}
  \label{fig:mapping_Ising-2DTExS}
\end{figure*}

Near a critical point, the thermodynamics is governed by universal scaling laws that depend only on symmetries and dimensionality. If the QCD critical point exists, it is expected to lie in the 3D Ising model universality class~\cite{Rajagopal:1992qz}. The \texttt{C++} implementation exploits this by introducing a non-universal map between the Ising variables, reduced temperature $r=(T-T_c)/T_c$ and magnetic field $h$, and the QCD coordinates $(T,\mu_B)$:
\[
(r,h)\quad \Longleftrightarrow \quad (T,\mu_B),
\]
with mapping parameters (in red in Fig.~\ref{fig:mapping_Ising-2DTExS}) ${\mu_B}_c$, $w'$, $\rho'$, and $\alpha'_{12}$ controlling the location, scale and orientation of the critical point in the $(T,\mu_B)$ plane. They are related to the linear-map parameters $({\mu_B}_c, w, \rho, \alpha_1, \alpha_2)$ used by the BEST collaboration~\cite{Parotto:2018pwx,An:2021wof}. Other works discuss further constraints on these parameters from either symmetries of QCD \cite{Pradeep:2019ccv} or causality and stability \cite{Mroczek:2022oga}.

\begin{figure*}[!htbp]
\begin{tabular}{cc}
   \includegraphics[width=0.5\linewidth]{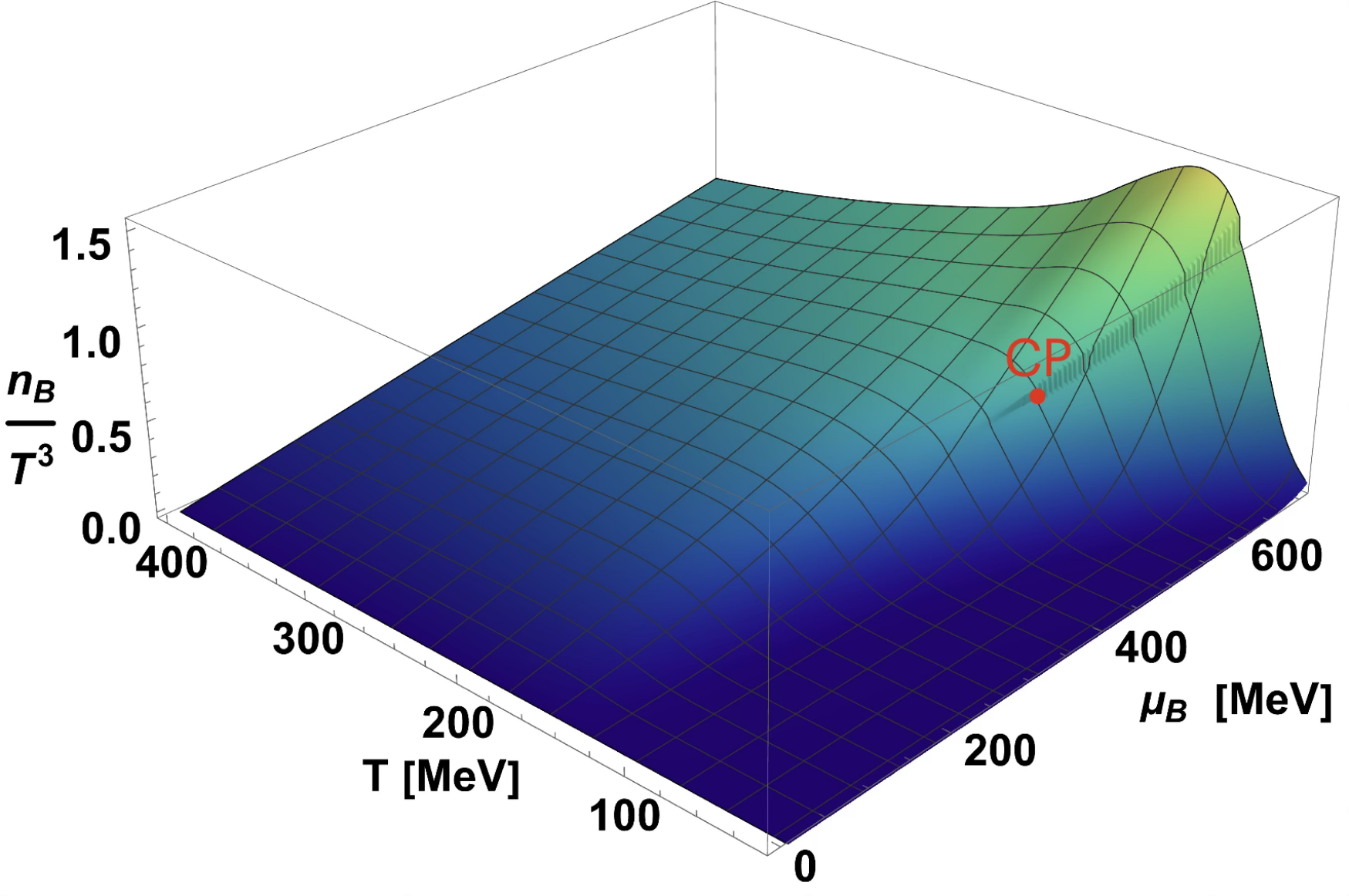}  &  \includegraphics[width=0.5\linewidth]{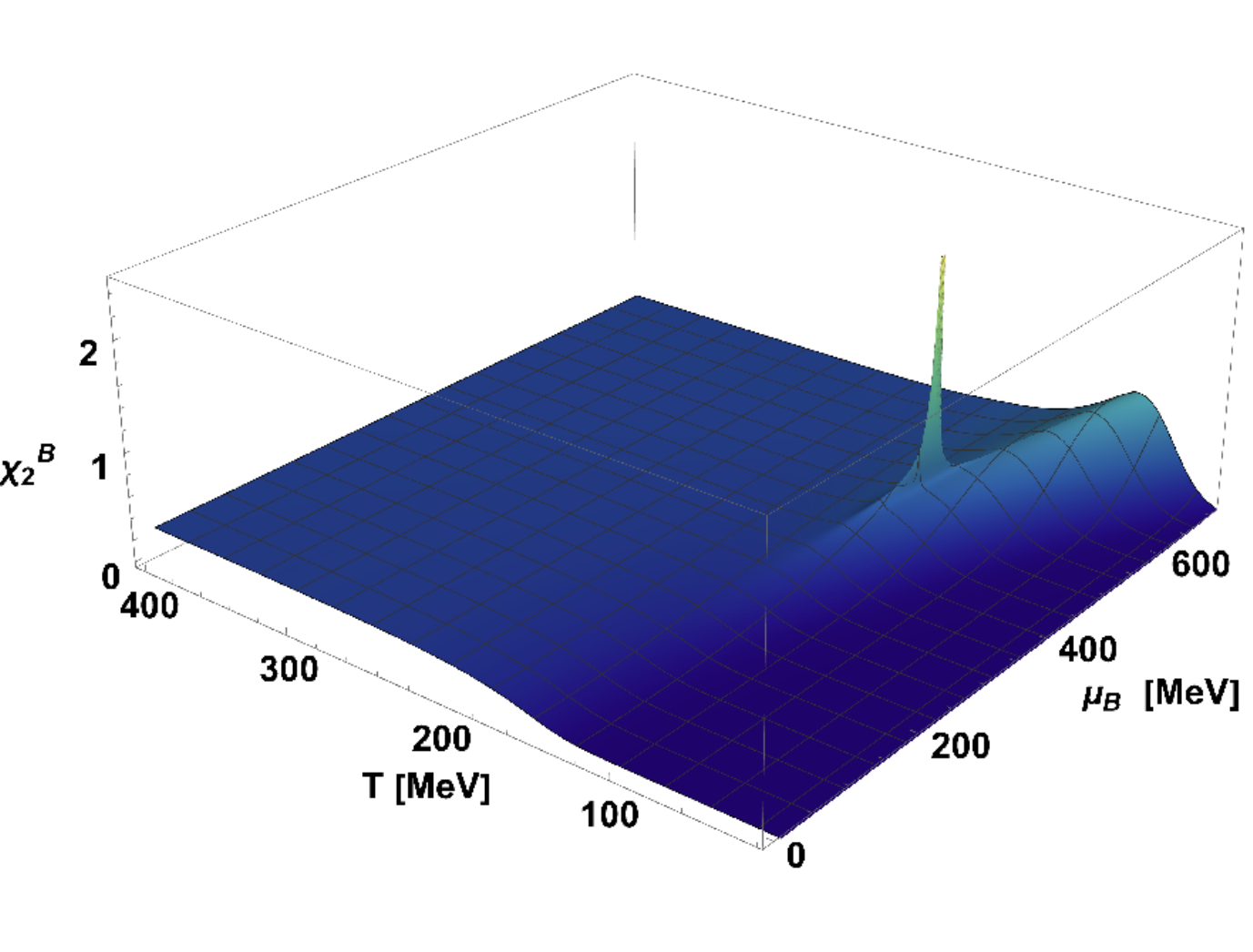}
\end{tabular}
\caption{ {Left:} Normalized baryon density $n_B/T^3$ as a function of $(T,\mu_B)$. A discontinuity appears for $\mu_B>{\mu_B}_c$ along the first-order line. {Right:} The second order susceptibility $\chi_2^B$ diverges at $(T_c,{\mu_B}_c)$. The parameters in this example are: ${\mu_B}_c=500~\text{MeV}$, $T_c=117~\text{MeV}$, $\alpha_1=11^\circ$, $w=15$, $\rho=0.3$, and $\alpha_{12}=\alpha_1$ (so $\alpha_2=\alpha_1-\alpha_{12}=0$), with $T_c$ and $\alpha_1$ fixed by the choice of ${\mu_B}_c$ while $w$ and $\rho$ are BEST Collaboration parameters \cite{Parotto:2018pwx} related to the mapping parameters $w^\prime$ and $\rho^\prime$ in Fig.~\ref{fig:mapping_Ising-2DTExS} as detailed in \cite{Kahangirwe:2024cny}. Both panels display calculations from the Ising 2D-TExS module, taken from Ref.~\cite{Kahangirwe:2024cny}. }
    \label{fig:thermo_Ising-2DTExS}
\end{figure*}

Because $\chi_2^B(T',0)$ is analytic across the QCD crossover, any non-analyticity in Eq.~\eqref{eq:Tprime_baryon_density} (critical point or first-order line) must enter through $T'(T,\mu_B)$. We therefore decompose
\[
n_B = n_B^{\rm reg} + n_B^{\rm crit},
\qquad
T' = T'_{\rm reg} + T'_{\rm crit}.
\]
Using Eq.~\eqref{eq:Tprime_baryon_density} and expanding $\chi_2^B$ around a reference pseudo-critical temperature $T_0$ (at $\mu_B{=}0$), the leading relation between the singular parts reads
\begin{equation}
  T'_{\rm crit}(T,\mu_B)
  \;=\;
  \left(\frac{\partial \chi_2^B(T,0)}{\partial T}\Bigg|_{T_0}\right)^{-1}
  \frac{n_B^{\rm crit}(T,\mu_B)}{T^3\,(\mu_B/T)}\,.
  \label{eq:Tcrit_relation}
\end{equation}
The regular component $T'_{\rm reg}$ is then fixed by matching the \emph{full} $T'$ to lattice constraints at $\mu_B{=}0$ up to the available order in $(\mu_B/T)$:
\begin{equation}
  T'_{\rm reg}(T,\mu_B)
  \;=\;
  T'_{\rm lat}(T,\mu_B) \;-\;
  \text{Taylor}_{n\le 2}\!\bigl[T'_{\rm crit}(T,\mu_B)\bigr],
\end{equation}
ensuring that the Taylor expansion of $T'$ reproduces the lattice-determined coefficients while all non-analytic behavior is confined to $T'_{\rm crit}$.
Substituting the full $T'(T,\mu_B)=T'_{\rm reg}+T'_{\rm crit}$ into Eq.~\eqref{eq:Tprime_baryon_density} guarantees that $n_B$ carries the correct critical singularity structure (see the left panel of Fig. \ref{fig:thermo_Ising-2DTExS}).

With $n_B(T,\mu_B)$ specified, the pressure follows from integration in $\mu_B$ at fixed $T$:
\begin{equation}
  \frac{P(T,\mu_B)}{T^4}
  \;=\;
  \frac{P_{\rm lat}(T,0)}{T^4}
  \;+\;
  \frac{1}{T}\int_{0}^{\mu_B}\!\! d\mu_B'\;
  \frac{n_B(T,\mu_B')}{T^3}\,,
\end{equation}
where $P_{\rm lat}(T,0)$ fixes the integration constant on the $\mu_B{=}0$ axis. All remaining observables are obtained by taking derivatives of $P(T,\mu_B)$ with respect to $T$ and $\mu_B$ such as $\chi_2^B$, shown in the right panel of Fig. \ref{fig:thermo_Ising-2DTExS}.

The \textit{Ising 2D-TExS} \texttt{C++} module computes thermodynamics with a 3D-Ising critical point while matching lattice constraints away from criticality. 
In the YAML configuration file, one can use \texttt{switchers} to select which thermodynamic observables to be printed, as well as choosing \texttt{LAT} (lattice mode without a critical point) or \texttt{PAR} (parametric mode with a critical point). 
In the second case, one can specify the critical-point location by providing either \texttt{T\_C} or \texttt{muB\_C} (choosing one uniquely determines the other such that the critical point lies on the transition line). The critical point mapping is controlled by tuning the parameters \texttt{w}, \texttt{rho}, and \texttt{alpha\_12} to set the size, shape, and orientation of the critical region.
In both cases, one can control the resolution via \texttt{T\_step} and \texttt{muB\_step} over $T\in[30,800]~\text{MeV}$ and $\mu_B\in[-700,700]~\text{MeV}$, as well as choosing the \texttt{grid\_type} among \texttt{RegularUniformGrid}, or \texttt{UniformT\_NonUniformmuB} / \texttt{UniformmuB\_NonUniformT} such that the transition line falls on the grid points. 
For the \texttt{UniformT\_NonUniformmuB} grid, the critical point location should be specified through \texttt{T\_C}, whereas for the remaining grid choices it should specified through \texttt{muB\_C}.
The computed thermodynamic observables, together with the transition line \texttt{muB, TC(muB)}, are written to the \textbf{output} directory as specified in the configuration. The \texttt{switchers} of the form \texttt{get\_y} are used to select the observables to be stored in the resulting EoS table, where $\texttt{y} \in [\texttt{P}, \texttt{E\_dens}, \texttt{s\_dens}, \texttt{B\_dens}, \texttt{Q\_dens}, \texttt{S\_dens}, \allowbreak \texttt{PressHessianMatrix}, \texttt{cs2}]$,. The corresponding data layout and variable definitions are also documented.

\subsubsection{Holography}
\label{sec:holography}

The \emph{Holographic} (\textit{NumRelHolo}) module is based on a bottom–up 5D gauge/gravity construction~\cite{Maldacena:1997re,DeWolfe:2010he}, designed to model strongly coupled QCD matter at finite $T,n_B$. Within this framework, the thermodynamics of the quark–gluon plasma is mapped to properties of charged black-brane solutions in asymptotically Anti–de Sitter (AdS$_5$) spacetime.
The model provides thermodynamic observables in the range $T \sim 20$–$800$~MeV and $\mu_B \sim \pm 1000$~MeV. It describes a homogeneous, isotropic plasma with a single conserved charge (baryon number), while $\mu_S,\mu_Q$ as well as external fields and anisotropies, are not included~\cite{Rougemont:2015wca,Rougemont:2023gfz}.

A key feature of this EMD construction is that it is calibrated to reproduce continuum-extrapolated lattice QCD thermodynamics at $\mu_B = 0$, and then extended to finite baryon density without introducing additional parameters. In the region where lattice results are available, the model quantitatively reproduces thermodynamic observables such as $s,P,n_{B}$~\cite{Grefa:2021qvt}. Furthermore, different realizations of the EMD model constrained by lattice data yield consistent predictions at finite $n_B$, including the location of a critical point as determined through Bayesian analyses~\cite{Hippert:2023bel}. This provides strong evidence that the model captures robust features of QCD thermodynamics beyond the calibration region.

The holographic framework is particularly well suited to describe the strongly coupled QGP near the crossover and at moderate $n_B$, beyond the reach of lattice QCD expansions. In addition, it enables the computation of transport coefficients within the same framework, as will be discussed with more details in Sec.~\ref{subsec:Holo_transcoef}, and naturally captures the nearly perfect-fluid behavior of the QGP~\cite{Grefa:2022sav}. As an intrinsically strongly-coupled construction, it does not incorporate asymptotic freedom at very high $T$, nor does it describe the confined hadronic phase at low $T$, where hadronic degrees of freedom become relevant. These considerations define its domain of applicability.

In this model, a charged AdS$_5$ black brane\footnote{Following~\cite{Rougemont:2015wca}, the term ``black hole'' is used generically; however, the Ansatz solved there corresponds to a charged, spatially isotropic black brane. By contrast, a global AdS$_5$ black hole has a spherical horizon (topology $S^3$) and would correspond to the dual theory on a finite-volume three-sphere.} represents the thermal state of the quark–gluon plasma. By black brane, we mean an asymptotically AdS$_5$ charged black-hole solution with a planar horizon (topology $\mathbb{R}^3$), appropriate for a translationally invariant plasma in $\mathbb{R}^{3,1}$. The dynamics is governed by the EMD action
\begin{align}\label{eq:holo-action}
S &= \frac{1}{2\kappa_5^{2}} \int_{\mathcal{M}_5} d^{5}x\sqrt{-g}\nonumber\\ &\times \Bigg[
R - \frac{1}{2}(\partial_\mu\phi)^2 - V(\phi) - \frac{f(\phi)}{4}F_{\mu\nu}F^{\mu\nu}
\Bigg]\,,
\end{align}
where $\phi$ is a real scalar dilaton field responsible for breaking conformal symmetry, and $A_\mu$ is a $U(1)$ gauge field dual to baryon number, with $F_{\mu\nu}=\partial_{\mu} A_{\nu}-\partial_{\nu} A_{\mu}$ and $2\kappa_5^2=16\pi G_5$, where $G_5$ denotes Newton's constant in five dimensions. Here, $R$ and $\sqrt{-g}$ are the Ricci scalar and the square root of the determinant of the metric of the 5D bulk spacetime manifold $\mathcal{M}_5$. The breaking of conformal symmetry is controlled by the potential $V(\phi)$, while $f(\phi)$ governs the coupling between the nonconformal dynamics and the baryon sector.

Thermodynamic quantities are extracted from charged black-brane solutions of the equations of motion. The Hawking temperature and black brane horizon area determine $T,s$, while the boundary value of the temporal component of the gauge field sets  $\mu_B$, and the conserved electric flux determines the baryon density $n_B$.

The pressure follows from thermodynamic consistency,
\begin{equation}\label{eqn:dP_gibbs}
dP = s\,dT + n_B\,d\mu_B,
\end{equation}
which implies that $P(T,\mu_B)$ is path-independent provided Maxwell relations are satisfied.\footnote{The Maxwell relations can be checked directly from the EoS table. 
We have verified that they hold to good precision. }
In practice, we compute the pressure by integrating Eq.~\eqref{eqn:dP_gibbs} along a 
path through the numerical grid in the $(T,\mu_B)$ plane. Since the Maxwell relations are 
satisfied, the result is path-independent. Setting $P(T_{\rm low},\mu_{B,\rm min}) = 0$ at 
the lowest available grid point, the pressure can be written as
\begin{align}
P(T,\mu_B) =& 
\int_{T_{\rm low}}^{T} s(T',\mu_{B,\rm min})\, dT' \\
& + \int_{\mu_{B,\rm min}}^{\mu_B} n_B(T,\mu_B')\, d\mu_B', \label{eq:BHrho}
\end{align}
where $T_{\rm low}$ and $\mu_{B,\rm min}$ are the lowest temperature and chemical potential 
available in the numerical grid, respectively. By setting $P(T_{\rm low},\mu_{B,\rm min}) = 0$, 
we neglect the absolute pressure at the starting point; this approximation improves as 
$T_{\rm low}$ and $\mu_{B,\rm min}$ decrease. When $\mu_{B,\rm min} = 0$, 
Eq.~\eqref{eq:BHrho} reduces to integrating first along the $\mu_B = 0$ axis and 
then at fixed $T$.

The \textit{NumRelHolo} MUSES module supports multiple parametrizations of the potentials $V(\phi)$ and $f(\phi)$, including the polynomial–hyperbolic (PHA) and parametric (PA) Ans\"atze calibrated to continuum–extrapolated $2+1$–flavor lattice thermodynamics at $\mu_B{=}0$~\cite{Hippert:2023bel}. Solving the coupled radial field equations for each black-brane configuration yields a thermodynamically consistent EoS, including $P$, $s$, $n_B$, second-order  susceptibilities, and $c_{s}^{2}$, across the deconfined and strongly interacting region of the QCD phase diagram. Multiple solution branches are compared via their pressure to identify stable and metastable phases and to map first-order transition lines and the critical point~\cite{Rougemont:2023gfz,Hippert:2023bel}.

In the \textit{NumRelHolo} module,
the user provides $(T,\mu_B)$ ranges, choice of model parametrization, and output paths in a single YAML configuration file. 
Specifically, \texttt{model\_type} selects either \texttt{polynomial\_hyperbolic} or \texttt{muses\_parametric}, which determines the functional forms of $f(\phi)$ and $V(\phi)$ in Eq.~\eqref{eq:holo-action}; \texttt{eos\_options} defines the scan via \texttt{eos\_stages}, where the user may choose to interpolate points and perform a Maxwell construction; \texttt{temperature\_options} sets \texttt{T\_min}, \texttt{T\_max}, and \texttt{T\_step}, which define the uniform grid in $T$, while \texttt{N\_phi0\_lines} controls the resolution of the underlying black–hole shooting in the $\phi_0$ direction; \texttt{chemical\_potential\_options} sets \texttt{mu\_B\_min}, \texttt{mu\_B\_max}, and \texttt{mu\_B\_step}, which define the uniform grid in $\mu_B$, while \texttt{N\_Phi1\_lines} controls the resolution in the $\Phi_1$ direction; \texttt{output\_options} sets \texttt{output\_path}, \texttt{yaml\_output\_file}, \texttt{format} and \texttt{muses\_extra\_columns}. 
Launched via the MUSES workflow or locally (\texttt{./bin/muses\_numrelholo.exec ./input/config.yaml}), the code constructs a uniform $(T,\mu_B)$ grid, with quantities in MeV units and dimensionless susceptibilities normalized by $T^2$. 
Note that because the holographic EoS does not contain hadronic degrees-of-freedom, it does not run into scaling issues when outputting dimensionless quantities, unlike what was seen in the other heavy-ion EoSs (because this EoS is not relevant in the low $T$, low $\mu_B$ regime). 

The module interpolates the EMD black–hole solutions, and if enabled applies a Maxwell construction, producing \texttt{phi0Phi1\_lines.csv} (raw BH shooting lines), \texttt{eos.csv} (uniform grid including stable/unstable/metastable branches), \texttt{stable\_eos.csv} (stable branch after the Maxwell construction), \texttt{transition\_line.csv}, \texttt{spinodal\_lines.csv}, and a YAML summary describing file formats; all filenames and directories follow the user’s YAML configuration.
Defaults for the numerical options can be overridden by providing a \texttt{numerical\_options} section in the YAML configuration file.

\subsection{Extension of the Synthesis module}\label{sec:ext_of_syn}
The MUSES EoS modules, developed for both heavy-ion and neutron-star applications, each have a limited regime of validity. Modules describing the deconfined phase are typically applicable at high $T$ and/or  $\mu_X$, while hadronic models are reliable at low $T$ and a range of $n_X$. The {\it Synthesis} module allows the merging of two different EoSs, enabling the construction of a single EoS, thereby extending their combined domain of applicability in temperature and/or densities.

In the initial version of the module (\texttt{v1.0.0}), the merging was implemented through standard thermodynamic constructions. In particular, a first-order phase transition could be enforced via a Maxwell construction, requiring equality of pressures and baryon chemical potentials between the two EoSs, $P^{(1)} = P^{(2)}$ and $\mu_B^{(1)} = \mu_B^{(2)}$, where the superscripts $(1)$ and $(2)$ label the two input EoSs. Alternatively, a Gibbs construction could be employed, reducing the multi-dimensional space (e.g., $\mu_B,\mu_Q$ or $n_B,n_Q$) to a single variable ($\mu_B$ or $n_B$) under conditions such as $\beta$ equilibrium. In this case, global charge neutrality is imposed and a mixed-phase region emerges instead of a sharp interface (see Ref.~\cite{ReinkePelicer:2025vuh} for details). The module also supports interpolation-based approaches, constructing smooth crossover transitions using sigmoid functions to match thermodynamic quantities~\cite{Masuda:2012ed,ReinkePelicer:2025ado}.

In the new release of the module \texttt{v1.1.0}, we implement a new procedure \cite{Yang:2026brr} that allows combining two EoSs based on stability and statistical mixing, and that is generic to functions of any dimension. 
The merging procedure is based on a two-fluid equilibrium statistical mixture in the grand canonical ensemble, in which a single merged grand potential is constructed from the input equations of state and minimized with respect to the fluid fractions at fixed $(T,\mu_B)$. This guarantees thermodynamic consistency and stability, while allowing for the emergence of crossover behavior, first-order phase transitions, and a critical point.
As a concrete example, we applied this method to merge the holographic EoS with the QvdW-HRG EoS across a broad range of $(T,\,\mu_{B})$. This new procedure can be generalized to a higher dimensional phase diagram with more chemical potentials (e.g., $\mu_{S}$ and $\mu_{Q}$). 

The method is based on constructing a mixed pressure from the individual pressures $P_1$ and $P_2$ of each EoS,
\begin{align}\label{eq:pressure_mix}
\begin{split}
    P = & \;p\,P_1 + (1-p)\,P_2 - \;p\,(1-p)\,\frac{2\,T_c}{\Delta V} \\
    &- \frac{T}{\Delta V}\,\left[p\,\ln p + (1-p)\,\ln(1-p)\right],
\end{split}
\end{align}
where $p$ is the mixing weight associated with $P_1$, and $(1-p)$ with $P_2$. The third term represents a Shannon-type entropy-of-mixing contribution, controlled by the parameter $\Delta V$, which regulates the stiffness of the crossover and fluctuations near the phase transition. Note that $-[p\,\ln p + (1-p)\,\ln(1-p)]\geq 0$, so this term always increases $P$, reflecting the thermodynamic preference for mixed configurations. The last term breaks the concavity of the pressure in $p$, allowing for the emergence of a first-order phase transition with a tunable critical temperature $T_c$.
The equilibrium mixing weight, $\overline{p}(T,\mu_B)$, is determined by maximizing $P$~\eqref{eq:pressure_mix}, which is equivalent to solving the equation
\begin{align}
\left.\frac{\partial P}{\partial p}\right|_{T,\mu_B} = P_1 - P_2 - \frac{T}{\Delta V}\,\ln\frac{p}{1-p} - (1-2\,p)\,\frac{2\,T_c}{\Delta V} = 0.
\end{align}
In equilibrium, the system selects the phase with the higher $P$, and the phase transition occurs when $P_1 = P_2$. For $T>T_c$ the transition is a crossover, for $T<T_c$ it is first-order, and for $T=T_c$ it becomes second-order.
By construction, the condition $\partial P/\partial p=0$ ensures thermodynamic stability, while thermodynamic consistency follows from the definition of $P$. Then, $(s,n_B)$ are obtained from Eq.~\eqref{eq:pressure_mix} as
\begin{align}
    n=&\;\overline{p}\,n_1+(1-\overline{p})\,n_2,\label{eq:n} \\
    s=&\;\overline{p}\,s_1 + (1-\overline{p})\,s_2 \nonumber \\
    &- \frac{1}{\Delta V}\left[\overline{p}\,\ln \overline{p} + (1-\overline{p})\,\ln(1-\overline{p})\right] .
\end{align}
The discussion of how to run Synthesis appears later in Sec.\ \ref{sec:workflow}.
 
 \section{Observables \& Tools modules}
\label{sec:observables}

In this section, we introduce the new observable modules added to the MUSES Calculation Engine in the \textit{Calliope} release, namely the \textit{Partial Pressures} module, and the holographic transport coefficients (which calculations have been directly integrated to the pre-existent \textit{Holographic} module).

\subsection{Partial pressures}

The pressure of the ideal HRG model can be expanded in the Boltzman approximation, which allows us to rewrite it by separating different hadronic contributions based on $B,S,Q$ content \cite{Noronha-Hostler:2016rpd} as
\begin{equation}
\begin{aligned}
P\left(\hat{\mu}_{B}, \hat{\mu}_{S}, \hat{\mu}_{Q}\right) 
&= P_{000} + P_{00|1|} \cosh \left(\hat{\mu}_{Q}\right)+P_{100} \cosh \left(\hat{\mu}_{B}\right) \\
&\hspace*{-30pt} +P_{101} \cosh \left(\hat{\mu}_{B}+\hat{\mu}_{Q}\right)+P_{10-1} \cosh \left(\hat{\mu}_{B}-\hat{\mu}_{Q}\right) \\
&\hspace*{-30pt} +P_{102} \cosh \left(\hat{\mu}_{B}+2 \hat{\mu}_{Q}\right)+P_{0|1| 0} \cosh \left(\hat{\mu}_{S}\right) \\
&\hspace*{-30pt} +P_{0|1||1|} \cosh \left(\hat{\mu}_{S}+\hat{\mu}_{Q}\right)+P_{1|1| 0} \cosh \left(\hat{\mu}_{B}-\hat{\mu}_{S}\right) \\
&\hspace*{-30pt} +P_{1|1| 1} \cosh \left(\hat{\mu}_{B}-\hat{\mu}_{S}+\hat{\mu}_{Q}\right) \\
&\hspace*{-30pt} +P_{1|1|-1} \cosh \left(\hat{\mu}_{B}-\hat{\mu}_{S}-\hat{\mu}_{Q}\right) \\
&\hspace*{-30pt} +P_{1|2| 0} \cosh \left(\hat{\mu}_{B}-2 \hat{\mu}_{S}\right) \\
&\hspace*{-30pt} +P_{1|2||1|} \cosh \left(\hat{\mu}_{B}-2 \hat{\mu}_{S}-\hat{\mu}_{Q}\right) \\
&\hspace*{-30pt}+P_{1|3||1|} \cosh \left(\hat{\mu}_{B}-3 \hat{\mu}_{S}-\hat{\mu}_{Q}\right),
\end{aligned}
\end{equation}
where each partial pressure $P_{B,S,Q}$ reflects its corresponding content in terms of each conserved charge. Absolute values are shown when the sign may be positive or negative. 

This formalism allows us to easily express the partial pressures in the above equation in terms of susceptibilities of conserved charges. Considering the susceptibilities up to fourth-order, we obtain a linear equation that can be expressed in the form:
\begin{equation}
    \mathcal{X} = M \mathcal{P},
\end{equation}
where $\mathcal{X}$ is the vector of the susceptibilities, $\mathcal{P}$ the vector of partial pressures, and $M$ is the $21 \times 13$ matrix that relates partial pressures to susceptibilities.

Given that the linear system is overdetermined, a unique inverse solution to express partial pressures in terms of susceptibilities does not exist. 

Within this formulation, one can obtain the partial pressures from the susceptibilities of different models. To select which solution to represent each partial pressure, we pick combinations that minimize the lattice QCD uncertainty band \cite{Gonzales:2026mfx}.

\begin{figure}
    \includegraphics[scale=1]{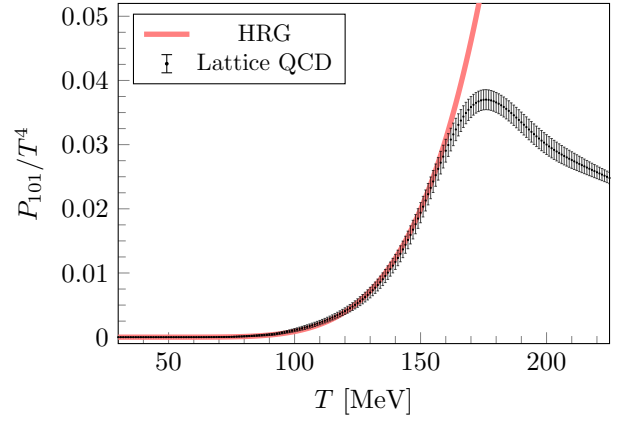}
    \caption{Normalized partial pressure $P_{101}/T^4$ as a function of the temperature at $\mu_X=0$. Results obtained through the ideal HRG model (red line) are compared to the ones from lattice QCD (black band).
    Figure adapted from Ref.~\cite{Gonzales:2026mfx}.
    }
    \label{fig:PartialPressure}
\end{figure}

Figure \ref{fig:PartialPressure} shows the normalized partial pressure $P_{101}/T^4$ as a function of $T$. The red curve is the HRG model result, while the dots are lattice QCD results obtained from the linear combination 
\begin{equation}
\begin{split}
    P_{101} / T^4  &= \frac{1}{24} \chi^{S}_{2} + \frac{1}{12} \chi^{BS}_{11} + \chi^{BQ}_{11} - \frac{1}{24} \chi^{S}_{4}  \\
    &\hspace{12pt} - \frac{1}{12} \chi^{BS}_{13}  - \frac{1}{2} \chi^{BQ}_{13} + \frac{1}{3} \chi^{SQ}_{13} + \frac{1}{6} \chi^{SQ}_{31}  \\
    &\hspace{12pt} + \frac{1}{2} \chi^{BQ}_{22} - \frac{1}{2} \chi^{SQ}_{22},
\end{split}
\end{equation}
calculated with the new continuum-estimated susceptibilities from Ref.\ \cite{Abuali:2025tbd}.
It is important to understand that the solution for the partial pressures can differ greatly depending on the model used to calculate susceptibilities. The lattice QCD results can serve as a benchmark against which to test the predictions of hadronic models.

To compute the partial pressures with the MUSES \textit{Partial Pressures} module, the user must provide one data file for each of the 21 susceptibilities, with columns organized as: $T$, susceptibility, and optionally the associated error values. 
The $T$ column can employ the user's unit of choice, where some variation in the values of the $T$ grid is allowed, based on sampling differences. However, it is essential that all files contain the same number of rows to ensure consistency across the dataset. 
The path to the folder containing the input files, which has to be located in the \texttt{input/} folder, is specified through the \texttt{suscept\_path}.
The file names for the susceptibility data must correspond to the specific susceptibility they represent. For example, the data file for $\chi^{BS}_{11}$ should be named \texttt{Chi11BS.dat}; it is essential that the file names follow this structure consistently for each susceptibility file. 
The file extension can be chosen by the user through the  \texttt{file\_extension} option, allowing for the use of different  susceptibility datasets within the same input folder. For example, the susceptibilities \texttt{Chi4S\_30-800-dT1.dat} and \texttt{Chi4S.dat} are distinguished based on the option \texttt{file\_extension} = \texttt{\_30-800-dT1.dat} and \texttt{file\_extension} = \texttt{.dat}, respectively. The file names may be formatted using either the $BQS$ or $BSQ$ notation, 
provided that all files with a similar extension use the same convention.

Once all susceptibility data is provided, the module computes the 13 partial pressures based on the fixed linear combinations selected in Ref. \cite{Gonzales:2026mfx}. 
The output is written into a file named \texttt{Partial\_Pressures.csv}, which contains information ordered as: temperature, normalized partial pressures, and the optional normalized partial pressure error, presented in columns 0, 1 to 13, and 14 to 26, respectively.

\subsection{Holographic transport coefficients}
\label{subsec:Holo_transcoef}

\begin{figure*}
    \centering
    \includegraphics[width=1\linewidth]{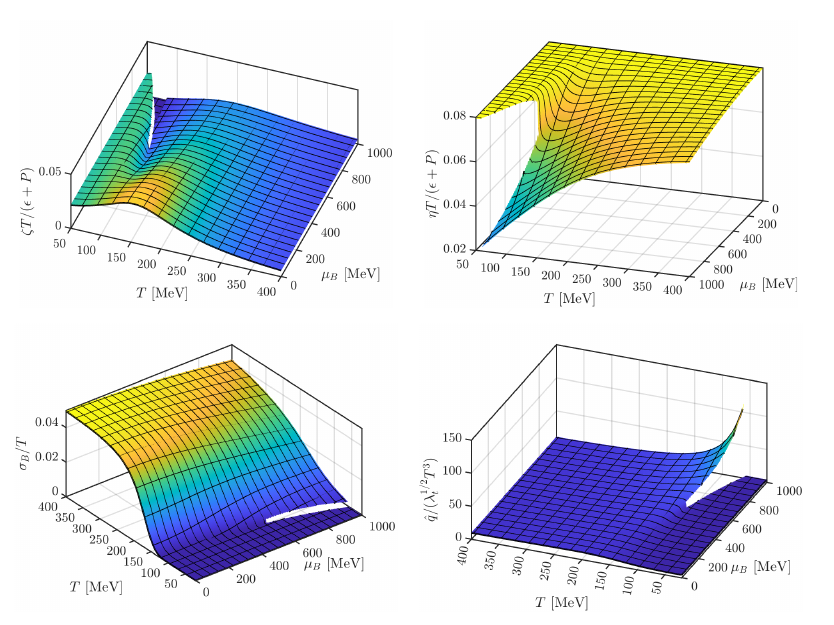}
\caption{
Einstein–Maxwell–Dilaton holographic transport coefficients as functions of $T$ and $\mu_B$, obtained by using the Polynomial–Hyperbolic Ansatz (PHA) from Ref.~\cite{Hippert:2023bel}. 
Top left: bulk viscosity $\zeta T/(\varepsilon + P)$. 
Top right: shear viscosity times temperature over enthalpy density, obtained from Eq.~\eqref{eq:shear_viscosity}. 
Bottom left: normalized baryon conductivity $\sigma_B/T$. 
Bottom right: scaled jet-quenching parameter $\hat{q}/(\lambda_t^{1/2} T^3)$.
Figures taken from Ref.~\cite{Khan:2026uqc}.    
}
\label{fig:BH_transpot}
\end{figure*}

In its initial implementation, the \textit{NumRelHolo} module only computed the EoS at finite $T$ and $\mu_B$, as reported in Ref.~\cite{Grefa:2021qvt,Hippert:2023bel,yang_2026_19154307}. 
In a recent work~\cite{Khan:2026uqc}, we have expanded the capabilities of the module  to compute the transport coefficients within the same holographic setup. 
This extension leverages the advantages of the gauge/gravity correspondence, which enables the evaluation of dynamical quantities both near and out of equilibrium. As a result, the EMD model now provides, in a consistent manner, both the EoS and the associated transport coefficients at finite $\mu_B$.

The module computes a broad set of transport coefficients within the holographic EMD framework, following the methodology established in Refs.~\cite{Rougemont:2015wca,Finazzo:2016mhm,Rougemont:2017tlu,Rougemont:2020had,Grefa:2022sav}. Specifically, we evaluate the baryon charge conductivity $\sigma_B$, the baryon diffusion coefficient $D_B$, the heavy-quark drag force and Langevin diffusion coefficients, as well as the jet quenching parameter $\hat{q}$, bulk $\zeta$ and shear $\eta$ viscosity. We note that the holographic model has the dynamical universality class of model-B \cite{Natsuume:2010bs,DeWolfe:2011ts}, so critical scaling is not identical to what we expect in QCD (i.e., model-H, see Ref.~\cite{Son:2004iv}), but one can still obtain useful insights from these calculations. 

The transport of charged baryons in a hot and dense QGP is studied by considering the baryon and thermal conductivities, as well as baryon diffusion coefficient \cite{DeWolfe:2011ts,Rougemont:2015ona}. Studying these observables at large $n_B$ is particularly crucial for understanding the dynamical behavior of strongly interacting matter under extreme conditions. 
Then, $\sigma_B$ is obtained by solving the equation of motion of vector perturbation, see Ref.~\cite{Rougemont:2015ona} for more details.

The normalized baryon conductivity $\sigma_B/T$ is shown in Fig.~\ref{fig:BH_transpot}. It increases smoothly with $T$, which means the transport of baryon number is enhanced in the deconfined QGP phase; however, it decreases as we increase $\mu_B$, reflecting reduced mobility of baryon charge in denser matter \cite{Rougemont:2015ona}. The critical point is located at $T_c=103$ MeV, ${\mu_B}_c = 598$ MeV: the discontinuity gap indicates a first-order phase transition region. 
In Ref.~\cite{Khan:2026uqc} and previously in Refs.~\cite{Grefa:2022sav,Rougemont:2023gfz}, the baryon diffusion coefficient $D_B $ is evaluated holographically using the Nernst–Einstein relation
\begin{equation}
D_B = \frac{\sigma_B}{\chi^B_2},
\label{eq:DB}
\end{equation}
where $\chi^B_2$ is the second-order baryon susceptibility. This quantity characterizes the fluid’s response to inhomogeneities in $n_B$. Plots and the detailed analysis are presented in Ref. \cite{Khan:2026uqc}.

Another important observable, which can be obtained using $\sigma_B$ and thermodynamic quantities at finite baryon chemical potential, is the thermal conductivity denoted by $\sigma_T$  computed using the equation
\begin{equation}
\sigma_T = \frac{D_B}{T} \, \chi_2^B 
\left( \frac{\varepsilon + P}{n_B} \right)^{\!2}
= T \, \sigma_B 
\left( \frac{s}{n_B} + \frac{\mu_B}{T} \right)^{\!2}.
\label{eq:sigmaT}
\end{equation}
The holographic thermal conductivity $\sigma_T$ quantifies the ability of baryon-rich matter to transport energy through baryon diffusion and is directly linked to the baryon conductivity $\sigma_B$. It reflects the interplay among $s,n_B,\mu_B$ \cite{Grefa:2022sav,Rougemont:2023gfz}.

The QGP is thought to be a nearly perfect fluid, characterized by its small shear viscosity to entropy density ratio, $\eta/s$, which is one of the most significant features of the deconfined QCD medium produced in heavy-ion collisions \cite{Heinz:2013th}. 
While in the current EMD model $\eta/s$ = $1/4 \pi$, regardless of $T$ and $\mu_B$, it is essential to acknowledge that the following dimensionless combination represents the true measure of fluidity in a baryon-rich medium \cite{Liao:2009gb}
\begin{equation}
\frac{\eta T}{\varepsilon + P} = 
\frac{1}{4\pi \left(1 + \frac{ n_B \mu_B}{sT}\right)}.
\label{eq:shear_viscosity}
\end{equation}
The normalized holographic shear viscosity is defined by   Eq.\ (\ref{eq:shear_viscosity}) and Fig.~\ref{fig:BH_transpot} shows the result as a function of $(T,\mu_B)$. 
Eq.~(\ref{eq:shear_viscosity}) reduces to $\eta/s = 1/4\pi$ for $\mu_B = 0$, whereas at finite $\mu_B$ the normalized shear viscosity exhibits both a minimum and an inflection point, as $n_B$ increases toward the critical point. 
At the critical point, we observe an infinite slope. However, as we move beyond the critical point in the $\mu_B$ direction, the normalized shear viscosity acquires a discontinuity gap at the first-order line. This is a consequence of the corresponding discontinuity gaps in $(s,n_B)$ in this region of the phase diagram. 

Another important observable is the bulk viscosity $\zeta$, which measures the medium resistance to deformations arising from compression or expansion of the fluid \cite{Rocha:2023ilf}. This quantity plays a significant role in relativistic heavy-ion collisions and it influences key observables such as the transverse momentum spectra, azimuthal anisotropy, and charged hadron multiplicities in heavy-ion collisions~\cite{Noronha-Hostler:2014dqa,Ryu:2015vwa}. In Fig. \ref{fig:BH_transpot} we show our result for $\zeta T/(\varepsilon+P)$  where one can clearly see a peak near the crossover transition, as expected. At the critical point, $\zeta$ diverges, while the first-order phase transition region is indicated by a discontinuity ~\cite{Noronha-Hostler:2008kkf, Bernhard:2016tnd, Bernhard:2019bmu}. 
The emergence of this divergence has important consequences for identifying the critical point and assessing the applicability of hydrodynamics.

The energy loss of heavy quarks traveling through the quark–gluon plasma can be studied through several holographic observables. These are included in our extended MUSES framework. Observables include the jet-quenching parameter $\hat{q}$, the drag force, and the longitudinal and transverse momentum-diffusion coefficients ($\kappa_{||}$, $\kappa_{\perp}$). All of these observables are calculated from the dynamics of a trailing string in the EMD background ~\cite{Gubser:2006bz,Herzog:2006gh,Gursoy:2009kk}. These quantities characterize, respectively, the rate of transverse momentum broadening, the energy loss per unit time, and the energy loss due to stochastic motion of the heavy quark. For a detailed analysis, see Ref. \cite{Khan:2026uqc}.

Calculation of the transport coefficients and energy loss in the \emph{NumRelHolo} module is controlled by the \texttt{transport\_options} block in the YAML configuration file. When \texttt{transport\_coefficient} is set to \texttt{true}, the code computes the transport coefficients $\hat{q}$, $\sigma_B/T$, $\zeta/s$, and the heavy-quark energy-loss observables (drag force and transverse/longitudinal momentum diffusion coefficients $\kappa_\perp$, $\kappa_\parallel$). These quantities are appended as additional columns to the output files \texttt{phi0Phi1\_lines.csv}, \texttt{eos.csv} and \texttt{stable\_eos.csv}.

The heavy-quark velocity used in energy-loss calculations is specified through the \texttt{velocity} option, which accepts either a single value or a list of values (e.g., \texttt{velocity: 0.5} or \texttt{velocity: [0.1, 0.5, 0.8, 0.99]}). For each velocity at zero-based index \texttt{N}, three dedicated columns are added to the output files: \texttt{drag\_N}, \texttt{kappa\_perp\_N}, and \texttt{kappa\_par\_N}. When running in MUSES output format (\texttt{format: muses} under \texttt{output\_options}), the extra columns included in the output—both thermodynamic derivatives and transport quantities—must be listed explicitly via the \texttt{muses\_extra\_columns} option (e.g., \texttt{muses\_extra\_columns: ["chi2\_B", "qhat", "sigma\_B", "zeta", "drag\_0", "kappa\_perp\_0", "kappa\_par\_0"]}).

\section{Workflow examples}
\label{sec:workflow}

\begin{figure}[!htbp]
    \centering
    \includegraphics[width=1\linewidth]{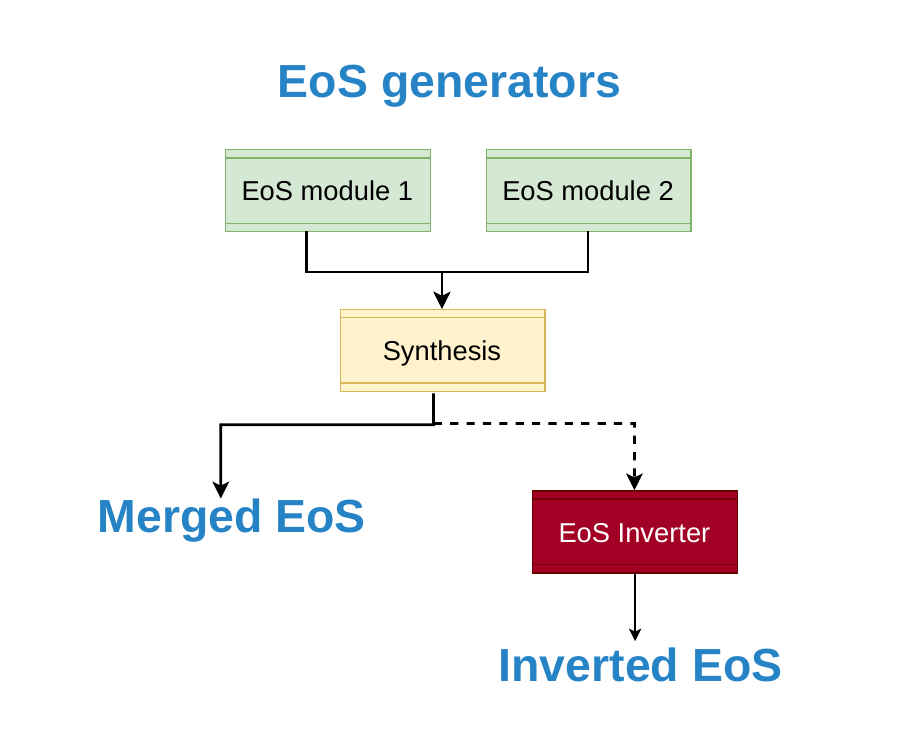}
    \caption{Workflow example that creates two EoSs which are then merged into one by the \textit{Synthesis} module. The merged EoS can eventually be inverted using the \textit{EoS Inverter} module.
    }
    \label{fig:flow_example}
\end{figure}

In this section, we discuss different possible workflows useful for heavy-ion phenomenology and notably used to produce the results shown in Sec.~\ref{sec:hydro}, made possible thanks to the new modules introduced in the \textit{Calliope} release of the MUSES Calculation Engine.
In particular, we present the new EoS merging method integrated into the \textit{Synthesis} module that enables now to use MUSES to produce larger-range EoS tables, and the \textit{Inverter} module necessary to produce ready-to-use EoS tables for numerical hydrodynamic simulations.
Note that EoS tables used in the present work and obtained thanks to these workflows can be found at Ref.~\cite{jahan_2026_20823256}.

In Figure \ref{fig:flow_example}, we show a generic example of a workflow within the MUSES Calculation Engine. We remark that many other workflows can be created by the user. In this example, each EoS module generates a table containing thermodynamic quantities such as $P,\varepsilon,s,n_X$ among others, as functions of temperature $T$ and the conserved-charge chemical potentials $\mu_X$. 
This output is then combined within the {\it Synthesis} module, which produces a table for the merged EoS, as explained in \ref{sec:ext_of_syn}. 
Alternatively, the table generated by the {\it Synthesis} module can be used as an input for the {\it EoS Inverter} module, which produces a table organized as a regular grid in entropy density $s$ and number densities $n_X$, instead of the usual $T$ and $\mu_X$ coordinates naturally employed when computing heavy-ion collision EoSs. 
Note that the line connecting the \textit{Synthesis} module to the \textit{Inverter} module is dashed, because the current implementation is solely designed to perform smooth interpolations for now, hence only treating correctly EoS with crossover transitions.

\subsection{Merging 2D EoSs: Holography with HRG}
\label{sub:merg-2D-holo-HRG}

\begin{figure*}[!htbp]
\begin{tabular}{cc}
   \includegraphics[width=0.5\linewidth]{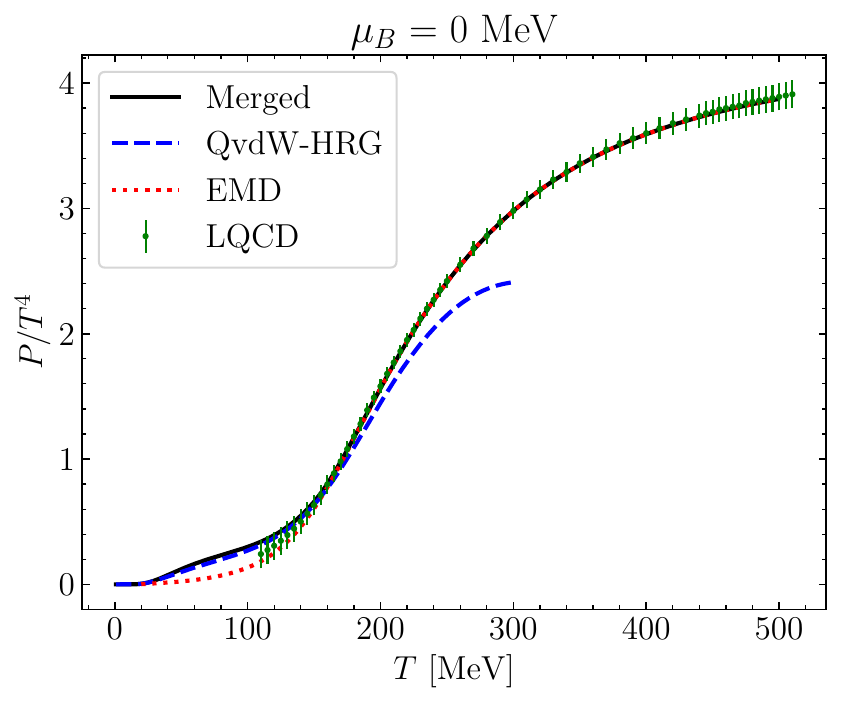}  &  \includegraphics[width=0.5\linewidth]{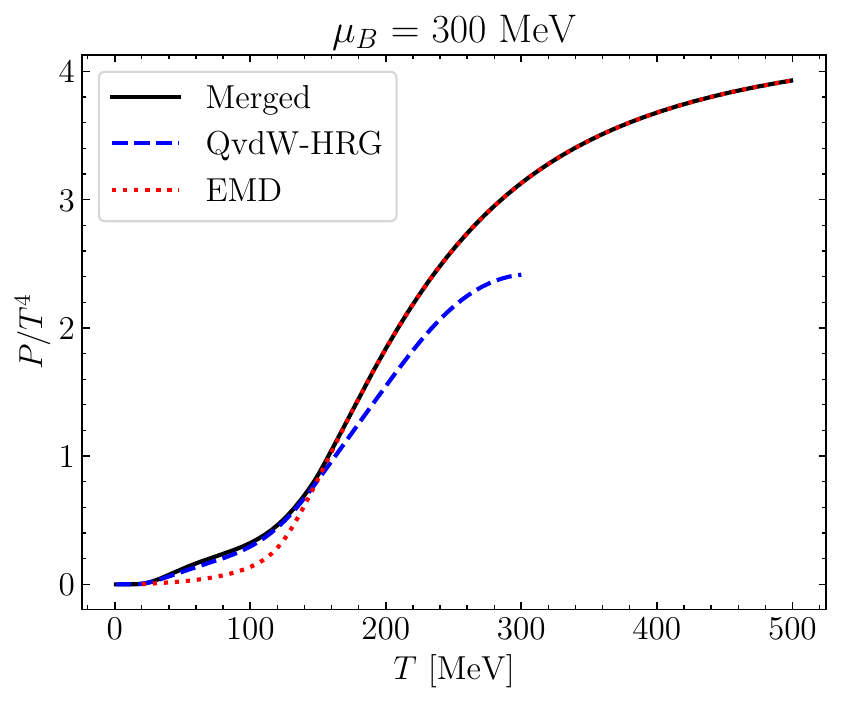} \\
    \includegraphics[width=0.5\linewidth]{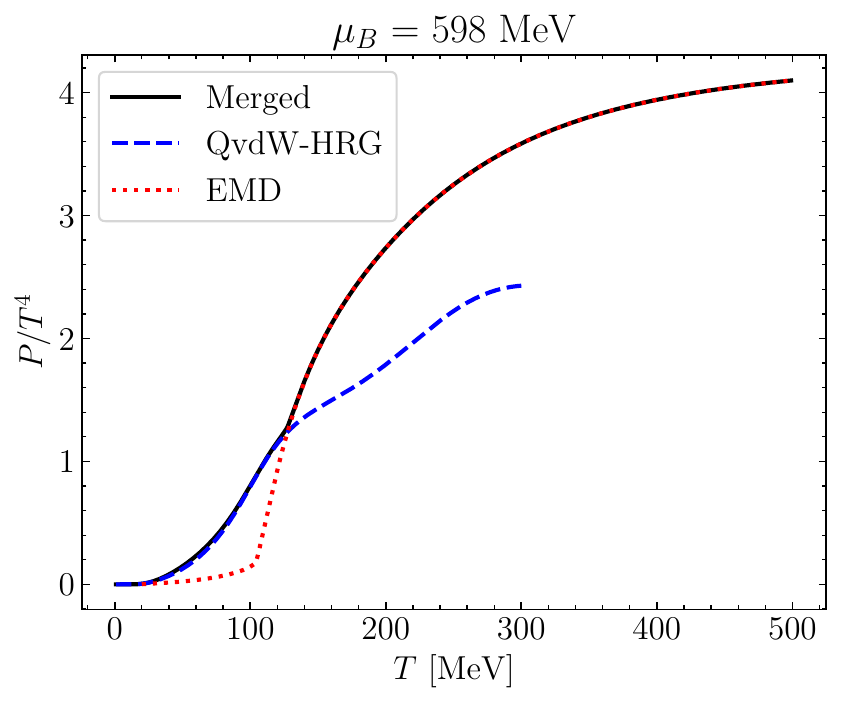}  & \includegraphics[width=0.5\linewidth]{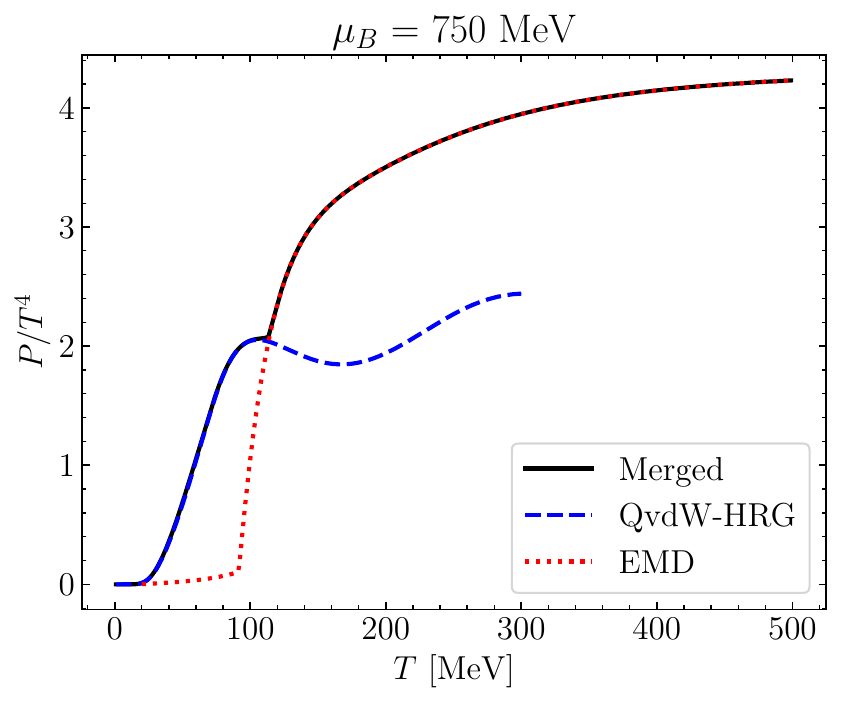}
\end{tabular}
    \caption{Pressure as a function of temperature for the merged 
    EoS (continuous lines), QvdW-HRG (dashed lines), and the EMD holographic model (dotted lines) for  different $\mu_B$ values, for the case $\mu_{B_c}=598$ MeV. At $\mu_B=0$ we compare with LQCD data from Ref.~\cite{Borsanyi:2013bia}.
}
    \label{fig:pressure_BH_HRG}
\end{figure*}

The holographic and the QvdW-HRG EoSs are valid in different regimes of $(T,\mu_B)$. 
While the QvdW-HRG describes the hadronic phase of matter and is valid in the $(T,\mu_B)$ regime below the deconfinement transition, the holographic EMD model describes strongly interacting matter for values of $(T,\mu_B)$ around and above the transition region. 
Consequently, one can create a workflow such as the one shown in Fig. \ref{fig:flow_example}, where the \emph{NumRelHolo} and the \emph{HRG} module outputs are combined in the {\it Synthesis} module that generates a merged EoS which is, in principle, valid for a broader range of $T$ and $\mu_B$ (with $\mu_Q=\mu_S=0$).

We merge the holographic EoS and the QvdW-HRG EoS with three different critical chemical potentials, ${\mu_B}_c=200, 598, 1000$ MeV, with corresponding matching parameter $\Delta V = 1.98\times 10^{-6}, 4.75\times 10^{-6}, 1.98\times 10^{-6}~\mathrm{MeV}^{-3}$ respectively. 
In general, $\Delta V$ must be manually tuned for each new pair of EoSs to be merged, with the appropriate value identified by testing a range of candidates and selecting the one that yields the best matching behavior. 
The QvdW-HRG EoS employed in this work was obtained by using the PDG2020 particle list (setting the parameter \texttt{particle\_list} to the corresponding path, i.e., \texttt{"/Thermal-FIST/input/list/PDG2020/list.dat"}), with van der Waals parameters set to typical values $a=0.329$ GeV.fm$^3$ and $b=3.42$ fm$^3$, and a finite meson excluded-volume parameter set to $b_M=0.1$ fm$^3$.

With the parameters for the {\it Synthesis} module set, the module produces a merged EoS. 
Here, we show results of the case $\mu_{B_c}=598$ MeV as examples. The pressure as a function of $T$ for different fixed $\mu_B$ values is shown in Fig.  \ref{fig:pressure_BH_HRG} for the holographic model, the QvdW-HRG model, and the merged pressure from the {\it Synthesis} module. 
At $\mu_B=0$, we also plot the LQCD results from Ref.~\cite{Borsanyi:2013bia} for comparison.
For all $\mu_B$ values, the merged pressure follows the QvdW-HRG below the transition temperature (where $P_{\rm HRG}>P_{\rm BH}$), while above the transition it follows the holographic model (where $P_{\rm BH}>P_{\rm HRG}$). 
For $\mu_B=0$ and $\mu_B=300$ MeV, the merged EoS evolves from the hadronic phase to the deconfined phase through a crossover, and one can observe a region of phase mixing near the transition, where the probabilities of the two phases are comparable. 
We also show the case at the critical chemical potential, ${\mu_B}_c=598$ MeV, where the transition is more abrupt.
Finally, for $\mu_B=750$ MeV, the transition is first-order, and we can see that the merged pressure exhibits a distinct kink at the transition.
The remaining merged thermodynamic quantities can be found in Ref. \cite{Yang:2026brr}.

To use the {\it Synthesis} module, the user only needs to specify a few parameters. Specifically, the user needs to set \texttt{synthesis\_type}, under \texttt{global}, to \texttt{statistical\_mixture} to use this method to merge. 
Then, the user needs to specify the free parameters of the method and the grid under \texttt{statistical\_mixture}: \texttt{deltaV} sets the value of $\Delta V$; \texttt{Tc} sets the critical temperature; \texttt{y\_bounds} sets the upper and lower bounds of $\ln\frac{p}{1+p}$, which sets the mixing weight $p$ to be approximately between 0 and 1; \texttt{n\_scan} sets the grid resolution while solving for $p$; \texttt{grid} sets the grid format that the merged EoS employs. Under \texttt{grid}, \texttt{mode} sets how the grid is constructed. If \texttt{mode} is set as \texttt{union}, the grid is the union of points of the input EoSs. If \texttt{mode} is set as \texttt{uniform}, the user can specify their desired grid with a uniform step size, and needs to specify \texttt{T\_min}, \texttt{T\_max}, \texttt{mu\_min}, \texttt{mu\_max}, \texttt{dT}, and \texttt{dmu}. 
The \textit{Synthesis} module code is archived on Zenodo at \cite{pelicer_2025_17584997}.

\subsection{EoS Inverter}

EoSs generated using heavy-ion collisions modules of the MUSES Calculation Engine are tabulated as functions of the thermodynamic variables, namely $(T,\mu_X)$. Quantities such as $s$ or $n_X$ can then be expressed as: 
\begin{equation}
    (s, n_X)  = F_{\text{EoS}}(T, \mu_X).
\end{equation}
In heavy-ion simulations, the natural variables are the \textit{dynamical} variables, $s$ (or $\varepsilon$) and $n_X$ \cite{Plumberg:2024leb,Schenke:2010rr,Karpenko:2013wva}. 
The choice between $\varepsilon$ or $s$ usually depends on the specific hydrodynamic code, wherein most grid-based codes currently use $\varepsilon$ and most smoothed particle hydrodynamics (SPH) codes use $s$. 
Since the proof-of-principle external solver used in this work is \texttt{CCAKE}, which is a mesh-free Lagrangian hydrodynamics code, our examples here are written in terms of $(s, n_X)$. However, within MUSES the user can of course choose a different set of target variables such as $(\varepsilon, n_X)$.
As a consequence, the inversion of an EoS is tethered to the ability to find the functions $G_{\text{EoS}}$ such that
\begin{equation}
    (T, \mu_X) = G_{\text{EoS}}(s, n_X),
\end{equation}
namely, expressing the thermodynamic variables as functions of the dynamic variables. The function $G_{\text{EoS}}$ is the inverse of $F_{\text{EoS}}$, namely, $G_{\text{EoS}} = F^{-1}_{\text{EoS}}$, assuming such an inverse exists. In practice, EoSs are given in the form of tables and, thus, finding $G_{\text{EoS}}$ requires performing a numerical inversion and interpolation.

There are two ways of handling that process in practice. The first way consists of constructing the numerical inverse of the function $F_{\text{EoS}}$ and doing the needed interpolations during the simulation. For this, the original output of the MUSES Calculation Engine is sufficient. The second way consists of performing the numerical inversion beforehand on the whole EoS table and only computing the interpolations during the simulations. 
% In the latter case, the MUSES Calculation Engine now provides an efficient numerical procedure to build the inverted table through the newly introduced \textit{EoS Inverter} module \cite{pelicer_2026_20721054}.
In the latter case, the MUSES Calculation Engine will provide an efficient numerical procedure to build the inverted table through the new \textit{EoS Inverter} module, to be part of the \textit{Calliope} release.

The inversion procedure follows the algorithm described in CCAKE \cite{Plumberg:2024leb}. It uses the GNU Scientific Library and, in particular, the \textit{gsl\_multiroot\_fsolver\_hybrids}, which implements a modified Powell hybrid method. This method combines two ideas: a robust global step that safely guides the search toward the root, and a fast local quasi-Newton update once the solution is close. In practice, the solver first stabilizes the search to avoid divergence, then switches to a Broyden-type step that converges rapidly. This approach provides both reliability and efficiency for the inversion.

In Fig.~\ref{fig:flow_example}, one can see how the \textit{EoS inverter} module is incorporated into an example MUSES Workflow schematically. In this workflow, a merged EoS is then inverted to be used within an external solve. One can also directly invert an EoS table generated from a single module, without merging it through the \textit{Synthesis} module. The output of the \textit{EoS Inverter} module is a table where the thermodynamic variables are expressed on a regular grid in the dynamical variables. 
Since we are using \texttt{CCAKE} in this paper, which uses the hydrodynamic variables of $(s,n_X)$, we will focus on quantifying the accuracy of the inversion in this basis as well as limitations in the range of $(s,n_X)$ after the inversion. 
As mentioned above, most grid-based codes use $(\varepsilon,n_X)$ as their natural hydrodynamic variables, but this change of variables would face the exact same challenges with the inversion and range of the tables as the ones discussed here. 

\begin{figure}
    \centering
    \includegraphics[trim={0 0 0 1.14cm},clip,width=\linewidth]{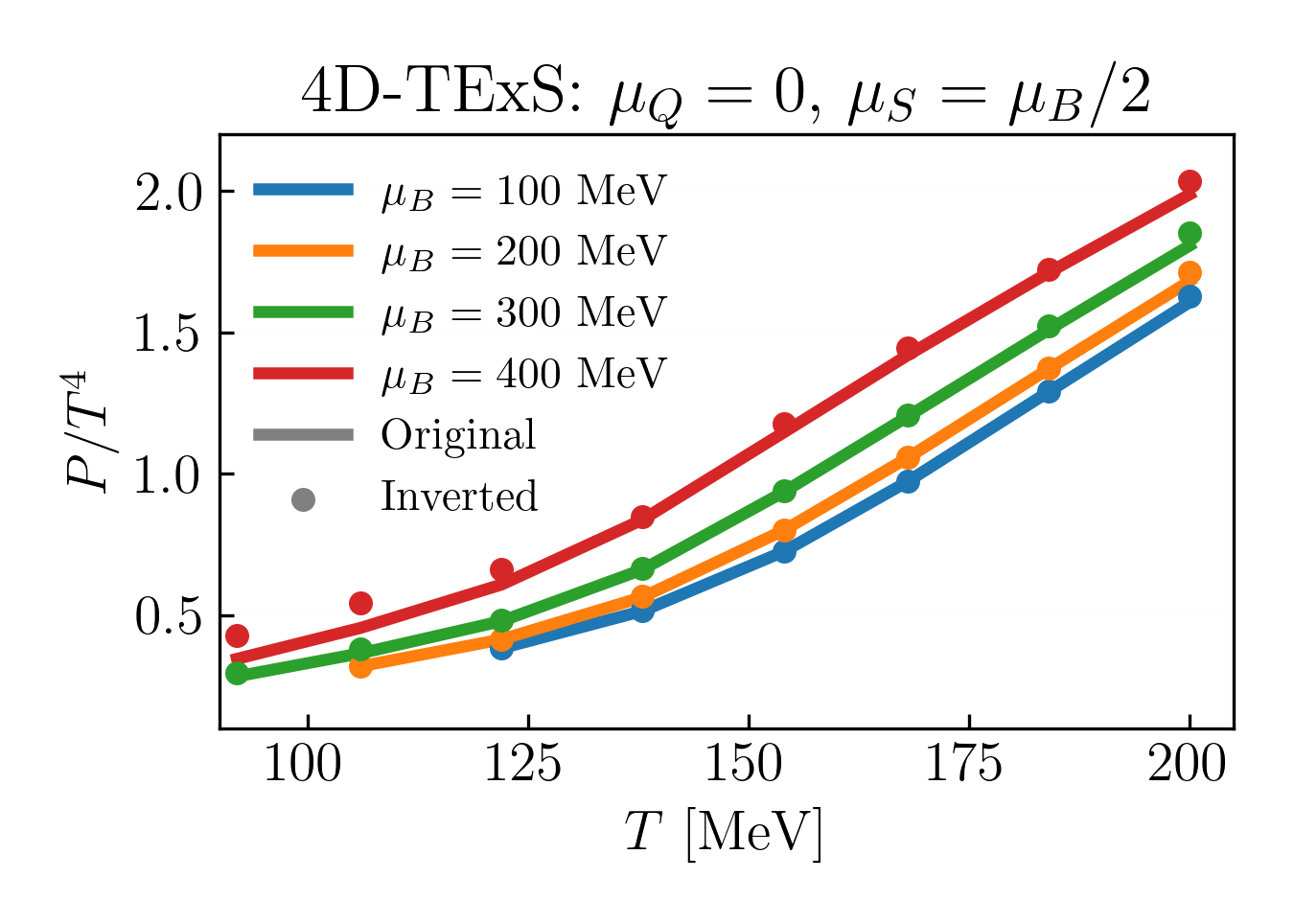}
    \caption{Pressure normalized by $T^4$ as a function of the temperature for the 4D-TExS EoS. The plot shows curves corresponding to the original table (lines) and inverted table (points) at $\mu_B=$ 100, 200, 300 and 400 MeV, with $\mu_Q=0$ and $\mu_S=\mu_B/2$.}
    \label{fig:EOSinverter}
\end{figure}

In Fig.~\ref{fig:EOSinverter}, we compare $P/T^4$ from the original 4D-TExS EoS to values reconstructed from the inverted table. The comparison is shown as a function of $T$ for fixed values of $\mu_B=100, 200, 300, 400$ MeV, $\mu_Q=0$ and $\mu_S=\mu_B/2$. 
In the region of temperatures above $T\gtrsim 100$ MeV, the inverted table reproduces the original EoS well along these trajectories. We have checked that the agreement holds across the $(\mu_Q, \mu_S)$ space. At low $T$ and large $\mu_B$, the inversion starts to break down, and the code resorts to a backup EoS (see Appendix \ref{sec:backupEOS} for details). This is also the region where the susceptibility-based approach employed to construct the original table is less reliable due to the $\mu_X/T$ expansion. We therefore interpret this comparison as a test of the numerical inversion in the region where the original 4D-TExS table itself is under better control.

We note that, even with an accurate algorithm in the \textit{EoS Inverter} module, the obtained table may not be sufficient for direct use in simulations for three reasons: 
\begin{enumerate}
    \item the range in the grid,
    \item the grid size spacing,
    \item the effects from a first-order phase transition \\
    (see Sec.\ \ref{subsec:firstorder}).
\end{enumerate}
Thus, a user should carefully test their inverted EoS in terms of range and grid-size spacing before direct use in simulations. Here we provide guidance on the challenges involved.

The range and grid size spacing in $(s, n_X)$ depend on the choice of range and grid size spacing in $(T,\mu_X)$, but the challenge here is that in general there is a very non-linear mapping between the $(T,\mu_X)$ space and the $(s, n_X)$ quantities. Essentially, a regular fixed grid in $(T,\mu_X)$ leads to an inverted triangle shape in $(s, n_X)$ (in other words, at high $T$ or $s$ we have a broad coverage, but at low $s$ the coverage is very narrow in the $n_B$ direction).  Thus, in order to get a regular grid in $(s, n_X)$ one requires significantly more points at low $T$ across a range of $\mu_B$, with significantly fewer points in the large $T$, large $\mu_B$ space. A very similar problem exists in astrophysics at $T\sim0$ when trying to convert a low $\mu_B$ grid into an $n_B$ grid.

\begin{figure*}
    \centering
    \includegraphics[width=0.99\linewidth]{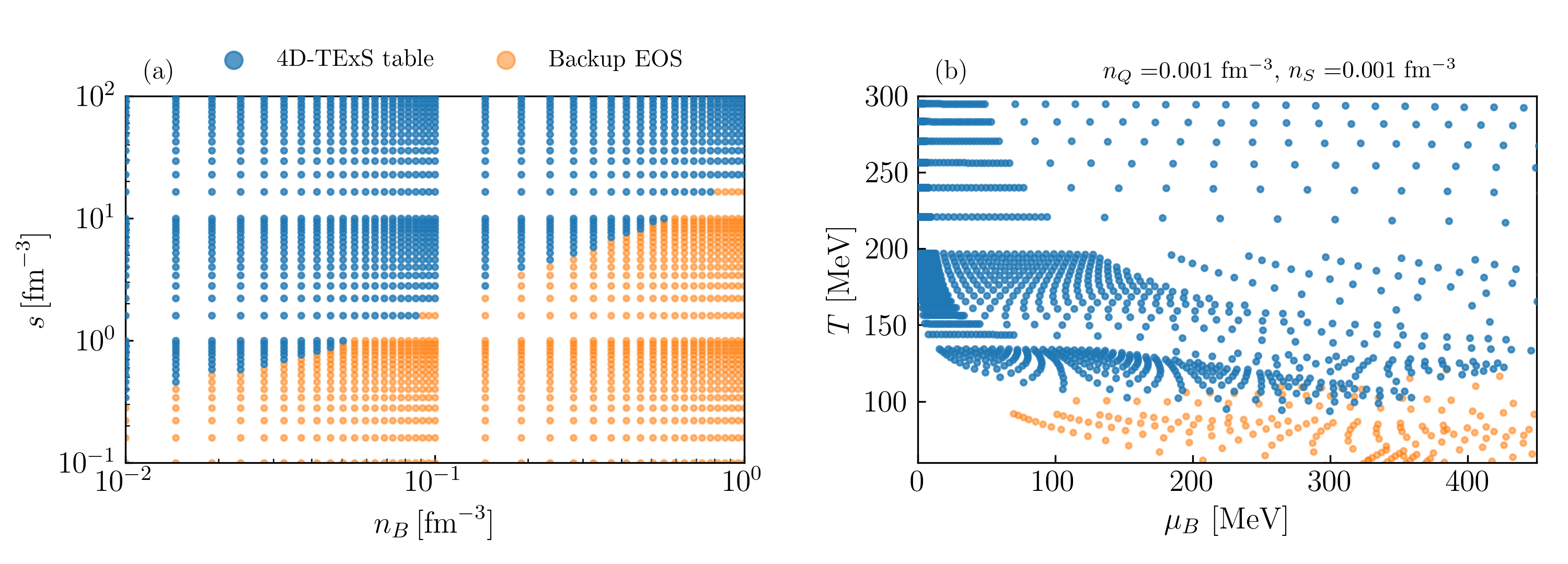}
    \caption{Illustration of the highly non-linear mapping between a fixed grid $\log_{10}-\log_{10}$ in $(s,n_B)$ variables into $(T,\mu_B)$ variables. The results shown are for the 4D-TExS EoS (in blue) combined with backup EoS that has already been inverted using the \textit{EoS Inverter} (in orange). Both panels show the inverted EoS at fixed $n_S = n_Q = 10^{-3}$ fm$^{-3}$.
    Left: Fixed points on a $\log_{10}-\log_{10}$ grid in the $(s, n_B)$ plane are shown. Each point represents an entry in the EoS table. Right: The exact same points in the left figure are now plotted on the $(T, \mu_B)$ plane. These points are no longer evenly spaced due to the highly non-linear mapping from the $(s, n_B)$ plane into the  $(T, \mu_B)$ plane. }
    \label{fig:table_points}
\end{figure*}

Fig.~\ref{fig:table_points} shows the coverage of the inverted table at fixed $n_S$ and $n_Q$, displayed both in the $(s,n_B)$ space (left panel) determined from inverting the 4D-TExS EoS and also after mapping the same points back to the $(T,\mu_B)$ plane (right panel). Because the mapping between $(T,\mu_B)$ and $(s,n_B)$ is highly nonlinear, especially at low temperature where $s$ and $n_B$ can vary rapidly, we use a $\log_{10}$--$\log_{10}$ grid in $(s,n_B)$. 
The corresponding points in the $(T,\mu_B)$ plane obtained this way are therefore not uniformly distributed: we find dense coverage at large temperatures and small chemical potentials, while the high-$\mu_B$, high-temperature region is more sparsely sampled. 
The color coding indicates whether each point is obtained from the original table or from the backup EoS. For the susceptibility-based table used here, most low-temperature points requested in $(s,n_B)$ lie outside the original table and therefore require the backup EoS.
In future applications, in which the \textit{Synthesis} can be used across the full 4D table, much of this low $T$ region can instead be supplied by the \textit{HRG} module. Nevertheless, a remaining challenge is the low-$s$, large-$n_B$ region: at low $T$, the HRG dependence of $n_B$ on $\mu_B$ is very soft. Therefore, increasing $\mu_B$ can produce only a small increase in $n_B$, making it difficult to reach densities above $0.2\, fm^{-3}$.

Returning to the spacing coverage of the inverted EoS, the user has multiple options to choose how best to handle this challenge. 
One option is that the user can invert several tables generated on different complementary ranges of $(s,n_X)$. This is possible within the MUSES Calculation Engine by generating several workflows. 
A second option is that the user can perform a single workflow but selecting different grid constructions: linear, logarithmic, and decade-spaced grids, which can improve coverage of the target domain.
A third (future) option would be to invert the tables on a non-regular grid in variables that can be expressed analytically as functions of the dynamic variables such as in Ref.~\cite{Monnai:2024pvy}. 
The latter will not be included in the \textit{EoS Inverter} module at first, but will be given as an option in future implementations.
Also, we note that this module complements, for heavy-ion collisions EoS modules, the inverter for neutron star EoSs embedded in the MUSES CompOSE package introduced in our previous work \cite{ReinkePelicer:2025vuh}. The CompOSE package produces multidimensional tables as functions of ($n_B$, $Y_Q$, $T$) in standard format for the CompOSE repository of EoSs for astrophysics~\cite{Typel:2013rza,CompOSECoreTeam:2022ddl,Dexheimer:2022qhn}.

\begin{figure}
    \centering
    \includegraphics[width=\linewidth]{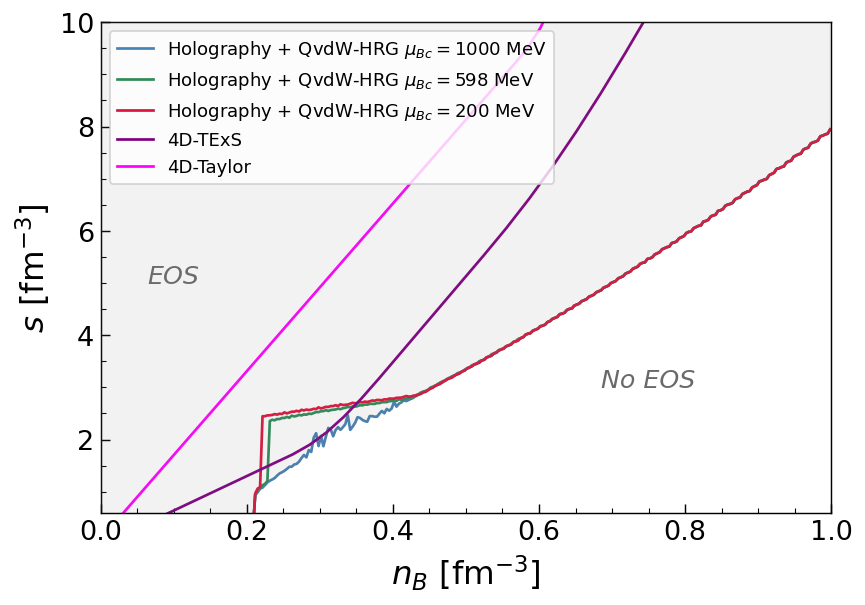}
    \caption{Boundary of the range of validity (shaded in gray),  in the natural hydrodynamic variables of entropy vs. net-baryon density, for the MUSES EoSs discussed in this section. 
    The white area shows the region in the $(s,n_B)$ plane where no MUSES heavy-ion collision EoS currently reaches.
    }
    \label{fig:snBrange}
\end{figure}

In Fig.~\ref{fig:snBrange} we show the range of validity in terms of $(s,n_B)$ of the various MUSES heavy-ion collision EoSs made available with this paper. This comparison is made at $\mu_S=\mu_Q=0$, and we ensure that the points shown in this diagram are both causal and stable.
\footnote{We note that another established EoS exists from Ref.~\cite{Monnai:2024pvy} that is based on a lattice QCD Taylor expansion up to 4th-order, supplemented by a subset of terms at 6th order, which is matched to the ideal HRG EoS using a smoothing function. At this time, we cannot include it in this comparison because it is not on a fixed grid in $(T,\mu_X)$ nor does it provide all the causality and stability information required for this comparison.} 
The holography EoSs shown here have a sharp first-order phase transition (with the exception of $\mu_{Bc}=1000$ MeV that is at the boundary of the EoS). 
We cut off the very low $s$ regime because not all EoSs cover the $T\rightarrow  0$ limit, and also to avoid the liquid-gas phase transition. Given that this regime is well below freeze-out, it is not relevant to accurately describing the heavy-ion collision observables. 

As expected from the range available in the chemical potentials, the 4D-Taylor EoS has the smallest range in $(s,n_B)$. The next best coverage comes from the 4D-TExS EoS that provides a significantly larger coverage in $(s,n_B)$ compared to the 4D-Taylor. The set of EoSs that have the largest range comes from Holography+QvdW-HRG, which fall onto a single boundary in the QGP phase and the HRG phase as well. Only around the phase transition do we see differences in the range of $(s,n_B)$ for the different Holography+QvdW-HRG EoSs, which occur because of the choice of having either a critical point followed by a first-order phase transition ($\mu_{Bc}=200$ MeV or $\mu_{Bc}=598$ MeV) or not ($\mu_{Bc}=1000$ MeV is at a large enough value to be outside of the relevant range of heavy-ion collisions). 
The points that would fall in this range across the first-order phase transition are all either metastable or unstable.

There is a narrowing at low $s$ in terms of the coverage at finite $n_B$ for both lattice-based expansions because they have not yet been matched to the HRG EoS (instead, only the susceptibilities are matched to HRG at low $T$). 
However, even in the case of an EoS matched to the HRG, this only gains a slightly larger range in $n_B$ in the low $s$ limit. This occurs because, at low $s$, we run into a challenge in the HRG EoS, where increasing to very large $\mu_B$ leads to almost no change in $n_B$. Thus, we reach a maximum of $n_B^{max,HRG}\sim 0.2\,fm^{-3}$ that we currently cannot go beyond. 
This boundary highlights the need for future work that matches effective models like CMF to the HRG EoS. 

\subsection{First-order Phase Transitions and Natural Hydrodynamic Variables}\label{subsec:firstorder}

\begin{figure}
    \centering
    \includegraphics[width=\linewidth]{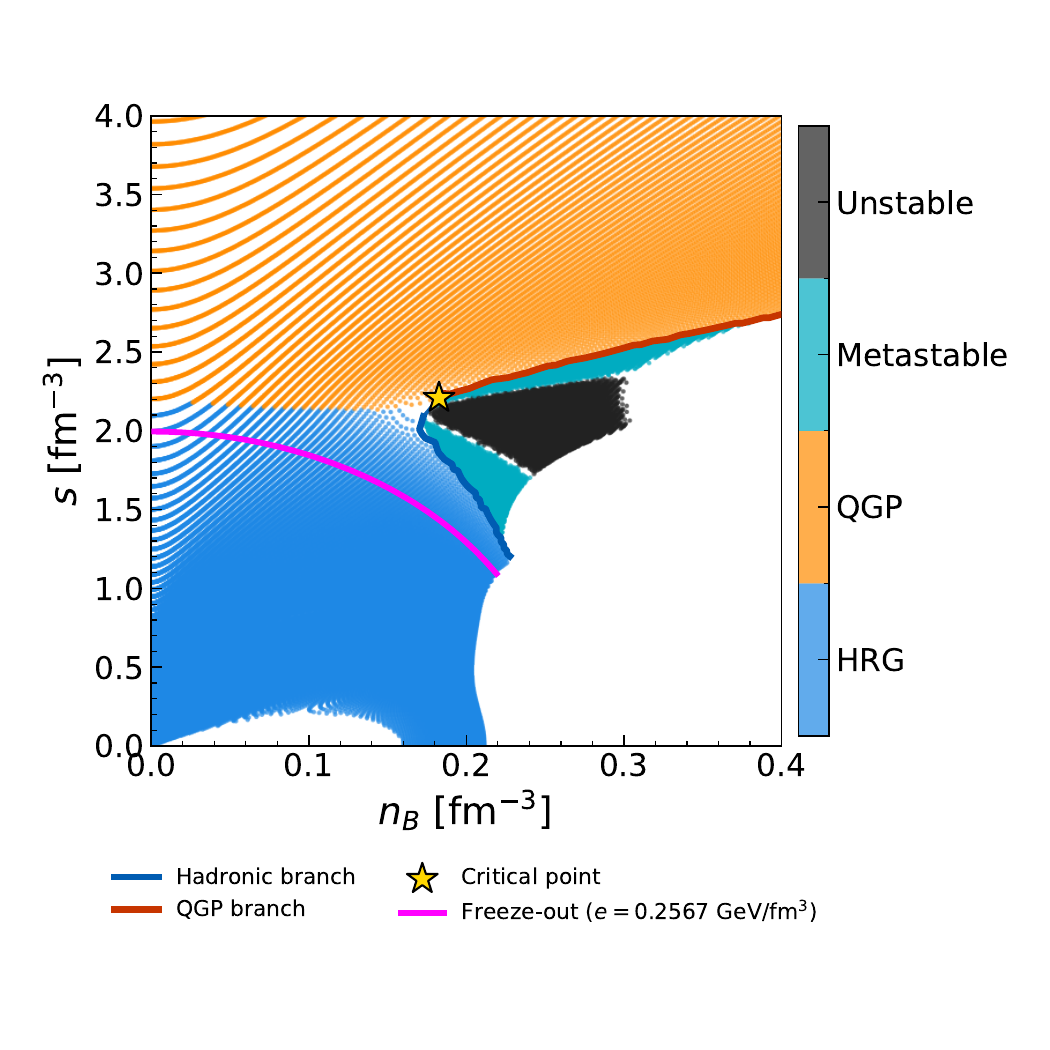}
    \caption{Phase diagram (in the natural hydrodynamic variables) of the Holography+QvdW-HRG EoS with a critical point at $\mu_B^{CP}=598$ MeV. 
    The individual points represent a fixed grid in the $(T,\mu_B)$ plane that has been mapped into the $(s,n_B)$ plane, and they are colored by the phase of matter that they represent. The critical point is shown with a star and the first-order line is outlined in red and blue (hot QGP phase is red and cold HRG phase is blue). Within the first-order phase transition, we show also the metastable and unstable regions. An example freeze-out line is shown assuming a constant energy-density at freeze-out.   }
    \label{fig:snB_meta}
\end{figure}

Let us now return to the discussion on the missing points across the first-order phase transition in Fig.~\ref{fig:snBrange}.  In an equilibrium Maxwell construction, the coexistence line is characterized by equal pressure and chemical potentials in the two phases, while extensive densities such as the entropy density $s$ and baryon density $n_B$ are discontinuous across the transition. 
Equivalently, when the EoS is represented as a function of entropy density along an appropriate hydrodynamic trajectory, the mixed-phase region may appear as an interval with a nearly constant pressure, corresponding to a vanishing or very small effective speed of sound, $c_s^2 \simeq 0$. This is a well-defined equilibrium thermodynamic construction, but it is not automatically suitable for dynamical out-of-equilibrium simulations. 
A relativistic fluid cell evolving through a first-order transition may involve metastability, nucleation, spinodal decomposition, finite-size domains, interface dynamics, and relaxation effects, none of which are captured by an instantaneous equilibrium Maxwell construction. Thus, using such an EoS directly in hydrodynamic simulations can lead to substantial numerical and physics challenges.

In Fig.~\ref{fig:snB_meta} we zoom in on the first-order phase transition regime to demonstrate this challenge. Here we use the Holography+QvdW-HRG EoS with the critical point at $\mu_B^{CP}=598$ MeV, which is shown as a star. The phase boundary of the first-order line separates into a hot (QGP) line shown in red and a cold (HRG) line shown in blue. In between these lines, one finds the metastable and unstable regime for this EoS.  While these surfaces can be tabulated explicitly via a post-processing step described below, relativistic viscous hydrodynamic simulations may face numerical issues when run with the unstable phase (see Sec.\ \ref{sec:quant_time}-\ref{sec:moveCP} for \texttt{CCAKE}-specific discussions).
The metastable and unstable surfaces shown in Fig.~\ref{fig:snB_meta} are obtained by a separate post-processing script applied to the \textit{Synthesis} module output, and provided alongside the equilibrium table as supplementary phase-2 rows; this is independent of the Synthesis module of Sec.~\ref{sec:ext_of_syn}, which returns only the equilibrium phase. The construction proceeds as follows. The merged pressure $P(p)$ defined by Eq.~\eqref{eq:pressure_mix} is non-concave for $T<T_c$ inside the coexistence band, and admits up to three stationary points for any given $(T, \mu_B)$ pair: the global maximum (the equilibrium phase $\overline{p}$, returned by \textit{Synthesis} module), a local but non-global maximum (the metastable phase), and a local minimum that separates them, which is locally convex in $p$ and is therefore the spinodally unstable point. Local stability follows from the curvature
\begin{align}\label{eq:meta_curvature}
\frac{\partial^2 P}{\partial p^2}\bigg|_{T,\mu_B} = \frac{1}{\Delta V}\!\left[4\,T_c - \frac{T}{p\,(1-p)}\right],
\end{align}
which is negative at the two maxima (stable/metastable) and positive at the local minimum (unstable). The spinodal lines are the places where Eq.~\eqref{eq:meta_curvature} vanishes. The solutions can be calculated analytically:
\begin{equation}
p_{\pm}(T) = \tfrac{1}{2}\big(1\pm\sqrt{1-T/T_c}\big),
\end{equation} 
for $T\le T_c$, with the unstable interval $p\in(p_-(T),~p_+(T))$. We attach a phase tag to each row of the final EoS table, with 0, 1, and 2 representing EoS 1, EoS 2, and the unstable region, respectively. 

The Holography+QvdW-HRG EoS is then passed to the \textit{EoS Inverter} module, which constructs an inverted table from the original coordinates $(T, \mu_B)$ into the target coordinates $(s, n_B)$. 
When metastable and unstable branches are included, the mapping between $(T, \mu_B) \rightarrow(s, n_B)$ is multivalued, and a single well-defined interpolator in the intensive variables cannot be constructed. To avoid this problem, our algorithm separates the table by phase (stable, metastable, and unstable) and builds a separate  interpolator for each. It also stores the values of $[s_{\rm min}, s_{\rm max}]$ and $[n_{B_{\rm min}}, n_{B_{\rm max}}]$ for each phase, so that the inversion is only attempted if the target values fall within the table domain. 

For a requested target point $(s, n_B)$, the code attempts the inversion branch by branch, and it may find solutions in different branches. We prioritize the unstable, metastable and stable branches, respectively. This convention preserves the non-stable branches, since spurious stable solutions may always be constructed within the non-stable phases due to the interpolation.

Because the metastable and unstable branches fall within the first-order phase transition, they are  not regular in the ($T, \mu_B)$ plane.  We use a Delaunay-triangulation algorithm to perform a piecewise-linear interpolation on the resulting triangular mesh, instead of the optimized multilinear interpolation used for regular grids (e.g., for producing \Cref{fig:table_points}). 
However, the Delaunay triangulation interpolates over the convex hull of the scattered points, and may include regions that are not part of the original table domain, leading to spurious inverse solutions. 

To avoid these artifacts in the 2D EoS case, we build an additional envelope in the $(s, n_B)$ plane. 
Here at each fixed  $s_i$, a width
$\Delta s$ can be set by the user, across which we store the maximum baryon densities $n_B^{\rm max}$ covered by the original table. Then, a target point $(s, n_B)$ is accepted only if its baryon density lies within the envelope. This additional check prevents spurious solutions, but it requires the user to fine-tune the size of $\Delta s$ that best fits the original table. For the Holography+QvdW-HRG EoS we used $\Delta s=0.05$~fm$^{-3}$.

\begin{figure}
    \centering
    \includegraphics[width=0.9\linewidth]{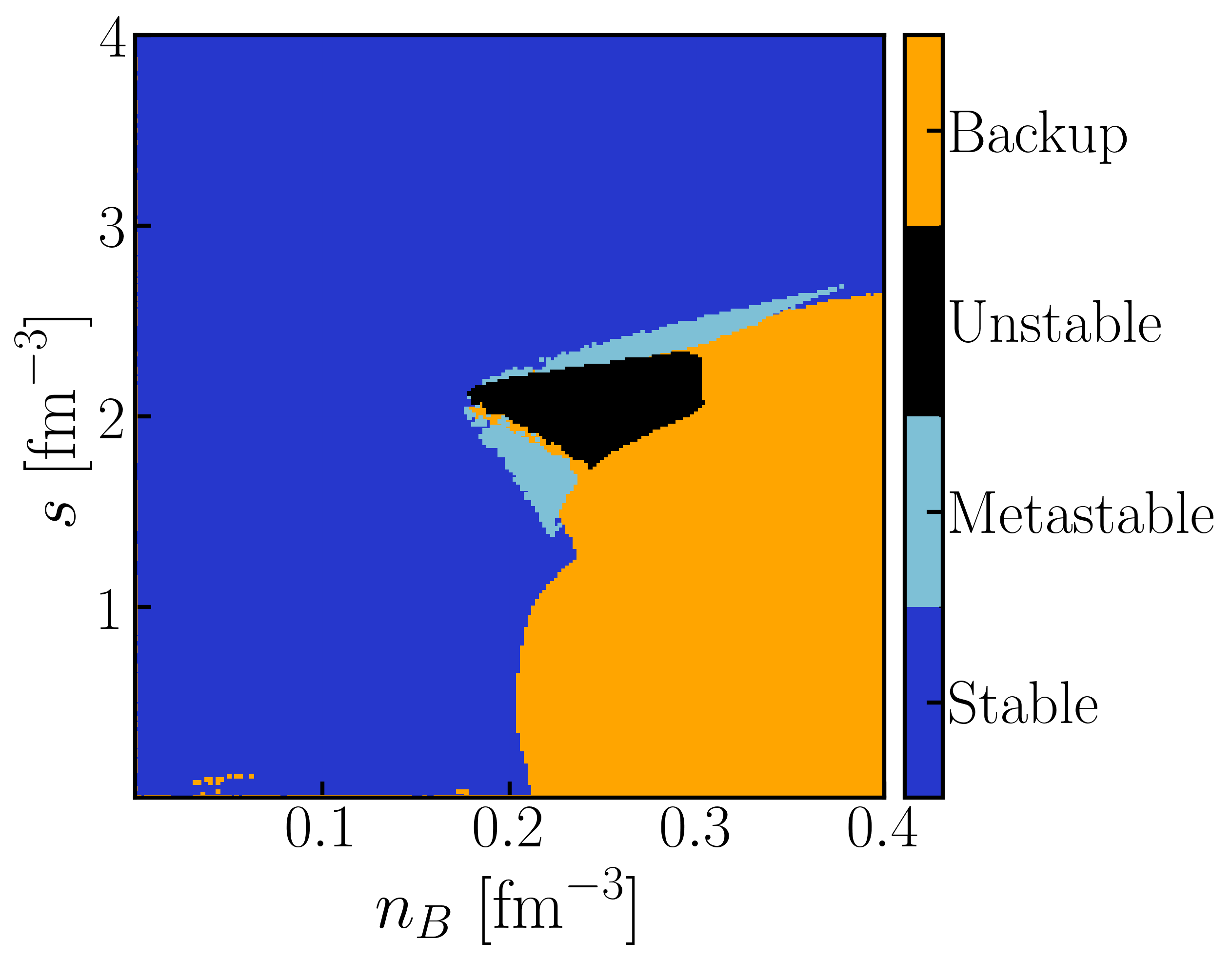}
    \caption{Points in entropy density and net-baryon density from the inverted Holography+QvdW-HRG  table. }
    \label{fig:holo_inverted}
\end{figure}

In \Cref{fig:holo_inverted} we show the resulting inversion of the table Holography+QvdW-HRG shown in~\Cref{fig:snB_meta}. For points outside of the original table domain, we use a backup EOS. 
We have checked the accuracy of reproducing our thermodynamic variables after applying the EoS Inverter and find that they are reasonable within the expected interpolation error of order $\mathcal O(1\%)$ for points far from the boundaries of the table, and not exceeding a few percent near the boundary (similar to what was found for 4D-TExS).

While the EoS used in simulations in Sec.~\ref{sec:hydro} is unstable (across the first-order phase transition line) in the sense that $\chi^B_2<0$ and/or $c_s^2<0$, it does not necessarily imply that the equilibrium pressure is negative (so $\varepsilon+P>0$). 
In fact, in our hydrodynamic simulations discussed in Sec.~\ref{sec:hydro} the pressure is positive and we do not observe numerical issues that could be associated with the lack of mechanical stability. 
However, this clearly is a topic that still requires further work (see, e.g., Ref.~\cite{Kapusta:2024nii}), but the holography EoS used in this work never reaches this point in our current applications.

The issue of how to handle the first-order phase transition is not specific to heavy-ion collisions. Closely related difficulties occur in
numerical-relativity simulations of neutron-star mergers with first-order phase transitions, where discontinuities in thermodynamic derivatives and regions with very small $c_s^2$ can create numerical instabilities or ambiguities in the hydrodynamic evolution
\cite{Ibanez:2017xrx,Aloy:2018jov,Legred:2023zet,Rivieccio:2024sfm,Chatterjee:2025zrh}.
Common strategies include smoothing or parameterizing the transition region, replacing the sharp Maxwell construction by a Gibbs or mixed-phase construction, or using piecewise representations that maintain thermodynamic consistency while avoiding pathological behavior in the numerical evolution. These prescriptions are useful regularizations of the problem, but they do not by themselves constitute a microscopic
dynamical theory of phase conversion. They instead provide controlled ways to reduce the numerical artifacts associated with evolving relativistic fluids through a sharp equilibrium first-order transition.
In this paper, we will use such an approach as a first step that will be discussed below. However, we emphasize that we consider this as a preliminary exploration, not the final answer. 

A handful of heavy-ion collision papers  have explored the inclusion of mixed phases and
phase-conversion dynamics in hydrodynamic simulations
\cite{Steinheimer:2013xxa,Kapusta:2024nii}. Such approaches are physically appealing
because they move beyond simply inserting an equilibrium coexistence construction into
a dynamical calculation, and instead begin to address how the EoS should influence the
real-time evolution of the fireball. 
Additional upgrades would be required within \texttt{CCAKE} to perform a mixed-phase study, which are straightforward but necessitate further testing. Moreover, the inclusion of a mixed phase would require significant changes within the \textit{Synthesis} module as well. Thus, we leave this to a separate study. 
Beyond that, other further validation tests would be required for a systematic study of the first-order phase transition such as
treating possible discontinuities or rapid variations in transport coefficients
\cite{Soloveva:2020hpr,Grefa:2022sav}, higher-statistics in terms of the number of events, and a careful tuning of hydrodynamic model parameters. We therefore defer a systematic treatment of
first-order phase transitions in dynamical heavy-ion simulations to future work.

\section{Heavy-ion collision simulations}
\label{sec:hydro}

\begin{figure*}
    \centering
    \includegraphics[width=\linewidth]{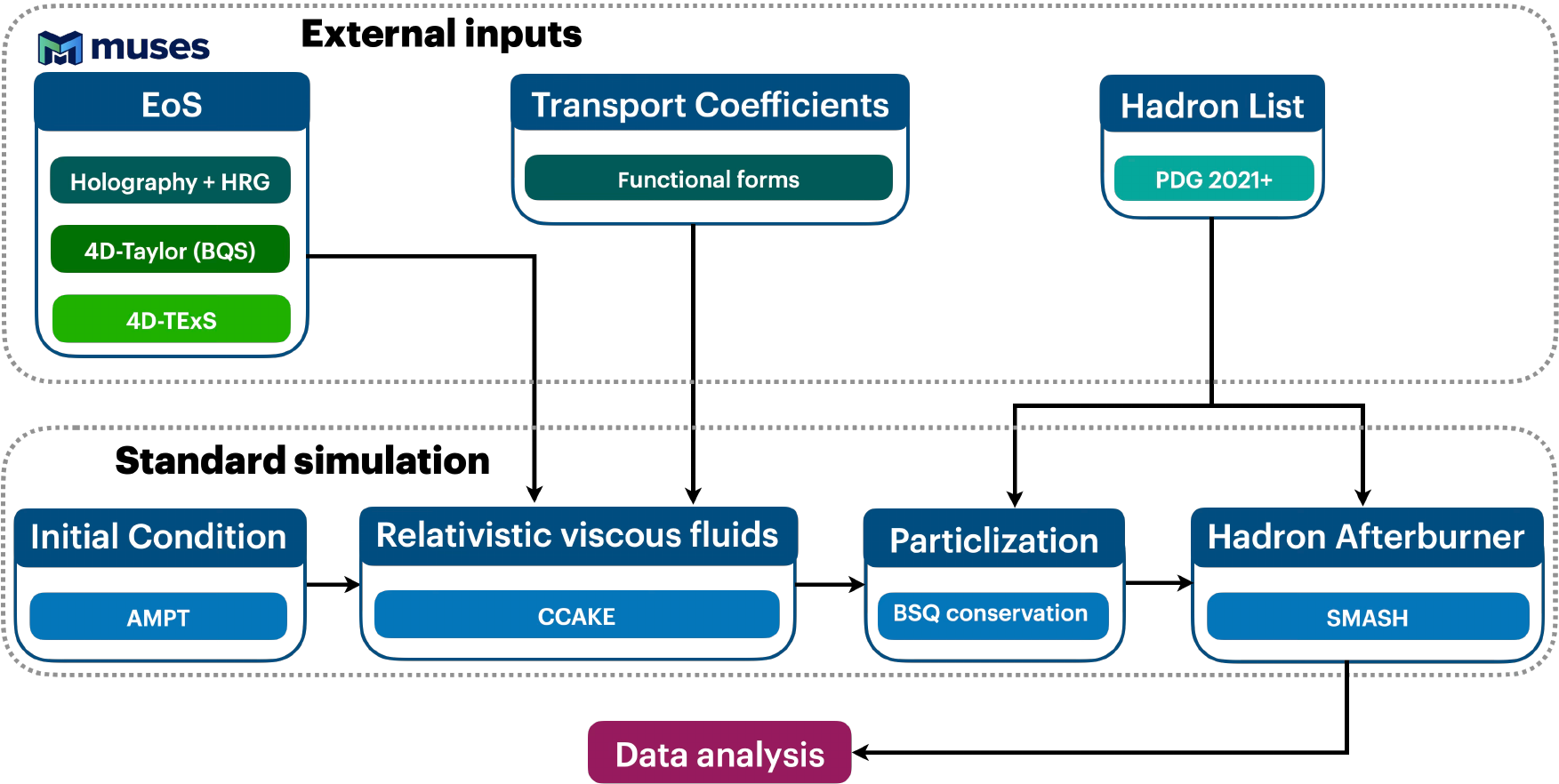}
    \caption{Simulation chain used in this work. The standard stages of heavy-ion collisions are shown, together with the model used for each stage. External input tables required by the simulation are also indicated. These inputs may come from different sources; for example, the EoS can be easily swapped to assess its impact on observables.
    }
    \label{fig:HICchain}
\end{figure*}

Theoretical constraints on the high-density region of the QCD phase diagram require reliable model-to-data comparisons, which in turn depend critically on reducing systematic uncertainties in the modeling. 
While the state-of-the-art description of relativistic heavy-ion collisions is fairly well defined in terms of modeling at high collision energies, some open questions still remain at large densities \cite{Dexheimer:2020zzs,An:2021wof,Almaalol:2022xwv,Lovato:2022vgq,MUSES:2023hyz,Sorensen:2023zkk}. In particular, for hydrodynamic simulations of the RHIC Beam Energy Scan (BES-I and BES-II), an additional layer of complexity arises from the need to initialize and dynamically evolve multiple conserved charge currents, namely baryon number, strangeness, and electric charge. This, in turn, requires an EoS that remains thermodynamically consistent over the region of phase space probed during the evolution.

A central reason for this complication is that, as the collision energy \(\sqrt{s_{NN}}\) decreases, the collision time increases and baryon stopping becomes more efficient. 
At very high \(\sqrt{s_{NN}}\), the baryons are not stopped and travel farther down the beam pipe, such that they leave a double hump structure at large forward and backward rapidity \cite{Garcia-Montero:2023gex}. As \(\sqrt{s_{NN}}\) is lowered, the baryon densities shift closer and closer to mid-rapidity until at some point the largest net-baryon density is seen at mid-rapidity \cite{Shen:2022oyg}.
These effects lead to regions where large $n_B$ or rather large $\mu_B$ can be reached in the forward and backward regime at high \(\sqrt{s_{NN}}\) \cite{Becattini:2007ci,Li:2016wzh,Li:2018fow,Brewer:2018abr,Du:2024pbd}, but this regime is not yet possible to measure experimentally. 

To quantify the \(\sqrt{s_{NN}}\) dependence, consider two nuclei colliding at nearly the speed of light. The velocity of each ion in the center-of-mass frame is determined by the beam energy per nucleon pair
\begin{equation}
    v_{\mathrm{ion}}=\sqrt{1-\left(\frac{2m_p}{\sqrt{s_{NN}}}\right)^2}\,,
\end{equation}
where \(m_p\) is the proton mass. Writing the contracted longitudinal size as
\begin{equation}
    R_{\mathrm{beam}}=\frac{2R_0 m_p}{\sqrt{s_{NN}}}\,,
\end{equation}
with \(R_0\) being the nuclear radius in its rest frame, the degree of longitudinal contraction may be characterized by the ratio \(R_{\mathrm{beam}}/R_0\). Over the BES-II energy range, this ratio is expected to lie approximately in the interval \([10^{-2},\,0.2]\). The physical significance of this contraction is that it controls the longitudinal overlap time of the two nuclei. A simple estimate for this timescale is
\begin{equation}
    t_{\mathrm{overlap}} \sim \frac{2R_{\mathrm{beam}}}{v_{\mathrm{ion}}}
    = \frac{4R_0 m_p}{\sqrt{s_{NN}}\,v_{\mathrm{ion}}}\,,
\end{equation}
which increases as \(\sqrt{s_{NN}}\) decreases. A longer overlap time enhances baryon stopping and leads to a larger deposition of baryon number in the fireball, corresponding to a larger midrapidity $\mu_B$. In addition to that, lower collision energies also imply lower initial temperatures, which creates the non-trivial combination of thermodynamics challenging the LQCD validity regime. For a deeper discussion on these details, see Refs.~\cite{Petersen:2008dd,Gale:2013da,Shen:2017bsr}.

The typical workflow of heavy-ion collision simulations 
is shown in Fig.~\ref{fig:HICchain}. For this work, we use the approach from the \texttt{NuclearConfectionary} \cite{ccakesite}:
\begin{itemize}
    \item An initial state module is used to describe the collision dynamics, and sometimes a short period following that collision. Here, we use a multi-phase transport \texttt{AMPT} \cite{Zhang:1999bd} for the initial state for Au$+$Au at collision energies $\sqrt{s_{NN}} = 7.7,\ 19.6,\ \text{and } 39~\mathrm{GeV}$ and set the normalization of the initial condition $\mathcal{N}=\{1.43,1.6,1.11\}$, transverse smearing length and longitudinal smearing length to be $\sigma_{\perp}= \sigma_{\eta}=\{0.8,1,1.2\}$, respectively. This provides 3+1D initial conditions across the BES-II and includes a modeling of the out-of-equilibrium initial state.

    \item After a time period $\tau_0= 1.0, 1.3, 2.0~\mathrm{fm}$ for $\sqrt{s_{NN}} = 39,\ 19.6,\ \text{and } 7.7~\mathrm{GeV}$, respectively, the initial condition is passed on to a relativistic viscous hydrodynamic module, where we use \texttt{CCAKE}~2.0 \cite{ccake2.0,Plumberg:2024leb}. \texttt{CCAKE} uses SPH  to solve the relativistic viscous fluid dynamic equations of motion. During this phase, the system expands and cools over time, until each individual fluid cell reaches the energy density freeze-out criterion $\varepsilon_{FO}$, fixed from the EoS as the value of $\varepsilon$ at $T=150$ MeV and vanishing chemical potentials.

    \item As an external input,  \texttt{CCAKE} requires a 2D, 3D, or 4D EoS as well as knowledge of the transport coefficients: shear viscosity $\eta T/w$, bulk viscosity $\zeta T/w$, and the $BSQ$ diffusion matrix $\kappa_{XX}/T^2$ where $X=B,S,Q$. We examine in this work the simulation choices listed in Sections \ref{subsub:model-eos} and \ref{subsub:model-transport}.

    \item After all fluid cells have reached $\varepsilon_{FO}$, their hypersurface is sampled (i.e., turning the fluid into hadrons) ensuring that $BSQ$ is conserved globally based on the PDG21+ particle list \cite{SanMartin:2023zhv}. The dynamics of the hadrons is described using \texttt{SMASH} \cite{SMASH:2016zqf}, which uses hadron transport to describe particle decays and interactions.
    \item The final output of \texttt{SMASH} is sent to our analysis scripts used to construct heavy-ion collision observables such as multiplicity, collective flow etc. 
\end{itemize}

For this work, we ran central Au+Au events at collision energies $\sqrt{s_{NN}} = 7.7,\ 19.6\ \text{and } 39~\mathrm{GeV}$, while varying the type of EoS used in the simulation, the dimensionality of the charge sectors, and whether a critical point is present or not. We summarize the free parameters chosen within the \texttt{NuclearConfectionary} in Table \ \ref{tab:hydropars}. While dynamical initialization  is possible within  the \texttt{NuclearConfectionary} \cite{Pala:2025qoa}, we instead chose to pick a later start time within \texttt{CCAKE} for the ease of direct comparisons. 

\begin{table}[h]
    \centering
    \begin{tabular}{|c|c|ccc|}
    \hline
          &   & \multicolumn{3}{c|}{$\sqrt{s_{NN}}$ [GeV]} \\
    Description     & Symbol & $39$  & $19.6$ & $7.7$ \\
    \hline
     Initial time    & $\tau_0$ & 1.0 & 1.3 & 2.0 \\
     Smearing radius    & $\sigma_{\perp} \mathrm{and} \sigma_\eta$ & 0.8 & 1.0 & 1.2 \\
      Normalization   & $\mathcal{N}$ & 1.43 & 1.6 & 1.11 \\
     Freeze-out energy    & $\varepsilon_{FO}$ & \multicolumn{3}{c|}{$\varepsilon(T=150,\mu_X = 0)$} \\
    Shear viscosity    & $\eta T/w$ & \multicolumn{3}{c|}{0.08}\\
     Bulk viscosity magnitude    & $A$ & \multicolumn{3}{c|}{1.67552} \\
    Bulk viscosity scaling    & $p$ & \multicolumn{3}{c|}{2} \\
    Diffusion   & $\kappa_{XX}$ & \multicolumn{3}{c|}{0} \\
    Dynamical initialization   &  & \multicolumn{3}{c|}{OFF} \\
    Hadron List   &  & \multicolumn{3}{c|}{PDG21} \\
     \hline
    \end{tabular}
    \caption{Chosen free parameters within the \texttt{NuclearConfectionary}. 
    }
    \label{tab:hydropars}
\end{table}

\subsubsection{EoS model choices}
\label{subsub:model-eos}
We make use of two different categories of EoSs: 2D EoSs defined in the $(T,\mu_B)$ plane where $\mu_S=\mu_Q=0$, and 4D EoSs defined in the  $(T,\mu_B,\mu_S,\mu_Q)$ phase diagram.
The EoSs from MUSES that are used in the heavy-ion collision simulation chain presented here, for which corresponding tables are also available online in Ref.~\cite{jahan_2026_20823256}, are:
\begin{itemize}
    \item {\bf 2D EoSs}: Holography (EMD) merged to QvdW-HRG with a movable critical point (${\mu_B}_c=200,598,1000$ MeV) defined by the merging procedure presented in Sec. \ref{sec:holography}. 
    When the first-order phase transition is within the range of the hydrodynamic simulations, we include the metastable and unstable regions across the first-order line (in this case, we relax the internal condition in \texttt{CCAKE} concerning the sign of $c_s^2$). 
    \item {\bf 4D EoSs}: the  4D-Taylor presented in Sec.\ \ref{subsubsec:BQS} and the 4D-TExS discussed in Sec. \ref{subsubsec:4D-TExS}. Neither one of them includes a critical point. 
\end{itemize}
It is important to note that the hydrodynamical evolution in \texttt{CCAKE} requires an EoS that is thermodynamically consistent down to $T=0$. 
This is mainly required for reasons of energy-momentum conservation and $BSQ$ charge conservation, but the exact details (as long as the EoS is thermodynamically consistent) for this low-$T$ EoS are not directly relevant since all fluid regions that reach it have long since frozen out.

We are of course aware that incorporating the unstable regime of the EoS within relativistic viscous hydrodynamics opens up the possibility of experiencing issues with mechanical stability \cite{Almaalol:2022pjc}. However, in this first exploratory work, we still perform simulations that include the unstable regime for a few different reasons. 
First, the range of $(s,n_B)$ probed in our EoSs never reaches a regime where the actual thermodynamic pressure becomes negative, only $c_s^2$, thus $\varepsilon+P$ is non-negative.
While we know  that $c_s^2<0$ will lead to loss of hyperbolicity (hence, to acausality) and possibly other problems in relativistic \emph{ideal} fluids \cite{Gavassino:2021kjm,Gavassino:2021owo,Gavassino:2023xkt}, once viscous corrections are included, the fluid characteristic velocities are not determined solely by the equilibrium $c_s^2$. In fact, it is known that they receive corrections from transport coefficients and also from the dissipative fluxes themselves in second-order hydrodynamics, as discussed in Refs. \cite{Bemfica:2019cop,Bemfica:2020xym}
\footnote{For example, consider a second-order theory with a bulk viscous scalar $\Pi$ at zero chemical potential. When $(\varepsilon+p+\Pi)>0$ and $0\leq \zeta/\tau_\Pi \leq (1-c_s^2) (\varepsilon+P+\Pi)$, with $\zeta$ being the bulk viscosity and $\tau_\Pi$ the bulk relaxation time, causality still holds in the out of equilibrium fluid even when the equilibrium $c_s^2<0$, see Ref.~\cite{Bemfica:2019cop}.}. Therefore, it is generally not well understood how a nonlinear viscous fluid behaves in the thermodynamically unstable regime, especially under the far-from-equilibrium conditions involved in heavy-ion collision simulations.
\footnote{Near global equilibrium, and in the linearized regime, violations of thermodynamic stability inequalities are expected to create instabilities, see Ref.~\cite{Gavassino:2021kjm}.} 

In our simulations, we observe that $c_s^2<0$ only occurs in a very small part of the fluid and possible effects from this are further weakened by the fact that SPH simulations naturally incorporate a coarse graining scale that smooths out gradients, acting as a UV filter that dampens out unstable modes. In other words, the influence of $c_s^2<0$ appears to be in practice too localized and too short-lived to affect the numerical simulations presented in this work. That being said, incorporating the metastable and unstable regimes allows us to at least explore time-dependent dynamics as one crosses the phase transition, and to begin to answer interesting questions regarding the amount of time it takes to cross the first-order line. Nevertheless, given the caveats above, simulations of relativistic viscous fluids containing thermodynamically unstable regions should be thoroughly investigated, as further theoretical and numerical work is needed to perform systematic comparisons with experimental data.  
Since our focus here is the equation of state rather than hydrodynamic simulations and comparison to data, we interpret the results in this section as  motivation for further dynamical studies of the effects from a first-order QCD phase transition in relativistic viscous fluid simulations. 

\subsubsection{Transport coefficient choices}
\label{subsub:model-transport}

In principle, one would like to use the same transport coefficients calculated by the \textit{Holographic} module within the fluid dynamic simulations. 
However, at the moment further developments are needed before that is possible. The \textit{Holographic} module only covers the deconfined phase so matching to hadronic transport coefficients would be needed (see,  e.g., Refs.~\cite{McLaughlin:2021dph,Danhoni:2024kgi}). 
Moreover, one would ideally like to include the appropriate critical scaling of the transport coefficients at the critical point; see, e.g., Ref.~\cite{Monnai:2016kud}.

For the runs described here, we use 
a constant $\eta T/(\varepsilon+P)=0.08$, and a bulk viscosity that is connected to the speed of sound squared via our postulated Ansatz 
to mimic the expected weak-coupling dependence of the bulk viscosity
\footnote{One could choose to use this Ansatz for $\frac{\zeta T}{w}$ instead of $\zeta/s$ but since we are only looking for qualitative effects and they both display peaks in the location of interest, this choice does not play a significant role here.} 
as well as a critical scaling component:
\begin{equation}
    \frac{\zeta }{s}\equiv A\left(\frac{1}{3}-c_s^2\right)^p\left[1+\left(\frac{\xi}{\xi_0}\right)^3\right],
\end{equation}
where the entropy density is $s$,  $A=1.67552$, $p=2$ (e.g. $p=2$ for weak coupling \cite{Arnold:2006fz}, $p=1$ for strong coupling within certain frameworks \cite{Buchel:2007mf}, and \cite{Czajka:2018bod} for a range of Ansaetze), and $\xi$ is the correlation length (see for critical scaling component \cite{Onuki:1997njj,Moore:2008ws,Monnai:2016kud,Dore:2020jye}).
At the moment, we are not able to obtain the correlation length directly from holography, but we use $\chi^B_2$ as a proxy for it, because it diverges at the critical point with $\chi_2^B\propto  \xi^2$. Thus, we can use $\xi \propto \left(\chi_2^B\right)^{1/2}$ as a way to mimic critical scaling.

Although baryon diffusion simulations are not included in this work, these EoSs were tested within \texttt{CCAKE} with diffusion turned on, and the simulation chain was confirmed to work. 
Results and effects on observables coming from baryon diffusion will be investigated in a separate, hydrodynamics-focused future work.

\subsection{Dynamical trajectories and heat maps across the phase diagram}

\begin{figure}
    \centering
    \includegraphics[width=.8\linewidth]{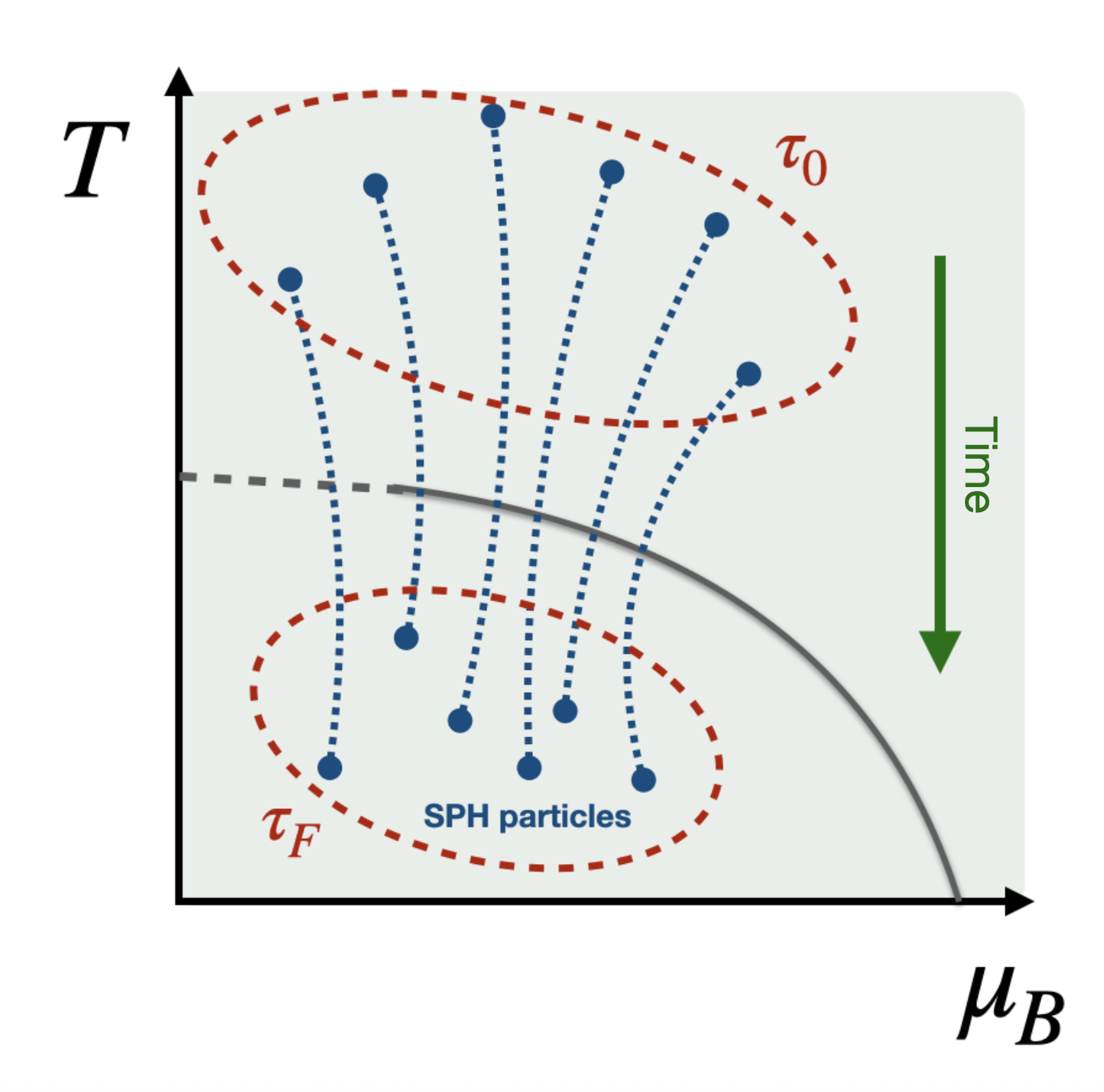}
    \caption{Cartoon of dynamical SPH trajectories through the $(T,\mu_B)$ plane. 
    In \texttt{CCAKE} the fluid is broken up into individual ``SPH particles'' using a Lagrangian method where these individual particles flow in time (they are not on a fixed grid). As such, we can track from the initial time $\tau_0$ how the SPH particles cool and trace out trajectories over time until their final time step $\tau_F$. Since our simulations are out-of-equilibrium, due to entropy production these trajectories do not follow isentropic paths.
    }
    \label{fig:SPHtraj}
\end{figure}

Previous studies have examined averaged distributions of thermodynamic quantities as functions of $\sqrt{s_{NN}}$ within grid-based hydrodynamic simulations \cite{Monnai:2024pvy,Goes-Hirayama:2025nls}. 
By contrast, the Lagrangian formulation of \texttt{CCAKE} gives the dynamical trajectories of individual SPH particles, making it possible to follow explicitly how a fluid element represented by the SPH particle cools through the QCD phase diagram in a strongly interacting, out-of-equilibrium medium. 
Since these simulations are carried out in full 3+1 dimensions, such trajectories can be resolved across a wide range of rapidities, and at each slice of rapidity, selectively.

\begin{figure*}
    \centering
    \includegraphics[width=\linewidth]{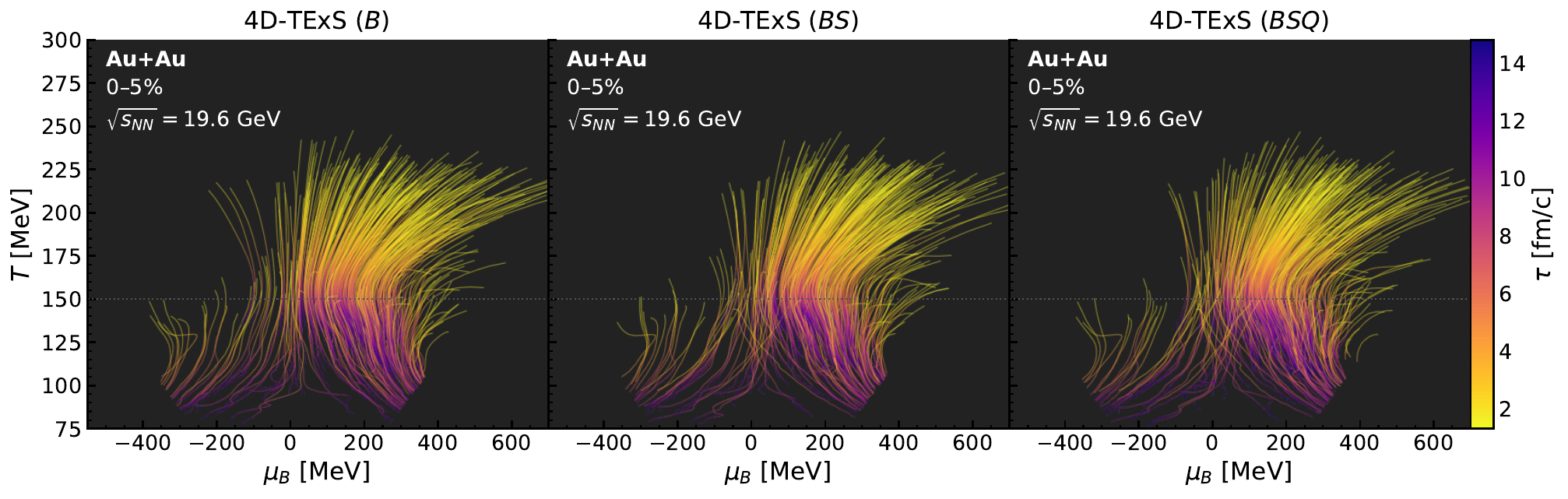}
    \caption{For a single central Au+Au event at $\sqrt{s_{NN}}=19.6$ GeV, we plot $5,000$ SPH trajectories (randomly selected) at mid-rapidity across the $(T,\mu_B)$ plane using the 4D-TExS EoS with different charges enabled (baryon density only, baryon and strangeness, or baryon, strangeness, and electric charge).  The initial times start at high $T$, and the particles cool over time such that the displayed trajectories end at the final time step of the fluid evolution.
    }
\label{fig:trajectories}
\end{figure*}

In Fig.~\ref{fig:SPHtraj} we show a cartoon of such SPH trajectories and how they may appear in the $(T,\mu_B)$ plane. The initial distribution in $(T,\mu_B)$ at $\tau_0$ comes from the initial condition itself, the normalization $\mathcal{N}$, and the EoS that maps $(s,n_B)\rightarrow (T,\mu_B)$. 
The hydrodynamic evolution dictates how the SPH particles move in time and space, and each individual SPH particle can resolve its local point in the EoS, such that we can observe their dynamics as the fluid expands and cools in time. As such, the trajectories in Fig.~\ref{fig:SPHtraj} depend on the EoS and transport coefficients. Since our simulations are out-of-equilibrium, due to entropy production these trajectories do not follow isentropic paths. Hydrodynamics runs until the last fluid cell reaches freeze-out and no fluid cells remain above freeze-out, which defines the time scale $\tau_F$. 
Because this is a field, by the time $\tau_F$ is reached, the vast majority of fluid cells have already long since frozen out and are already well-below the freeze-out line.  
These trajectories are useful to explore, because they provide insight into the range of thermodynamic variables required for the EoS in simulations, the influence of out-of-equilibrium effects, and whether critical lensing\footnote{Critical lensing \cite{Stephanov:2004wx,Nonaka:2004pg,Asakawa:2008ti,Dore:2022qyz} is the bunching of isentropic trajectories that can occur at a critical point. 
Dynamical critical lensing \cite{Dore:2022qyz} is when this effect is enhanced in an out-of-equilibrium environment.} takes place. Here we also show comparisons to the more typical heat-maps across all space-time points in the fluid, which are more comparable to what grid-based codes show. 
We note that the use of tracer particles in numerical relativity simulations from grid-based codes has enabled the study of similar ``SPH-like'' features for the analyses of binary neutron star merger simulations \cite{Bovard:2017dfh,Bovard:2017mvn}.

\subsubsection{4D-TExS (no critical point)}

\begin{figure*}
    \centering
    \includegraphics[width=\linewidth]{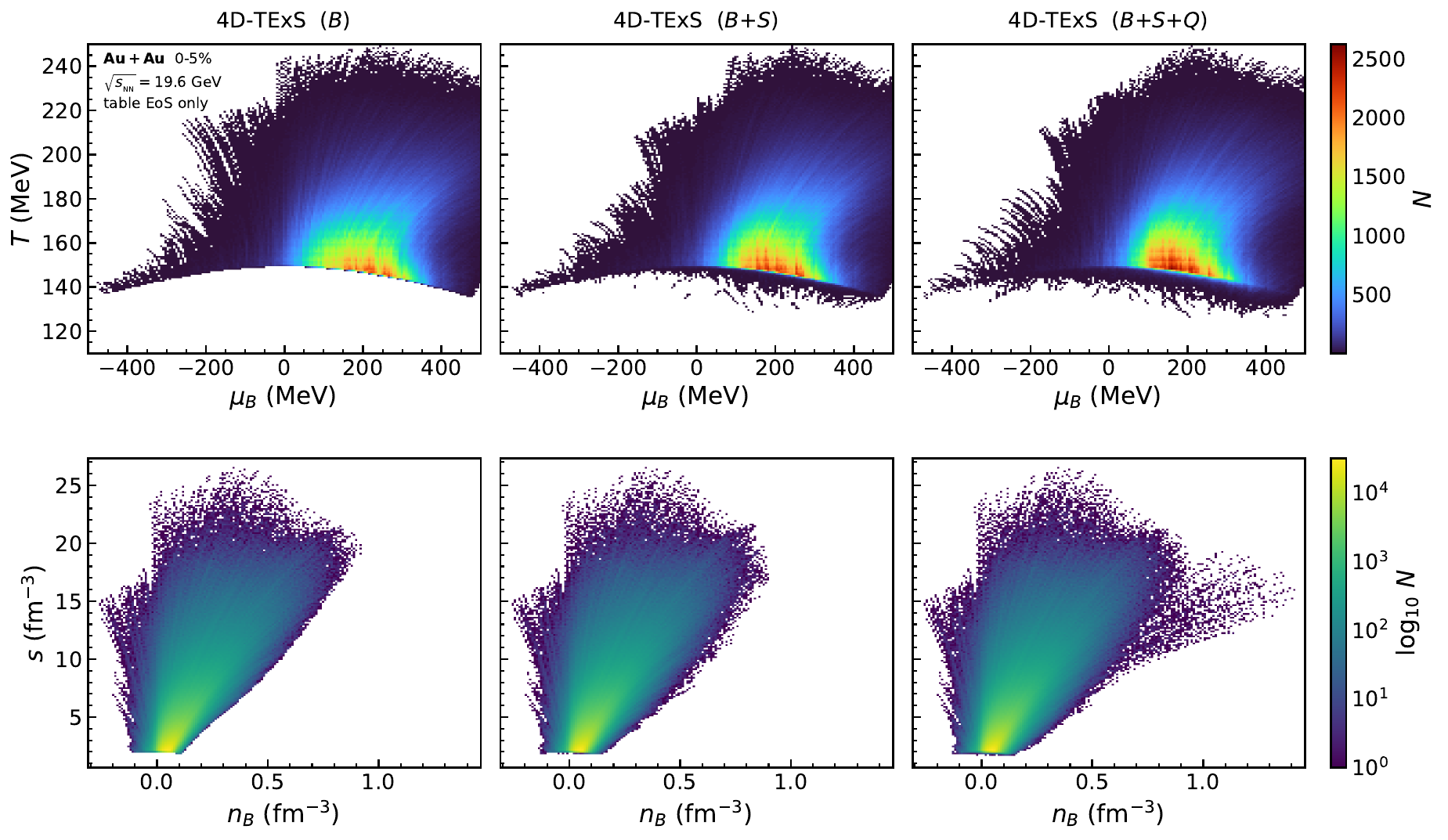}
    \caption{For a fixed, single initial condition at $\sqrt{s_{NN}}=19.6$ GeV, we plot a heat map over all SPH particles across all time steps using the 4D-TExS EoS with $B$ conservation, $BS$ conservation, or $BSQ$ conservation.  We only include regimes of the simulation above freeze-out. The top row shows the distribution in the $(T,\mu_B)$ plane and the bottom row shows the distribution in the $(s,n_B)$ plane. }
    \label{fig:heatmapBSQ}
\end{figure*}

In Fig.~\ref{fig:trajectories} we show the SPH trajectories using the 4D-TExS EoS at mid-rapidity within $|\eta| < 0.5$,
where we show results in the left panel for only baryon conservation, in the middle panel for baryon and strangeness conservation, and in the right panel all 3 conserved charges of baryon number, strangeness, and electric charge. 
Since \texttt{AMPT} initial conditions include all three conserved charges, in order to run a reduced number we simply set the initial densities to zero for the other conserved charge(s). For instance, our baryon-only conservation runs have an initial $n_B^0$ but sets $n_S^0\rightarrow0$ and $n_Q^0\rightarrow0$.  Within \texttt{CCAKE}, the full 4D EoS is always ran such that $\mu_S,\mu_Q$ are not equal to zero because the code then finds the strangeness neutral trajectory. 
We do not show large forward/backward rapidity results for this EoS because significantly fewer fluid cells would be obtained from the original EoS table itself (later results obtained with the Holography+QvdW-HRG EoS are plotted across all rapidities). 
Finally, we highlight that for 3+1D \texttt{CCAKE} simulations, we use a large number of SPH particles (around $10,000$) to ensure numerical accuracy \cite{Pala:2025qoa}, but plotting all trajectories in the entire phase space would not allow one to easily visualize the expansion. Therefore, here we only plot randomly sampled $5,000$ SPH particle trajectories instead for better visualization. 

In Fig.~\ref{fig:trajectories} we see the general behavior that we expect of fluid cell trajectories that pass through the phase diagram. They start (yellow colors) at high temperatures (and often high $\mu_B$), then they cool over time (mid-times are shown in pink) leading to a decrease in $\mu_B$. Finally, around freeze-out, the fluid cells begin increasing in $\mu_B$ again (although the temperature continues to drop). Note that fluid cells are shown way beyond the freeze-out point because they must be kept in the fluid dynamic simulations for consistency. However, after the point of freeze-out they no longer contribute to the calculation of observables. 
Since we include viscous effects in our simulations, we find that the SPH trajectories deviate from isentropes, which highlights the importance of using realistic simulations to determine how the fluid scans the different regions of the QCD phase diagram. 

One might wonder why we do not see a strong dependence of our trajectories on the number of conserved charges in the hydrodynamic simulations shown in Fig.~\ref{fig:trajectories}. That is likely because we have not included the full $BSQ$ diffusion matrix in this work.
Using ideal currents generally keeps these conserved charges separate, leading to less mixing (see, e.g., Ref.~\cite{Pala:2025qoa}). Because of the large uncertainties in the values in the diffusion matrix coefficients, we leave a careful study of their influence to future work. 
Nevertheless, we remark that the current MUSES EoSs already provide the necessary information on  equilibrium quantities, such as the $BSQ$ susceptibilities, needed in future simulations of $BSQ$ diffusion. 

In Fig.~\ref{fig:heatmapBSQ}, we plot heat maps for the same simulations shown in Fig.~\ref{fig:trajectories}. We see that the fluid simulations probe the expected range in $(T,\mu_B)$, close to what one would expect from thermal fits for $\sqrt{s_{NN}}=19.6$ GeV, with SPH particles concentrating on the red region centered around $\mu_B\sim 200$ MeV. 
While the fluid traverses a very wide region of the $(T,\mu_B)$ plane, it is clear that both the  $\mu_B<0$ and the high $T$ regimes are accessed significantly less than the low $T$ region close to freeze-out and $\mu_B\sim 200$ MeV. 
Nonetheless, we stress that a large coverage of the EoS in the phase diagram is required even for $\sqrt{s_{NN}}=19.6$ GeV. 

Looking more closely at the heat map in Fig.~\ref{fig:heatmapBSQ}, we do see some subtle differences between $B$, $BS$, and $BSQ$.  For only $B$ conservation, we see that the fluid is not quite as tightly concentrated around $\mu_B\sim 200$ MeV as in the case of $BS$ and $BSQ$. 
If we look at the same event but shown in the natural hydrodynamic variables of $(s,n_B)$, we see nearly linear lines across the $(s,n_B)$ plane without significant deviations between $B$, $BS$, and $BSQ$. 
However, the case of $BSQ$ does have some small set of fluid cells that reach to very large $n_B$ that is not seen for $B$ or $BS$. 
The differences between the $(T,\mu_B)$ plane and the $(s,n_B)$ plane show that hydrodynamic simulations without particle diffusion at this collision energy are not necessarily sensitive to the included number of conserved charges, although the coverage in $(T,\mu_B)$ looks very different from that observed in the $(s,n_B)$ plane. 

Finally, the mid-rapidity trajectories clearly illustrate the challenge posed by fluid cells that exceed the bounds of the EoS table. Since the 4D-TExS EoS has not yet been matched directly to the HRG model at low $T$, relying instead on the extrapolation of susceptibilities for the time being, a sharp cutoff for $\mu_B/T>3.5$ was implemented in the CCAKE simulation for this EoS. 
This affects the low $T$, large $\mu_B$ region in the lower-right part of these plots. Additionally, some fluid cells require combinations of entropy density and $BSQ$ charge densities that lie outside the domain of the original EoS table. In such cases, \texttt{CCAKE} switches to conformal backup EoSs to ensure energy-momentum conservation; see Ref.~\cite{Plumberg:2024leb} and Appendix~\ref{sec:backupEOS} for more details. 
While densities map smoothly onto these backup EoSs, the mapping into the $(T,\mu_B)$ plane does not. Thus, the backup EoS trajectories tend to shift the fluid cells to much higher temperatures (and often lower $\mu_B$ as well). We do not show these backup EoS trajectories here because their interpretation can be misleading, but phase-diagram trajectories determined from conformal EoSs are exemplified and discussed in Ref.~\cite{Ingles:2025yrv}. 

\subsubsection{First-order phase transition: metastable+unstable regime}\label{sec:quant_time}

As previously discussed in Sec.\ \ref{sec:workflow}, we compare two scenarios for the first-order line: a smoothed phase transition such that all points have $c_s^2>0$, and one containing both the metastable and unstable regime such that $c_s^2$ can become negative in certain parts of the EoS. 

We are able to run all of our hydrodynamic simulations until completion without experiencing any crashes, even though we do have some fluid cells that briefly experience $c_s^2<0$. 
As mentioned before, we never reach a regime of negative pressure and the region in the simulations where $c_s^2<0$ is extremely small and also very short-lived, so they do not seem to influence the stability of the numerical solver. 
We find that at $\sqrt{s_{NN}}=7.7$ GeV for the Holography+QvdW-HRG EoS with a critical point at $\mu_{Bc}=598$ MeV, there are only $3.8\%$ of all fluid cells that ever hit a point where $c_s^2<0$. 
Thus, this is a rare occurrence within a single event. 
Additionally, even the tiny fraction of fluid cells that experience $c_s^2<0$ quickly go back to a positive $c_s^2$ shortly afterwards. 

To further understand this point, 
it is convenient to define time here either in terms of a \emph{numerical time} i.e., the number of time steps $\Delta N_t$ spent with $c_s^2<0$ or in terms of a \emph{physical time} i.e., the amount of proper time $\Delta \tau$ the fluid spent in the negative $c_s^2$ regime. 
We quantify how much numerical time is spent in the $c_s^2<0$ regime via
\begin{equation}
  \langle \Delta N_t\rangle  \equiv \frac{\sum_i^{N_{SPH}}\sum_t^{N_t}\theta(-c_s^2)}{N_{SPH}N_t},
\end{equation}
where $\theta(-c_s^2)$ is the Heaviside step function which returns $1$ when $c_s^2<0$ and $0$ otherwise.
Similarly, we can quantify how much physical time is spent in the $c_s^2<0$ regime via
\begin{equation}
  \langle \Delta \tau\rangle  \equiv \frac{\sum_i^{N_{SPH}}\Delta \tau_i }{N_{SPH}N_t},
\end{equation}
where $\Delta \tau_i$ is the amount of physical time when a SPH particle $i$ has a negative $c_s^2$.
In our calculations, we obtain $\langle \Delta N_t\rangle= 3\cdot 10^{-4}$ and $ \langle \Delta \tau\rangle=1.4\cdot10^{-5}\,\rm{fm/c}$, which explains why we do not see any concerning numerical artifacts within our simulations. 
We point out that these two quantities can be related via $\langle \Delta \tau\rangle\approx dt \langle \Delta N_t\rangle$ where $dt$ is the time step size. 
Now we can proceed and compare the influence of a smoothed vs. genuine first-order transition in the simulations.

\subsubsection{Holography+HRG with critical point at different locations including metastable and unstable regimes}\label{sec:moveCP}

\begin{figure*}[!ht]
    \centering
    \includegraphics[width=\linewidth]{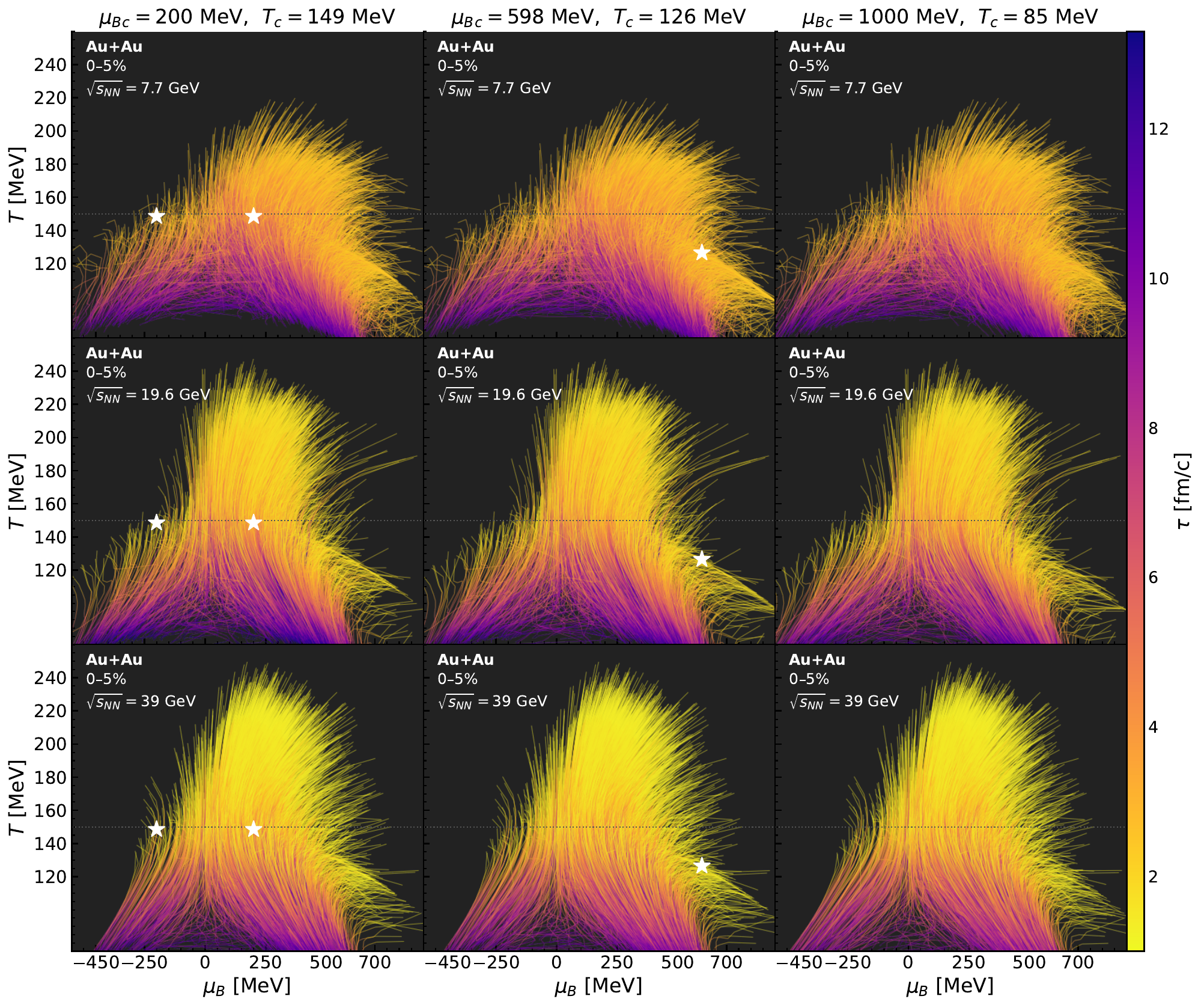}
    \caption{For a fixed, single initial condition at each $\sqrt{s_{NN}}=\left\{39,19.6,7.7\right\}$ GeV, we plot the SPH trajectories of $5,000$ SPH particles across the $(T,\mu_B)$ plane   using the Holography+QvdW-HRG EoS with different critical point positions at ${\mu_B}_c=\left\{200,598,1000\right\}$ MeV and $T_c = \left\{149,126,85\right\}$ MeV, respectively (here the metastable and unstable regions are included in the simulations).  The initial times start at high $T$, and the particles cool over time such that the trajectories end at the finite time step of the fluid (after the last SPH particle freezes out). White stars mark critical points.
    }
    \label{fig:holotrajectories}
\end{figure*}

In Ref.~\cite{Dore:2022qyz}, it was found that the existence of a critical point, the shape of the critical region in the $(T,\vec{\mu})$ plane, and the value of the transport coefficients can influence how a fluid passes through the QCD phase diagram, as well as how it approaches the critical point. Those initial simulations were done with a hydrodynamics toy model in 0+1D only. Here, we explore the influence of out-of-equilibrium effects, the location of the critical point, and the kinematics of the system on its passage through the QCD phase diagram. 

Fig.~\ref{fig:holotrajectories} shows an illustration of the influence of the critical point and the first-order phase transition line on the trajectories of SPH particles in the $(T,\mu_B)$ plane for the Holography+QvdW-HRG EoS. In that figure, we explore 3 different Holography+QvdW-HRG EoSs. The key difference is that these EoSs have critical points at different values of temperature and, more importantly, chemical potential. 
We choose first a critical point located at $T_c=149$ MeV and ${\mu_B}_c=200$ MeV which, despite being excluded by recent lattice QCD analysis \cite{Borsanyi:2025dyp}, will help us to study how the critical point location affects the spacetime evolution of the system at different rapidity ranges and beam energies of the RHIC BES program.
Our next choice is to place the critical point at $T_c=126$ MeV and ${\mu_B}_c=598$ MeV, which is 
compatible with both current lattice QCD-based prediction from Ref.~\cite{Shah:2024img} and Holographic QCD from Ref.~\cite{Hippert:2023bel}, and overall consistent with many other recent theoretical predictions \cite{Koch:2025cog}. 
The final choice places the critical point at $T_c = 85$ MeV and ${\mu_B}_c=1000$ MeV, to provide an EoS that effectively does not have a critical point in reach of the BES-II program in collider mode (we do not explore fixed-target results in this work). Furthermore, in Fig.~\ref{fig:holotrajectories} we explore three different beam energies to demonstrate the interplay between the location of the critical point and the value of $\sqrt{s_{NN}}$.

One can clearly see in Fig.~\ref{fig:holotrajectories} that, due to the range of phase space explored by different collision energies, there is less chance to pass through the critical region as one goes to higher collision energy and increases the location of the critical point in $\mu_B$. 
Considering first the
critical point at ${\mu_B}_c=200$ MeV, it would have led to strong effects at $\sqrt{s_{NN}}=39$ GeV, as SPH trajectories actually pass through the critical point both at positive and negative values of $\mu_B$ in this case. Additionally, one should also still see some effects at $\sqrt{s_{NN}}=19.6$ GeV.
However, 
for $\sqrt{s_{NN}}=7.7$ GeV collisions still with that same critical point location, the system has spread itself across a much wider range in $\mu_B$'s such that the critical point would play a smaller role in the evolution of the fluid (although still not negligible). 
We can then compare this to our other extreme
with the critical point located at ${\mu_B}_c=1000$ MeV,
that is entirely unprobed by the simulations and beam energies explored here. 
Our final case concerns 
the location of the critical point at ${\mu_B}_c=598$ MeV, the most supported by recent theoretical predictions.
We find that there are always SPH trajectories passing near the critical point across all beam energies, though only at  $\sqrt{s_{NN}}= 7.7$ GeV we see a significant number of them approaching the critical point and first-order line. We note that, at the higher $\sqrt{s_{NN}}\geq 19.6$ GeV, the SPH trajectories that pass near the critical point are predominantly at large forward and backward rapidities.

Furthermore, we note in Fig.~\ref{fig:holotrajectories} that at low $\mu_B$, the trajectories are nearly vertical, whereas trajectories that begin at larger $\mu_B$ are more distorted, displaying the bending behavior we described previously. However, we do not find significant effects from critical lensing or even significant differences in the trajectories if we move the location of the critical point. In Ref.~\cite{Dore:2022qyz}, it was explained that the degree of critical lensing strongly depends on the orientation of the critical region within the phase diagram. If the critical region spreads more along the $\mu_B$ axis, there is almost no critical lensing effect whereas a critical region that spreads across the $T$ direction can experience strong critical lensing. 
The tools to calculate kurtosis from the merged Holography+QvdW-HRG EoSs in this regime are still under development, so we currently cannot directly study the spread of the critical regime for this type of EoS across the QCD phase diagram. However, our results suggest that the critical regime is likely more tilted horizontally along the $\mu_B$ axis, not as strongly along the $T$-axis, given the lack of critical lensing effects seen here. 

While we cannot yet directly study the kurtosis for this EoS near the critical point, we can investigate the possibility of critical lensing using isentropes.  In ideal fluids both entropy and baryon number are conserved such that $s/n_B=const$. Thus, looking at different choices of constant-valued $s/n_B$ trajectories (the so-called isentropes) across the $(T,\mu_B)$ plane, can provide insight into how an ideal fluid would expand and cool over time. 
We have out-of-equilibrium effects in our simulations, so these should only be treated as qualitative guidance, but they can at least give an idea if we expect strong critical lensing effects to occur. Basically, in this scenario, one looks for ``bunching'' effects, or focusing of these trajectories around the critical point. 

\begin{figure}
    \centering
    \includegraphics[trim={0 0 0 0},clip,width=\linewidth]{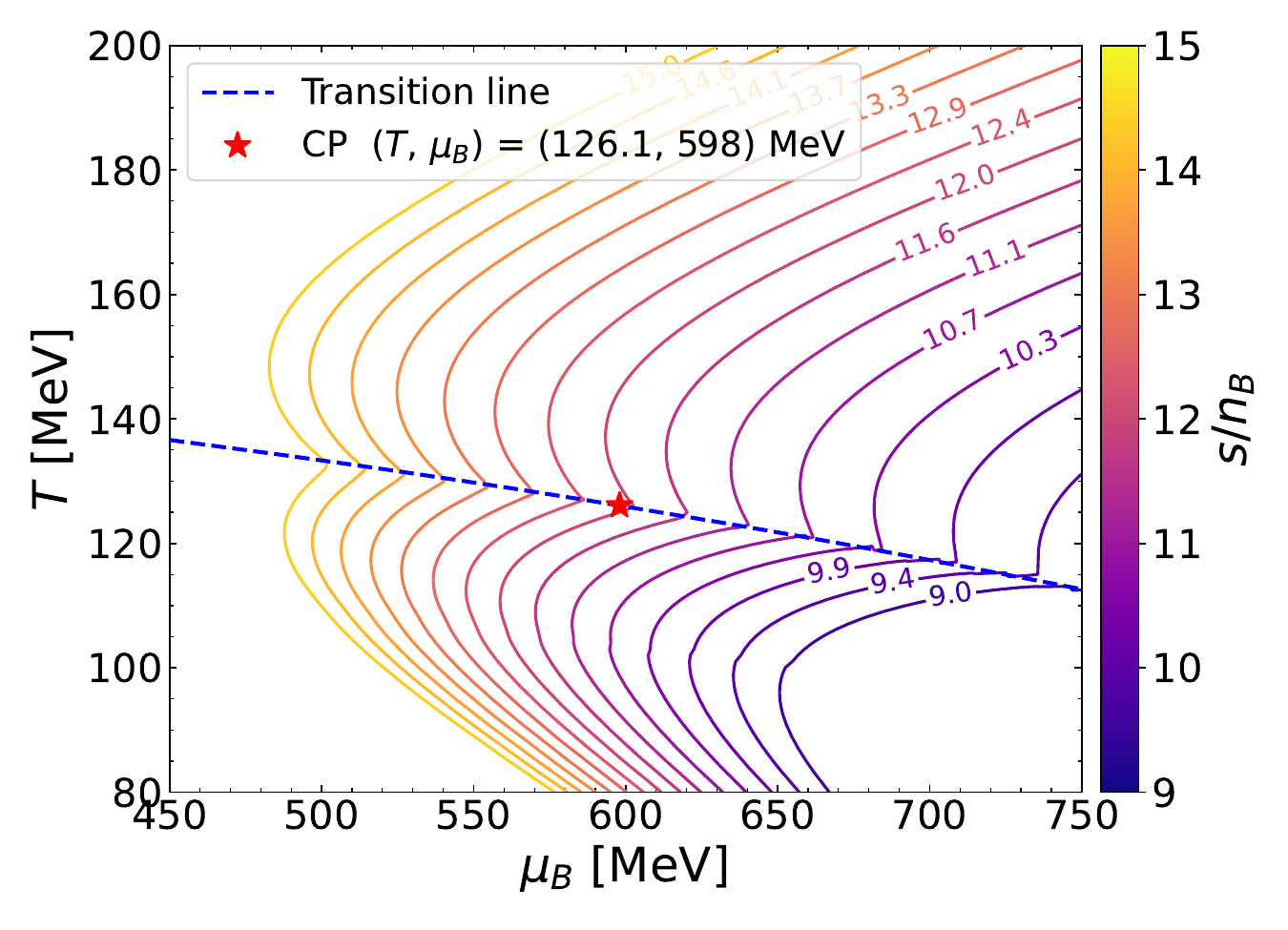}
    
    \caption{Isentropic trajectories computed using the Holography+QvdW-HRG EoS, emulating the path that an ideal fluid would travel across the QCD phase diagram. Here we zoom in to show the critical region for the holography EoS with ${\mu_B}_c=598$ MeV.
}
    \label{fig:isentropes}
\end{figure}

In Fig.~\ref{fig:isentropes}, we plot these trajectories close to the critical point for the ${\mu_B}_c=598$ MeV Holography+QvdW-HRG EoS. While there is a small kink in the trajectories as the phase transition is crossed, there does not appear to be any bunching of the trajectories, which leads us to conclude that this EoS does not demonstrate any significant critical lensing effects. 
Additionally, this figure helps us understand why the trajectories do not show a pronounced critical-lensing pattern, even when the critical-point location is varied.

Even though critical lensing is not strongly seen in our simulations, we do notice more trajectories that bunch across the first-order phase transition line. All plots shown here include the metastable and unstable regime in the EoS, and it is still unclear how much influence that has, compared to a mixed phase. We further note that the lack of critical lensing does not imply that a critical point does not necessarily affect observables computed in hydrodynamic simulations. Nevertheless, since we do not have strong guidance at the moment from first-principles about the precise orientation of the critical region, it is very hard to gauge the relevance of critical lensing effects in the hydrodynamic simulation of the matter formed in heavy-ion collisions. 

\begin{figure*}
    \centering
    \includegraphics[width=\linewidth]{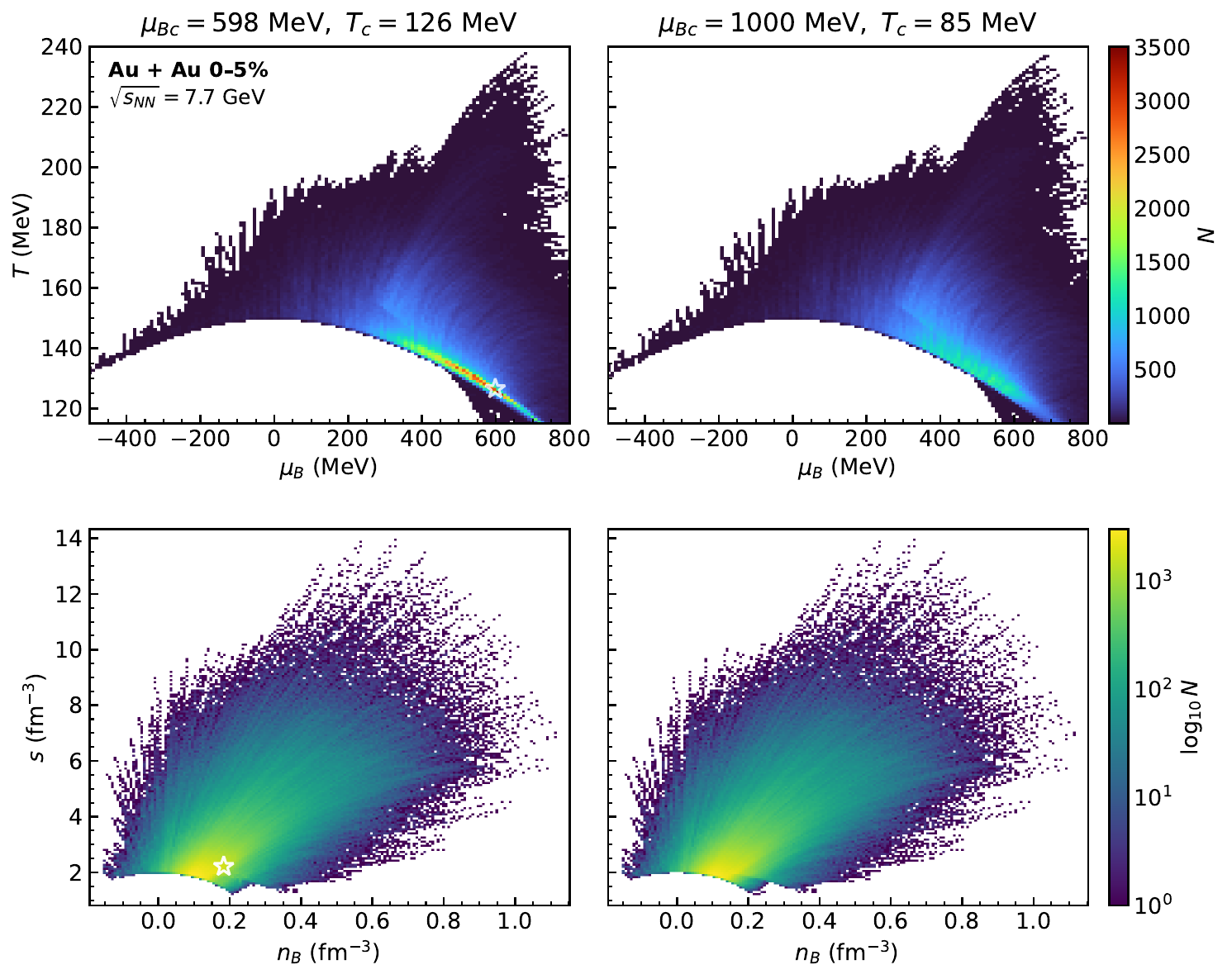}
    \caption{For a fixed, single initial condition at $\sqrt{s_{NN}}=7.7$ GeV, we plot the heat map over all SPH particles across all time steps, using the Holography+QvdW-HRG EoSs (including the metastable and unstable regimes) with different critical point positions at ${\mu_B}_c=\left\{598,1000\right\}$ MeV (left, right respectively). We only include regimes of the simulation above freeze-out. The critical point is indicated by a star.   
}
    \label{fig:heatmap}
\end{figure*}

Fig.~\ref{fig:heatmap} shows the distribution of SPH particles as heat maps across the $(T,\mu_B)$ plane (top row) and $(s,n_B)$ plane (bottom row). Only points at, or above freeze-out are considered.  Here, we only plot the results from the $\sqrt{s_{NN}}=7.7$ GeV beam energy run and compare the two critical point locations defined by ${\mu_B}_c=598$ MeV and ${\mu_B}_c=1000$ MeV (left and right panel, respectively). Looking first at the $(T,\mu_B)$ plane, we see that the fluid spends most of its time close to the freeze-out line where the dominant $\mu_B$ range is about $\mu_B\sim [300,600]$ MeV. Temperatures tend to remain relatively low, only about $20-30$ MeV higher than the freeze-out line at most. 
If we then compare to the bottom row in Fig.~\ref{fig:heatmap}, we see almost no difference caused by the presence of the critical point. However, it is clear that we are probing the positive $n_B$ regime significantly more. For ideal fluids, this plot would only contain straight lines across $(s,n_B)$ because no entropy is produced. Instead, we see clear deviations from a linear behavior in $(s,n_B)$, especially looking at the right of these plots where one can see non-linear bending in the heat map due to entropy production. 

One important feature that comes out of the heat map plot in Fig.~\ref{fig:heatmap}, top panel is the confirmation of the bunching in trajectories across the first-order line. We see up to three times more SPH fluid cells that pass across the first-order line for the ${\mu_B}_c=598$ MeV EoS compared to the ${\mu_B}_c=1000$ MeV EoS. This may hint at subtle signs of \emph{dynamical} critical lensing \cite{Dore:2022qyz}, even though it was not clear from the isentropic trajectories (i.e., critical lensing in equilibrium). We leave further investigations of its consequences for future work. 

\begin{figure*}
    \centering
    \includegraphics[width=\linewidth]{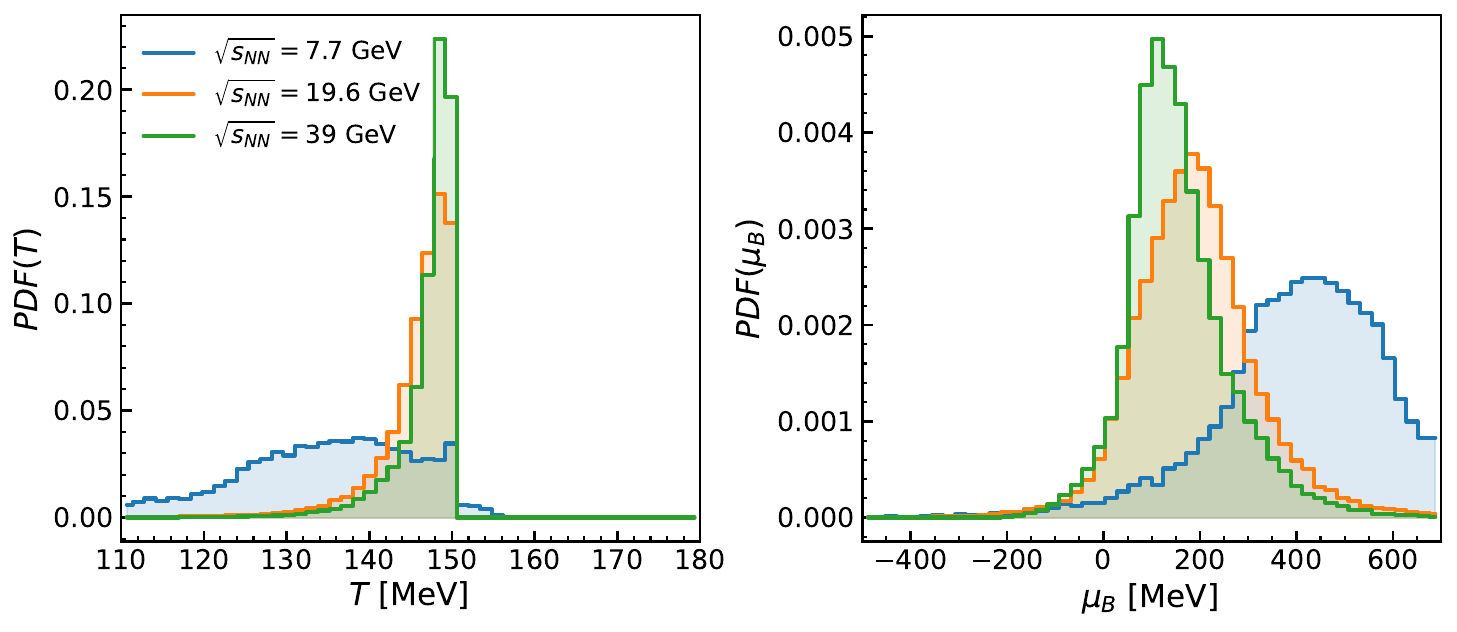}
    \caption{Probability distribution function (PDF) of $T$ (left panel) and $\mu_B$ (right panel) at freeze-out, for a single event at different collision energies, using the Holography+QvdW-HRG EoS with a critical point at $(T,\mu_B)=$(126, 598) MeV (with metastable and unstable branches included in th simulations). }
    \label{fig:freezeout}
\end{figure*}

One may wonder how the spread in the fluid cell trajectories influences the final freeze-out across $(T,\mu_B)$.  In approaches like thermal models, they obtain the average $\langle\langle T^{FO},\mu_B^{FO}\rangle\rangle$, averaged over all fluid cells in a given event and averaged over many events as well \cite{STAR:2017sal}. Here, we can explore this distribution in a single event instead, to see how broad of a probability distribution we obtain, and also if our averaged $\langle T^{FO},\mu_B^{FO}\rangle$ across all fluid cells in a single event is close to what one would obtain from thermal fit analyses. 

In Fig.~\ref{fig:freezeout}, we show the probability distribution of the freeze-out points for all the SPH fluid cells, for the same events as shown in Figs.~\ref{fig:holotrajectories} and \ref{fig:heatmap}. In the left panel of Fig.~\ref{fig:freezeout}, we find that the temperature range is quite sharply peaked around $T\sim 150$ MeV (note that our constant energy density was chosen to reproduce $T^{FO}=150$ MeV at $\mu_B=0$) for the higher beam energies. 
As we lower the beam energy to $\sqrt{s_{NN}}=7.7$ GeV, we see significantly more spreading in temperature, which is natural because the $\mu_B$ increase shifts the freeze-out to lower $T$ as well. We note that $\sqrt{s_{NN}}=7.7$ GeV has a very small number of fluid cells that freeze out at $T^{FO}>150$ MeV, which arises from the small handful of fluid cells that have frozen out on backup EoSs.   

Concurrently, we show the $\mu_B^{FO}$ distribution of these freeze-out points in the right panel of Fig.~\ref{fig:freezeout}. We find a significantly wider distribution in $\mu_B^{FO}$ (as compared to $T^{FO}$) across all beam energies. As expected from thermal fits, the peak of the distribution shifts to larger $\mu_B^{FO}$ as $\sqrt{s_{NN}}$ decreases. 
The differences between $\sqrt{s_{NN}}=39$ GeV and $\sqrt{s_{NN}}= 19.6$ GeV are relatively small (and their distributions themselves have significant overlap), whereas at $\sqrt{s_{NN}}=7.7$ GeV we see a genuine shift to much larger $\mu_B^{FO}$. Comparing to thermal fit results from the STAR beam energy scan \cite{STAR:2017sal}, the average thermal fit values for $0-5\%$ central are  $\langle\langle \mu_B^{FO}\rangle\rangle\sim 100$ MeV at $\sqrt{s_{NN}}\gtrsim 39$ GeV, $\langle\langle\mu_B^{FO}\rangle\rangle\sim 190$ MeV at $\sqrt{s_{NN}}\gtrsim 19.6$ GeV, and $\langle\langle\mu_B^{FO}\rangle\rangle \sim 400$ MeV at $\sqrt{s_{NN}}\gtrsim 7.7$ GeV, which are all consistent with the location of the peak of our probability distributions. 
A major takeaway from Fig.~\ref{fig:freezeout} is the need for an EoS used that can probe a wide region of the QCD phase diagram in hydrodynamic simulations. In fact, we see that the mean of the distribution does not cover the entire range of chemical potentials reached at freeze-out. Thus, one should not determine the range covered by an EoS in the $(T,\mu_B)$ plane needed in hydrodynamic simulations solely from thermal fit analyses.

\subsection{Rapidity dependence}

\begin{figure}
    \centering
    \includegraphics[width=\linewidth]{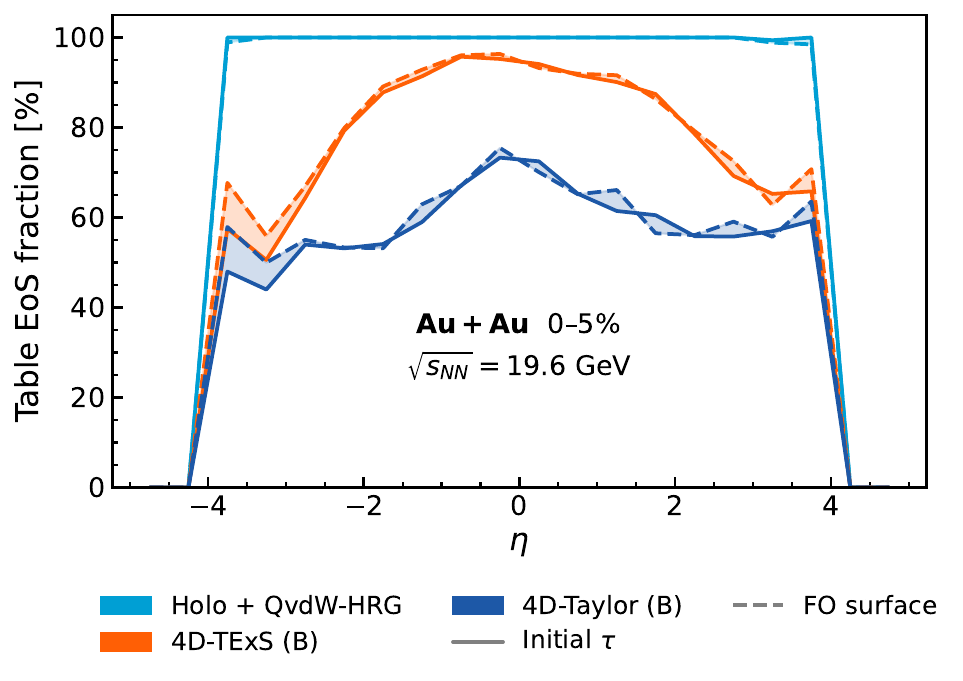}
    \caption{Spatial rapidity distribution of fluid cells that are on the EoS table validity range, where all fluid cells at tracked until they freeze out. We use the Holography+QvdW-HRG EoS (with the metastable and unstable regimes) with a critical point at $(T,\mu_B)$=(126,598) MeV, compared to 4D-Taylor and 4D-TExS EoSs for baryon charge only. The initial times (above freeze-out) are shown in solid lines while the freeze-out surface is shown with dashed lines.  } 
    \label{fig:eta_table}
\end{figure}

Baryon stopping is fundamentally a longitudinal phenomenon; its consequences are thus most naturally discussed in terms of the rapidity dependence of the deposited baryon number \cite{Kapusta:1982vp,Busza:1983rj,BRAHMS:2003wwg}. Careful examination of the robustness of the model results across rapidity range is an important checkpoint for comparisons to experimental data \cite{Shen:2022oyg,Du:2023efk,Elfner:2025ojd,Goes-Hirayama:2025nls,Constantin:2026hwh}. 
It is generally understood that larger baryon densities are reached at larger rapidities (and we have seen that in the plots obtained from our simulations throughout this Section) \cite{Biedron:2006vf,Becattini:2007qr,Li:2016wzh,Li:2018fow,Begun:2018efg,Brewer:2018abr,Du:2023gnv,Li:2023kja}.  
This larger range of $n_B$ reached at forward and backward rapidities has a number of consequences for the EoSs, that we will explore here. 

In Fig.~\ref{fig:eta_table}, we compare the percentage of fluid cells (at freeze-out or above) that fall into the range covered by our 2D EoS tables: 4D-Taylor with only baryon conservation (orange), 4D-TExS with only baryon conservation (navy blue), and Holography+QvdW-HRG (including the metastable and unstable regimes) with the critical point at $(T,\mu_B)=$(126,598) MeV (light blue), for collisions at $\sqrt{s_{NN}}=19.6$ GeV. 
For each EoS, the band represents the percentage of fluid cells on the table for both the initial and final times, where we use their respective rapidity at that specific time step.
Within \texttt{CCAKE}, the backup EoS does not allow one to return to the original table EoS, which implies that the initial time will always have the same or greater \% of fluid cells on the table compared to the late time. However, fluid cells can shift in rapidity over time so the solid and dashed lines in Fig.~\ref{fig:eta_table} do not have a fixed hierarchy due to this effect. 

As expected, results computed using the Holography+QvdW-HRG EoS are nearly entirely from the original EoS table itself; only a small fraction of fluid cells at very forward and backward rapidities are out-of-bounds. These fluid cells predominantly come from the low $T$, large $n_B$ regime where the QvdW-HRG EoS simply cannot reach a large enough $n_B$ to accommodate it (at low $T$, the $n_B(\mu_B)$ relation increases very slowly with $\mu_B$ MeV, such that one would require an extremely large $\mu_B > 3500$ MeV to reach $2n_{sat}$). Rather, one would need to switch to a hadronic EoS relevant for dense matter, such as for examples EoSs determined using relativistic mean-field theory models, to accommodate this regime. We also do not find nearly any time dependence for the percentage of the fluid cells coming from the holography EoS, since the vast majority can be found directly on the table even until late times. 

Now comparing the previous scenario to the results found using the 2D lattice QCD-based EoS, we find that the underlying expansion used to create the EoS (either the standard Taylor expansion or the $T^\prime$-Expansion Scheme) can affect the \% of fluid cells that appear on the table versus rapidity. 
As expected from its smaller range of validity, at $\sqrt{s_{NN}}=19.6$ GeV the 4D-Taylor expansion has the least coverage, with only approximately $50-70\%$ of its fluids cells that can be accounted for directly on the corresponding EoS table at mid-rapidity. Mid-rapidity is the most optimistic regime, as forward/backward rapidity (i.e., $|\eta| > 2$) only has approximately $40-60\%$ of fluid cells within the table's range. 
Because we use backup EoSs in \texttt{CCAKE}, this result implies that our simulation at $\sqrt{s_{NN}}=19.6$ GeV employing the Taylor expansion EoS is heavily influenced by the backup EoSs, indicating that one should not use it for hydrodynamic simulations of these beam energies (or lower), and only restrict its use to collisions at $\sqrt{s_{NN}} \leq 39$ GeV.
Thankfully, the 4D-TExS EoS has a significantly larger coverage, such that initial times have around $90-95\%$ of fluid cells at mid-rapidity on the corresponding EoS table.  
At forward / backward rapidity, more fluid cells resort to backup EoSs, but in our simulations this only becomes a significantly problem for $|\eta|>2$.
Hence, we conclude that most experimental observables could reasonably be studied using the 4D-TExS EoS for collision energies $\sqrt{s_{NN}}\gtrsim 19.6$ GeV, without getting too much influence from the backup EoSs.

\begin{figure}
    \centering
    \includegraphics[width=\linewidth]{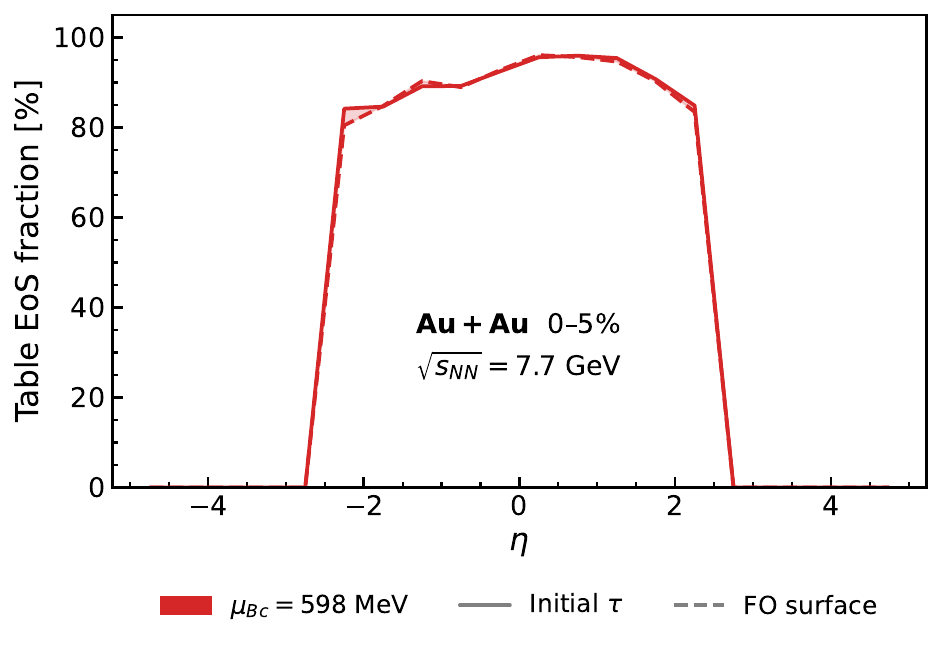}
    \caption{Spatial rapidity distribution of fluid cells that are on the EoS table validity range where all fluid cells at tracked until they freeze out. We use the Holography+QvdW-HRG EoS (with the metstable and unstable regimes) with a critical point at $(T,\mu_B)$=(126,598) MeV. The initial times (above freeze-out) are shown in solid lines while the freeze-out surface is shown with dashed lines.}
    \label{fig:eta_table_7p7}
\end{figure}
\begin{figure}
    \centering
    \includegraphics[width=\linewidth]{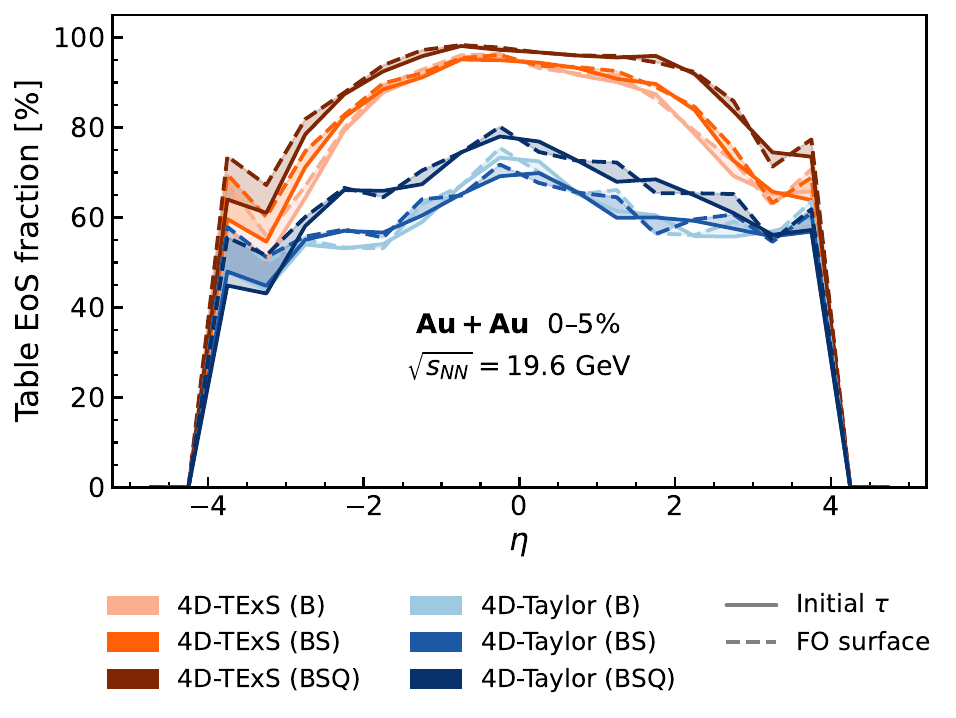}
    \caption{Spatial rapidity distribution of fluid cells that are on  the EoS table validity range where all fluid cells at tracked until they freeze-out. We use our two 4D EoSs, 4D-Taylor, and 4D-TExS, while systematically varying the number of conserved charges we include in the hydrodynamic evolution. The initial times (above freeze-out) are shown in solid lines while the freeze-out surface is shown with dashed lines. }
    \label{fig:eta_table2}
\end{figure}
\begin{figure*}
    \centering
    \includegraphics[width=\linewidth]{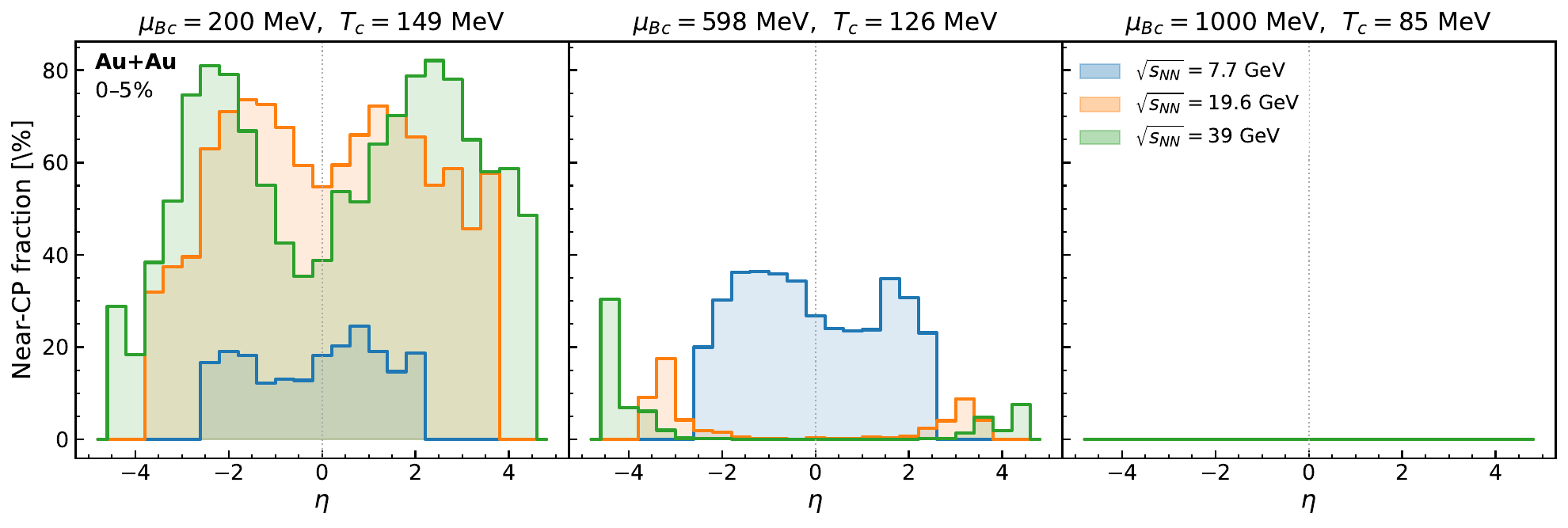}
    \caption{Spatial rapidity distribution of fluid cells that are ``near the critical point'' (defined inside the ellipse with semi-axes $\sigma_T = 25$ MeV and $\sigma_{\mu_B} = 100$ MeV)  for different critical point positions and beam energies at freeze-out.  }
    \label{fig:nearCP}
\end{figure*}

In Fig.~\ref{fig:eta_table_7p7}, we show the same plot for the Holography+QvdW-HRG EoS in the case of a $\sqrt{s_{NN}}=7.7$ GeV simulation. For this collision energy, the percentage of fluid cells covered by the corresponding EoS table is lower than in Fig.~\ref{fig:eta_table}, due to the larger $\mu_B/T$ ratios. 
At mid-rapidity, we find that only approximately $90\%$ of fluid cells comes directly from the table EoS. 
This indicates that one must be careful when using the Holography+QvdW-HRG EoS at lower energies. 

Next, in Fig.~\ref{fig:eta_table2}, we study the effect of using a 2D, 3D, or 4D EoS table with lattice QCD-expanded EoSs; in other words, we switch on only 1 conserved current in hydrodynamics (baryon number), 2 conserved currents (baryon number and strangeness), or 3 conserved currents (baryon number, strangeness, and electric charge). 
Once again, it is clear that the 4D-TExS EoS does substantially better than 4D-Taylor expansion for $|\eta|<2$. One interesting effect is that including the full $BSQ$ table with all three conserved charges turned on actually \emph{improves} the percentage of fluid cells on the table, compared to using only 1 or 2 conserved charges within hydrodynamics. 
An important point here is that the $\mu_B$ range of the trajectories is strongly correlated to the rapidity of the fluid. In fact, fluid cells with lower $\mu_B$ tend to be at mid-rapidity whereas the larger $\mu_B$ fluid cells tend to appear at larger forward or backward rapidity. 
There are two important consequences of this effect: first that the fluid cells passing across the critical point appear at large forward/backward rapidity, and second that most of the fluid cells that are not in the table's range tend to be at forward/backward rapidity. 

On Fig.~\ref{fig:nearCP}, we show the distribution of all fluid cells across rapidity versus those that are near the critical point (as defined as within $2\sigma$ of its location) for a single event in Au+Au collisions at the three different beam energies, using the Holography+QvdW-HRG EoS (including the metastable and unstable regimes) with critical points at the three different locations.
Using the unrealistically low critical point of ${\mu_B}_c=200$ MeV, we find that a large percentage of the fluid cells pass near the critical point ($\sim 40\%-60\%$ at mid-rapidity) across nearly all rapidities, which highlights once again that if the critical point was at such low ${\mu_B}_c$, its effects on hydrodynamic simulations would have been substantial.
We also find that, for the lowest beam energy of $\sqrt{s_{NN}}=7.7$ GeV, only about $20\%$ of fluid cells at mid-rapidity pass near the critical point. 
As a control simulation, we also study the critical point that is at a very large value of ${\mu_B}_c=1000$ MeV, for which no fluid cells passing near in our simulations, regardless of the beam energy. 

Moving on to our more realistic critical point at ${\mu_B}_c=598$ MeV, we observe a complicated interplay between $\sqrt{s_{NN}}$ and rapidity. For our highest beam energies of $\sqrt{s_{NN}}=19.6$ GeV and $\sqrt{s_{NN}}=39$ GeV, we find fluid cells that pass near the critical point at large forward and backward rapidity, but not at mid-rapidity. At $\sqrt{s_{NN}}=7.7$ GeV, approximately $\sim 20\%-40\%$ cells at mid-rapidity pass near the critical point. This illustrates that, when investigating the possible effects from a critical point, is its important to use a broader rapidity range of $|\eta|<2$. 

\begin{figure}
    \centering
    \includegraphics[width=\linewidth]{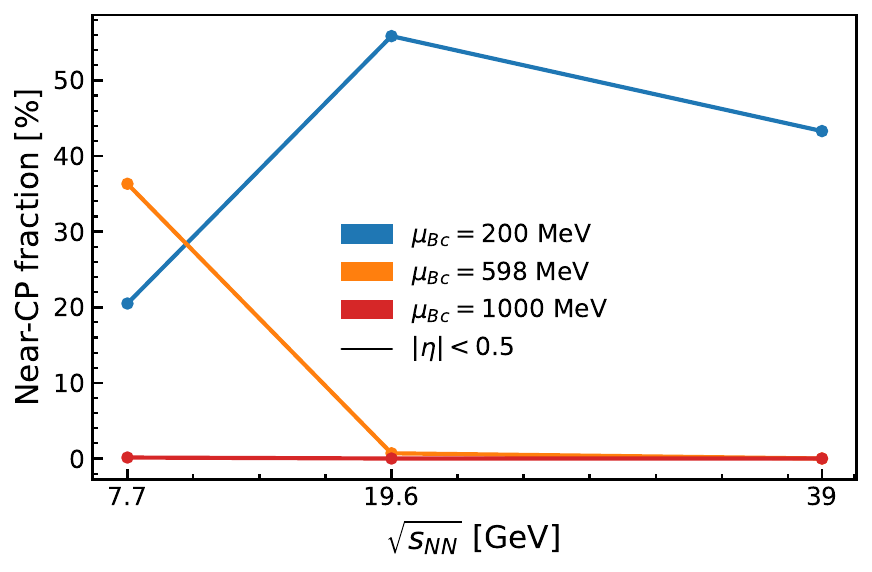}
    \caption{Percentage of fluid cells that enter the near-CP region at any time step of their dynamical lifetime. Results are shown for $0-5\%$ Au+Au collision simulations performed using Holography+QvdW-HRG EoSs with critical points at different locations (including the metastable and unstable regimes). 
    }
    \label{fig:cellwcut}
\end{figure}

We can summarize these results by averaging over the rapidity range covered by STAR, as shown in Fig.~\ref{fig:cellwcut}. 
There, we compare the percentage of fluid cells that enter the near-critical point region defined by the ellipse with semi-axes $\sigma_T = 25$ MeV and $\sigma_{\mu_B} = 100$ MeV, for the three different CP locations (${\mu_B}_c = 200$, $598$, and $1000$ MeV) and three collision energies ($\sqrt{s_{NN}} = 7.7$, $19.6$, and $39$ GeV) discussed until now, within the rapidity window $|\eta| < 0.5$. 
Note that we include fluid cells that reach this region both above and below freeze-out, with the idea that there may even be potential signatures as one approaches the critical point from below. We verified (although not shown here) that restricting the selection only to fluid cells that are still above freeze-out does not significantly change this result. 

For ${\mu_B}_c = 598$ MeV, only the lowest collision energy $\sqrt{s_{NN}} = 7.7$ GeV shows sensitivity, with almost $40\%$ of fluid cells entering the near critical point region, dropping to nearly zero at higher energies. 
For ${\mu_B}_c = 1000$ MeV, no sensitivity is observed across all collision energies, as expected. 
In contrast, when the CP is placed at ${\mu_B}_c = 200$ MeV, all three collision energies exhibit significant sensitivity, with fractions ranging from $\sim 20\%$ at $\sqrt{s_{NN}} = 7.7$ GeV to a peak of $\sim 55\%$ at $\sqrt{s_{NN}} = 19.6$ GeV, and $\sim 45\%$ at $\sqrt{s_{NN}} = 39$ GeV. This shows how the knowledge about the location of the critical point is important for heavy-ion simulations, as it determines which beam energies are most sensitive to its presence.

\subsection{Multiplicity across beam energies}
\label{sub:rapidity-dep}

Formalisms used to model initial conditions of heavy-ion collisions often include a free parameter known as the normalization constant that determines the overall energy scale of the system (see, e.g., Ref.~\cite{Moreland:2014oya}), which is typically fit to multiplicity data. In this exploratory work, which is focused on the EoS, we do not perform a detailed study involving systematic comparisons to experimental observables, because that would require more fine tuning with the hydrodynamic free parameters (e.g., transport coefficients, initial times, etc.) and substantially larger computational time. 
Thus, for the sake of illustration, we pick a single normalization per $\sqrt{s_{NN}}$ that is held constant and  provides reasonable results, and use the same value for all EoSs shown here. Therefore, it should be clear to the reader that a more systematic comparison, going beyond the initial illustrative study done here, is needed to draw stronger conclusions about the behavior of experimental observables in hydrodynamic simulations with EoSs covering a wide range of the phase diagram, which feature a tunable critical point. We leave such a detailed study to a future publication.

\begin{figure}[h!]
    \centering
    \includegraphics[width=0.99\linewidth]{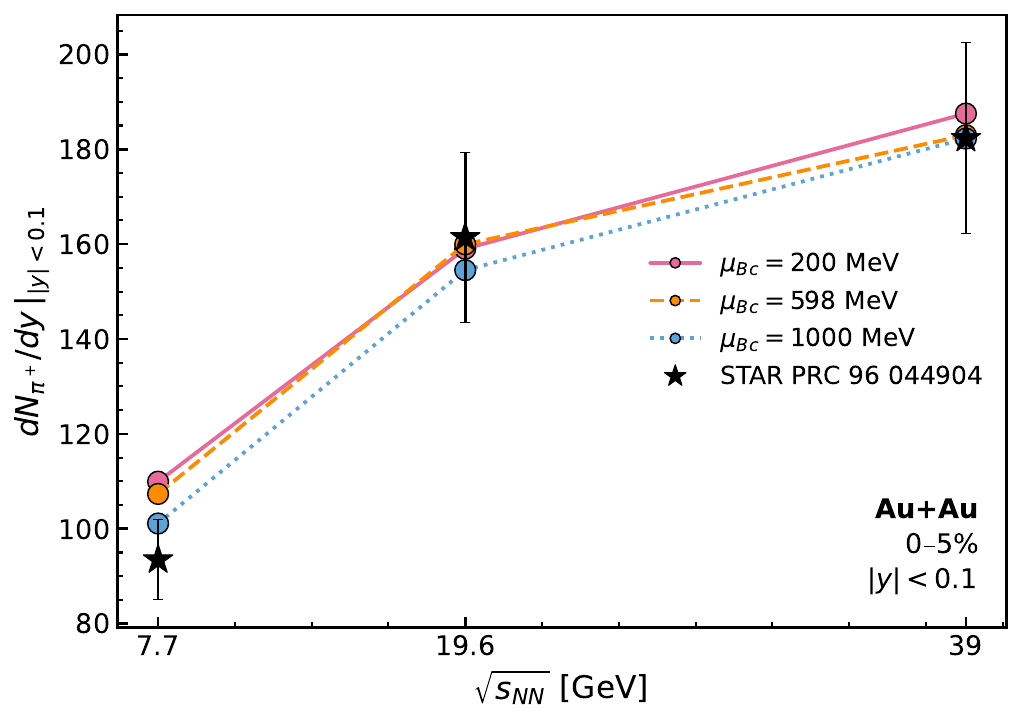}
    \caption{Collision energy dependence of the $dN/dy$ for $\pi^+$ at mid-rapidity ($|y|<0.1$), using Holography+QvdW-HRG EoSs with different critical point locations.}
    \label{fig:dndy}
\end{figure}

Keeping the caveat above in mind, we present in Fig.~\ref{fig:dndy} the multiplicity of $\pi^+$ at mid-rapidity using the Holography+QvdW-HRG EoSs with different critical points. We tune our normalization per $\sqrt{s_{NN}}$ to obtain pion data (see Tab.~\ref{tab:hydropars}), and observe that our results describe the data reasonably well across the full range of beam energies considered. The critical point located at ${\mu_B}_c=1000$ MeV yields a somewhat lower multiplicity of pions in this setup; however, the agreement with the data remains acceptable. Furthermore, the choice of critical point location has only a small effect on the total pion yield, indicating that the EoS does not strongly influence the overall pion multiplicity in our simulations. This suggests that the normalization is mostly insensitive to details of the phase structure encoded in the EoS. However, small adjustments will be required from EoS to EoS to make precise comparisons to data in future dedicated work. 

\begin{figure*}
    \centering
    \includegraphics[width=\linewidth]{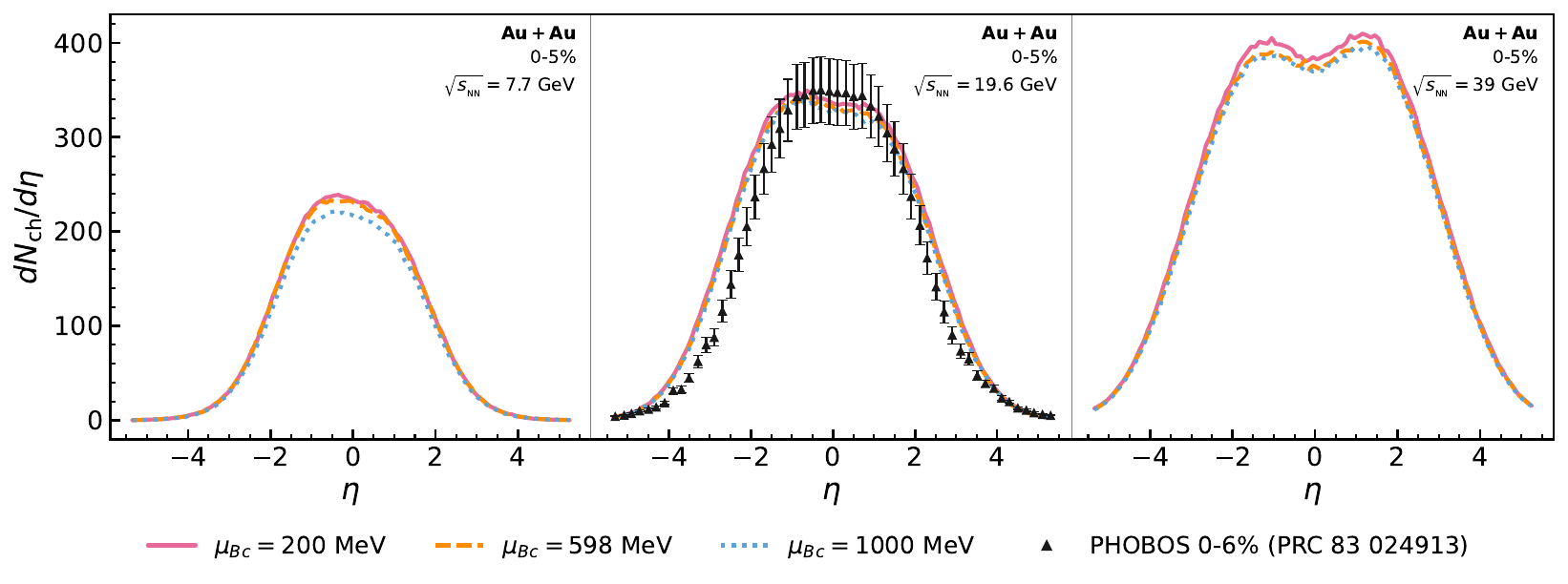}
    \caption{Results for the $dN/d\eta$ vs. $\eta$ distribution for all charged particles in Au+Au collisions at $\sqrt{s_{NN}}=19.6$ GeV, compared to PHOBOS data \cite{Back:2002wb} for $0-5\%$ centrality class (middle panel). Results for $\sqrt{s_{NN}}=7.7,39$ GeV in the $0-5\%$ centrality class are also shown in left and right panel, respectively, though no experimental data is available in these cases. Results of the simulations performed using Holography+QvdW-HRG EoSs with different critical points are compared.
    }
    \label{fig:dNdy_196}
\end{figure*}

In Fig.~\ref{fig:dNdy_196}, we show our results for $dN/d\eta$ for all charged particles in Au+Au collisions at $\sqrt{s_{NN}}=7.7,19.6, 39$ GeV using our Holography+QvdW-HRG simulations (single event) varying the location of the critical point. We find that all EoSs provide a reasonable descriptions of the PHOBOS data from Ref.~\cite{Back:2002wb} at $\sqrt{s_{NN}}=19.6$ GeV (no data is available yet for the other beam energies). We see very little deviations between the different EoSs for the width of the distribution, but slight differences exist at mid-rapidity that are consistent with our findings in Fig.~\ref{fig:dndy}. 
Here we note that the width of this distribution is strongly dependent on the underlying choice of the smearing length in AMPT, which we chose here to be $\sigma=1.2$ fm for $\sqrt{s_{NN}}=19.6$ GeV. Generally, smaller $\sigma$ decreases the width of the distribution while a larger $\sigma$ broadens the width.

\subsection{Consequences of limited EoS range on hydrodynamic observables}\label{sec:consequence}

Given the challenges we have discussed above concerning the direct use of the EoS obtained from lattice QCD in hydrodynamic simulations, we would like to quantify now how the different approaches used to reconstruct the EoS from the lattice may affect hydrodynamic observables.  
One can see in Fig.~\ref{fig:eta_table} that the 4D-TExS EoS provides significantly better coverage in the phase diagram than the 4D-Taylor EoS. However, even with the improved coverage of the 4D-TExS EoS, we found in our simulations that approximately 5-10\% fluid cells at mid-rapidity for $\sqrt{s_{NN}}=19.6$ GeV are out-of-bounds in central collisions, and lower beam energies further enhance this problem. 

In order to quantify this issue, we use the Holography+QvdW-HRG EoS that has nearly full coverage across the $\sqrt{s_{NN}}$ and rapidities explored here. For a single central event of $0-5\%$ centrality, we run it with all free parameters fixed to the exact same value (using the same initial condition) with the only difference that we compare the results obtained from the full Holography+QvdW-HRG EoS to the results obtained from the same EoS but with a cut that imposes $\mu_B/T<3.5$. 
This value is chosen to correspond to the evaluated limit of validity for the $T^\prime$-Expansion Scheme at finite $\mu_B$ \cite{Borsanyi:2021sxv}.
For this study, we use the EoS with a critical point located at $\mu_B^{CP} =1000$ MeV, to avoid any contamination effects from the critical point and first-order phase transition on our results. 
For this cut EoS, all out-of-bound fluid cells go to the backup conformal EoS (see Appendix~\ref{sec:backupEOS} for more details). While the backup EoS provides a thermodynamically consistent treatment everywhere, it is not identical to the Holography+QvdW-HRG EoS. Generally, for a given value of $(s,n_B)$, the backup conformal EoS leads to a higher $T$ and lower $\mu_B$ than what the Holography+QvdW-HRG EoS would give. 

\begin{figure*}
    \includegraphics[width=0.99\linewidth]{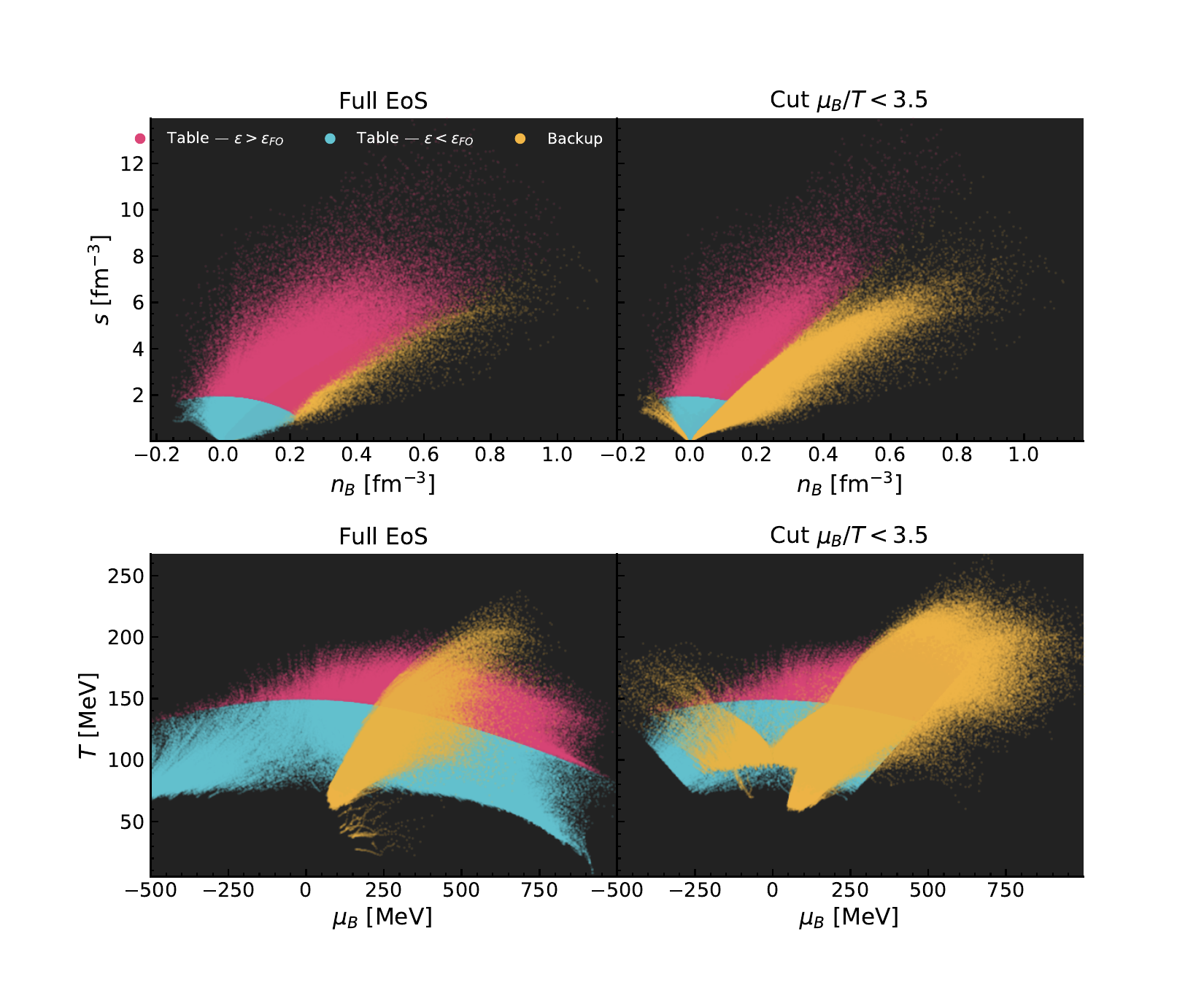}
    \caption{Display of the regions probed in the QCD phase diagram for a single Au+Au collision event at $\sqrt{s_{NN}}=7.7$ GeV in the $0-5\%$ centrality class. 
    Top row: entropy vs. net-baryon density reached within a single \texttt{CCAKE} simulation using the full Holography+QvdW-HRG with $\mu_B^{CP}=1000$ MeV (left) vs. the same EoS but implementing a cut at $\mu_B/T<3.5$ (right).
    Bottom row: the temperature and baryon chemical potential regime reached in the same simulations. 
    The fluid cells with thermodynamic quantities accessible from the table above freeze-out are shown in red and below freeze-out are shown in blue. Cells coming from the backup EoS are shown in orange.
    }
    \label{fig:snB_TmuB}
\end{figure*}

In Fig.~\ref{fig:snB_TmuB}, we investigate the case of a single Au+Au collision event at $\sqrt{s_{NN}}=7.7$ GeV in the $0-5\%$ centrality class.  In the top row in Fig.~\ref{fig:snB_TmuB} we plot the natural hydrodynamic variables $(s,n_B)$ both for the full EoS (Holography+QvdW-HRG with $\mu_B^{CP}=1000$ MeV) versus the cut EoS (Holography+QvdW-HRG with $\mu_B^{CP}=1000$ MeV but constraining $\mu_B/T<3.5$ such that out-of-bound fluid cells go to the backup EoS). 
We can see that, for the full EoS (left), there are still a handful of out-of-bound fluid cells, marked in orange, that are entirely concentrated in the low $s$, large $n_B$ regime. These fluid cells are out-of-bound due to intrinsic limitations of the Holography+QvdW-HRG EoS in terms of its range in $n_B$. Essentially, here one can continue to increase $\mu_B$ to very large values but $n_B$ remains nearly identical. 
Future work will try to  merge QvdW-HRG to CMF to extend the regime of validity of the hadronic phase. 
However, in the current paper, these fluid cells are nearly all well below freeze-out and/or at very large forward/backward rapidity.  As such, they do not significantly affect our simulations. 

However, once we apply the cut in $\mu_B/T<3.5$, we see that significantly fewer fluid cells have thermodynamic quantities that can be found on the EoS table (as expected) and a significant number of them are now above freeze-out, see Fig.~\ref{fig:snB_TmuB} (top, right). Additionally, even for  $n_B<0$ we begin to see EoS contamination due to resorting to the backup EoS at low $s$ that was not there when the $\mu_B/T<3.5$ cut was not implemented.  
In the $(s,n_B)$ phase space, the effects of this backup EoS contamination predominately appear to spread the range of the phase space of the event to slightly lower $s$ and larger $n_B$. 

The other challenge that we find here is the non-trivial mapping between $(s,n_B)$ (our natural hydrodynamic variables within \texttt{CCAKE}) vs. $(T,\mu_B)$ (our EoS variables). 
Here, the blue region maps SPH fluid cells \emph{below} freeze-out that are within bounds of the EoS table, and the pink region maps fluid cells \emph{above} freeze-out from the EoS table. 
Comparing the top row to the bottom row of Fig.~\ref{fig:snB_TmuB}, we observe that the cells above freeze-out span a much larger region in the $(s,n_B)$ coordinate systems, while the region covered by cells below freeze-out largely dominates in $(T,\mu_B)$ coordinates.
The reason for this discrepancy is that at low temperatures or entropies, one requires a very large $\mu_B$ to add or subtract a baryon whereas at high temperatures or entropies this process is comparatively easier. 

Coming back to the discussion on the contamination from the backup EoS, we already see some effects from EoS contamination in the full EoS case, looking at the orange fluid cells in the $(T,\mu_B)$ plane (Fig.~\ref{fig:snB_TmuB} bottom, left plot). These backup conformal EoSs have higher $T$ and lower $\mu_B$ than the table, which is problematic if there are too many of them at freeze-out. 
Once we apply the $\mu_B/T<3.5$ cut in the EoS, we find a significant amount of EoS contamination  that affects both the negative and positive $\mu_B$ regimes. 
Thus, in this case, these simulations include a significant amount of backup EoS contamination, which can affect future comparisons to experimental data. 

\begin{figure}
    \centering
    \includegraphics[width=1\linewidth]{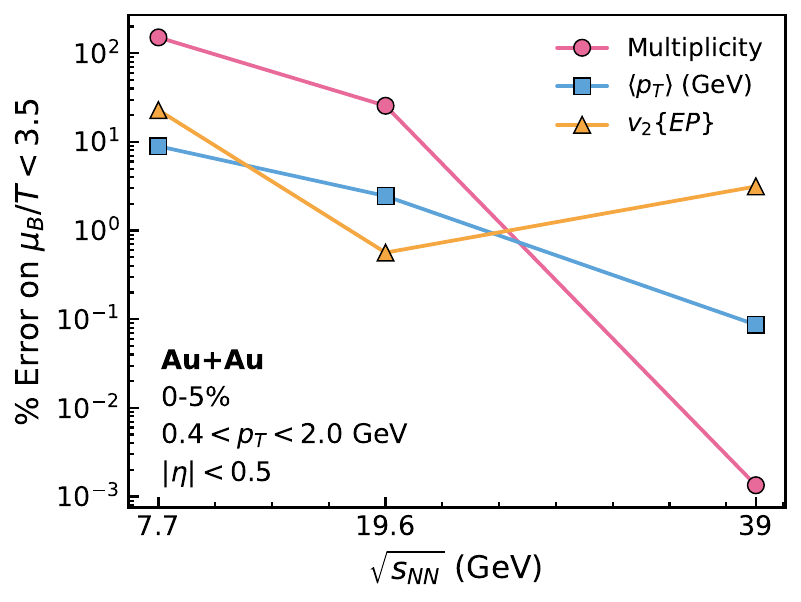}
    \caption{
    Percentage of error due to backup EoS contamination on different hydrodynamic observables as a function of center-of-mass collision energy, due to limitations of the lattice QCD-based expansion schemes 
    at $\mu_B/T<3.5$.
    We compare total multiplicity, average transverse momentum and elliptical flow in a single event, using STAR's rapidity cuts of $|\eta|<0.5$ and the Holography+QvdW-HRG EoS with a critical point at $\mu_B^{CP}=1000$ MeV.
    }
    \label{fig:EOScut}
\end{figure}

To quantify the effect of backup EoS contamination, 
we calculate the percentage of change for common hydrodynamic observables due to our implemented cut of $\mu_B/T<3.5$ on the EoS. It helps estimating how much the observables are affected by regions in the simulation that are not from the original EoS table. We use the following formula:
\begin{equation}\label{eqn:EOS_err}
    \frac{|\mathcal{O}_{full}-\mathcal{O}_{cut}|}{\mathcal{O}_{full}} \times 100,
\end{equation}
where $\mathcal{O}_{full}$ is the observable calculated using the full EoS and $\mathcal{O}_{cut}$ is the value of the observable using the cut of $\mu_B/T<3.5$ within the Holography+QvdW-HRG setup. 
We remind the reader that all other parameters are fixed in our simulation such that the effect only comes from implementing the  $\mu_B/T<3.5$ cut in the EoS itself. 
Since we perform this calculation for a single event, we only study the absolute value of this percentage error because this uncertainty may have non-trivial fluctuations on an event-by-event basis. Further work is needed in this regard.  

In Fig.~\ref{fig:EOScut}, the effect on the hydrodynamic observables coming from backup EoS contamination is shown. 
We consider as hydrodynamic observables the total multiplicity $dN/dy$, the average transverse momentum $\langle p_T\rangle $, and the elliptical flow $v_2\left\{EP\right\}$ (since this is for only a single event we use the event-plane method), all measured within $|\eta|<0.5$. 
As expected, we find a larger error coming  from backup EoS contamination due to out-of-bound fluid cells as one moves towards lower collision energies. 
The elliptic flow error is not larger than $20\%$ across all beam energies, but given that this is a single event it would likely require more events to determine more precisely the error from EoS contamination. The mean transverse momentum is the least sensitive observable, with error below $10\%$ across the different energies.

The observable most sensitive to the EoS contamination is the overall multiplicity. As mentioned before, in hydrodynamic simulations, the overall multiplicity can be tuned by varying the normalization $\mathcal{N}$ (see, e.g., Tab.\ \ref{tab:hydropars}). However, if there is a genuine issue due to the limitations in the range of the EoS that causes EoS contamination, and this affects the choice in $\mathcal{N}$, such an effect can lead to simulations that run either significantly longer or shorter than they should.
The reason for this is that $\mathcal{N}$ sets the overall temperature scale of the system, such that larger $\mathcal{N}$ implies that the system reaches larger $T$, while smaller $\mathcal{N}$ leads to smaller $T$. In these simulations, we found that the effect of EoS contamination leads to a significantly larger multiplicity. Thus, those  simulations would then be forced to choose a significantly smaller $\mathcal{N}$ in order to reproduce experimental data, which implies that the hydrodynamic phase would be significantly shorter as well.  This shorter hydrodynamic phase would then lead to a suppression of flow observables. 

In summary, our simulations indicate that limitations coming from lattice QCD EoSs expanded to finite $\mu_B$ can affect the calculation of hydrodynamic observables, which must be kept in mind and resolved before more accurate, and systematic, model-to-data comparisons can be made in the future.

\section{Summary and Outlook}
\label{summary}

In this work, we introduced the latest MUSES Calculation Engine version, \textit{Calliope}, to be released publicly upon acceptance of this manuscript for publication. This version is a major step forward from the initial release that focused primarily on cold, $\beta$-equilibrated neutron-star physics. \textit{Calliope} will include a total of 17 MUSES modules, as well as upgrades to previous modules, together with the implementation of a new GUI to make the use of the MUSES Calculation Engine more accessible. 
We also demonstrate compatibility of the MUSES EoSs with an external solver. This latest version of the Calculation Engine provides a comprehensive and modular description of the QCD EoS across a wide region of the phase diagram relevant to heavy-ion collisions and dense-matter applications. 
In the hadronic regime, the new \textit{HRG} module, incorporated via Thermal-FIST, offers three EoS options (ideal gas, excluded volume, or QvdW), all of which provide a reasonable comparison of the hadronic phase EoS to lattice QCD calculations in the transition region. 
In the deconfined sector, MUSES provides lattice-QCD-based expansion schemes (including the \textit{4D-Taylor} and \textit{4D-TExS} modules), which are extrapolations to finite chemical potentials that maintain consistency with continuum lattice data and the Stefan–Boltzmann limit; and the holographic \textit{NumRelHolo} module (based on the EMD model) that reproduces lattice QCD results, includes a critical point followed by a first-order line -- constrained by a Bayesian analysis -- and can extend to large $\mu_B$.  Together, these modules provide multidimensional, thermodynamically consistent EoS tables in a unified format, facilitating their direct use in hydrodynamic simulations and multi-module workflows within the MUSES Calculation Engine.

The long-term goal of the MUSES Collaboration is the ability to produce unified, multi-regime descriptions of strongly interacting matter. A key development towards this goal is the extension of the \emph{Synthesis} module, which introduces a generic, thermodynamically consistent framework to combine EoSs in 1D-2D across different regions of the phase diagram, while accommodating different types of phase transitions (crossover, critical point and first-order).
As a first demonstration, we combined the QvdW-HRG and holographic EoSs featuring critical points at three different locations, illustrating how hadronic and deconfined descriptions can be consistently connected into a single global EoS, while preserving thermodynamic stability and allowing for critical behavior. 
Together with the \emph{EoS Inverter}, these developments enable multidimensional EoSs to be generated in forms directly usable by hydrodynamic solvers.
At the same time, our discussion on how the EoSs produced by MUSES in the $(T,\mu_B)$ plane map into the $(s,n_B)$ plane highlights the present limitations of existing finite-density EoSs. 
While the new expansion method employed by the \textit{4D-TExS} module  substantially enlarges the region accessible to lattice-QCD-based EoSs compared to conventional Taylor expansions as produced by the \textit{4D-Taylor} module, neither these approaches nor the current holographic EoS alone cover individually the full thermodynamic domain explored during low-energy heavy-ion collisions. Even the Holography+QvdW-HRG EoSs fail to cover high-enough baryon density as reached in the lowest BES program collision energies. 
Mapping the EoSs into hydrodynamic variables, we made these limitations explicit and demonstrates that extending the available coverage in the $(s,n_B)$ plane remains thus an essential challenge. 
Within the MUSES framework, these limitations naturally motivate future synthesis with effective models of dense matter, such as CMF, in order to provide unified EoSs spanning the full range from lattice QCD to high-density nuclear matter.
Furthermore, the holographic EoS contains a gap along the first-order phase transition line if one does not include metastable and unstable points.
The MUSES \textit{Calliope} release will offer a solution, by making this regime available for users, so that they can decide on the physics they want to use to fill in this gap in $(s,n_B)$; we remark that, thanks to the \textit{EoS Inverter} module, one will also be able to explore the phase diagram and generate EoSs in the ($\varepsilon,n_B$) plane as well.

To make direct comparisons to heavy-ion collisions, one requires a dynamical framework, for which the EoS is an input. 
In this work, we have demonstrated the compatibility of our MUSES multi-dimensional EoS with the external solver \texttt{CCAKE}~2.0, which is part of the \texttt{NuclearConfectionary} dynamical framework. 
The MUSES EoSs are now publicly available and can thus be used in other external solvers to determine the hydrodynamic evolution of heavy-ion collisions.
In particular, the EoS tables used in this work can be found online in Ref.~\cite{jahan_2026_20823256}.
Using this setup, we performed the first systematic studies of lattice-QCD-based and critical-point EoSs within realistic $3+1$D viscous hydrodynamics. These simulations demonstrate that the thermodynamic coverage of current lattice-QCD expansion schemes remains a limiting factor already at intermediate Beam Energy Scan energies, specifically at $\sqrt{s_{NN}} = 19.6$ GeV as shown here. 
Although the use of the 4D-TExS EoS substantially reduces the number of fluid cells leaving the tabulated region by a factor 3 as compared to Taylor-expanded EoS, backup EoSs are still required, in particular at large forward/backward rapidity. This introduces sizable systematic uncertainties in bulk observables at low beam energies, up to a $20\%$ error for elliptic flow at $\sqrt{s_{NN}}\lesssim 19.6$ GeV, and nearly $100\%$ error for the multiplicity at $\sqrt{s_{NN}}\lesssim 7.7$ GeV, when limiting the available EoS to the region of $\mu_B/T<3.5$ covered by the $T^\prime$-Expansion Scheme. 
These results show that future studies employing lattice-QCD-based EoSs must explicitly quantify out-of-domain contamination before drawing quantitative conclusions, particularly in searches for critical phenomena.
An interesting outcome of these simulations is that evolving the full set of conserved charges can actually improve the effective coverage of multidimensional EoS tables. Allowing baryon number, strangeness, and electric charge to evolve dynamically gives fluid cells access to a larger region of thermodynamic phase space, reducing the need to resort to backup EoSs. 
All of these findings reinforce the importance of developing finite-$T$ EoSs of dense matter in a multidimensional phase space that consistently connect lattice QCD to hadronic descriptions, and that also manages to reproduce $\chi$EFT results and astrophysical constraints.

At last, we explored realistic EoSs containing movable critical points within the same dynamical framework. To our knowledge, these constitute the first $3+1$D relativistic \textit{viscous} hydrodynamic simulations capable of evolving through metastable and unstable regions of a first-order phase transition (previous work looked at ideal fluids with certain adaptations to handle the unstable regime i.e.,  including finite-range/capillarity terms \cite{Steinheimer:2012gc,Steinheimer:2013gla} or stochastic order-parameter dynamics coupled to an ideal-fluid \cite{Herold:2013qda}; or simplified dynamical models with dissipative effects such as those presented in Refs. \cite{Randrup:2010ax,Kapusta:2024nii}). 
One particular advantage of the SPH-based hydrodynamic simulations is that we obtain $3+1$D spacetime evolutions of individual fluid cells, so we can track their trajectories through the $(T,\mu_B)$ plane for a single event. This makes it possible to move beyond averaged hydrodynamic histories and directly study how out-of-equilibrium fluid cells cool down, propagate in rapidity, and approach candidate critical regions. 
By tracking individual fluid-cell trajectories through the $(T,\mu_B)$ plane, we quantified through our simulations how different beam energies probe candidate critical regions and found that the sensitivity depends strongly on the assumed critical-point location. While collisions at $\sqrt{s_{NN}}=7.7$~GeV provide the greatest sensitivity to a critical point near current theoretical expectations, our simulations also indicate that forward and backward rapidities can probe critical regions even at higher beam energies. This suggests that extending experimental rapidity coverage may significantly enhance the discovery potential of future critical-point searches.
Finally, we did not find significant effects from critical lensing in the holographic EoS through the SPH trajectories. The lack of critical lensing may be due to the mean-field nature of EoSs used, though it is more likely that this stems from the orientation of the critical region in the QCD phase diagram (i.e., smeared along the $\mu_B$ direction \cite{Dore:2022qyz}). The best check would be an analysis of the sign of kurtosis within the holographic EoS near the critical region. This is left for future work, as it requires a numerically accurate algorithm within MUSES to calculate susceptibilities from a multidimensional table (other modules such as \textbf{CMF++} use thermodynamic Jacobians to avoid this issue \cite{Gholami:2025cfq,Cruz-Camacho:2024odu}), which is still under development.

\subsection*{Outlook}

Beyond the developments discussed above about merging EoSs across the entire QCD phase diagram, other future developments are still necessary. Some of the EoSs require more conserved charges to improve their comparison to experimental data (e.g., Holography).  Additionally, it would be extremely beneficial to accurately describe critical scaling and include higher-order fluctuations, both in terms of the EoS and as well in fluid dynamics. 
Furthermore, it is clear that we need a much wider range on $(s,n_B)$ with a specific focus on the first-order phase transition regime. 
Further developments on including first-order phase transitions within relativistic viscous hydrodynamics that preserves causality and stability would be useful for such a study.

The dynamic simulations presented here are only the first step toward using the MUSES framework to constrain the existence and location of the QCD critical point using heavy-ion collisions. Direct comparisons to experimental data will require a full parameter estimation program, including systematic variation of the initial conditions, calibration of transport coefficients, and significantly larger statistics to control event-by-event fluctuations. Before proceeding to high-statistics production runs, we plan to perform a systematic analysis of the hydrodynamic model parameters and their correlations with the EoS input.

Transport properties, and diffusion in particular, remain a central challenge. The full $BSQ$ diffusion matrix has not yet been constrained in Bayesian analyses of multi-component fluids, and its behavior near a critical point is poorly understood. Moreover, the effect of criticality on the dynamical trajectories cannot be isolated without first quantifying how diffusion redistributes baryon number, strangeness, and electric charge throughout the evolving fireball. In this sense, the present work establishes the EoS and workflow infrastructure needed for the next stage: controlled, multidimensional, data-facing simulations of the QCD phase diagram in the context of heavy-ion collisions.

\vspace{0.1\linewidth}
\section*{Acknowledgments}

We would like to thank the wider MUSES collaboration for many discussions during our collaboration meetings and all colleagues who helped with testing the MUSES Calculation Engine.  
This work was supported in part by the National Science Foundation (NSF) within the framework of the MUSES collaboration, under grant number OAC-2103680.
This work used the Delta system at the National Center for Supercomputing Applications through allocation PHY250117 from the Advanced Cyberinfrastructure Coordination Ecosystem: Services \& Support (ACCESS) program, which is supported by National Science Foundation grants $\#$2138259, $\#$2138286, $\#$2138307, $\#$2137603, and $\#$2138296.
This work used Jetstream2 at Indiana University and Open Storage Network at NCSA through allocation PHY230156 from the Advanced Cyberinfrastructure Coordination Ecosystem: Services \& Support (ACCESS) program \cite{NSFACCESS}, which is supported by National Science Foundation grants \#2138259, \#2138286, \#2138307, \#2137603, and \#2138296.

\appendix

\section{Backup EoS}\label{sec:backupEOS}

In \cite{Plumberg:2024leb}, the concept of a backup EoS was developed in order to handle out-of-bound fluid cells in a thermodynamically consistent manner based on a lattice QCD expanded EoS, originally the 4D-Taylor EoS from \cite{Noronha-Hostler:2019ayj}. The reason why this is necessary is that every single fluid cell in a hydrodynamic simulation must have a thermodynamically consistent EoS in order to conserve energy and momentum, as well as the conserved $BSQ$ charges. Even fluid cells that are well below freeze-out and/or in a kinematic regime well outside the reach of detectors still require a thermodynamically consistent EoS. One cannot simply discard these fluid cells in the simulation because doing so would violate energy-momentum conservation.

Moreover, the EoS is necessary in fluid dynamics in order to close the equations of motion. Hydrodynamics is only an exactly determined system once an EoS is specified; otherwise, the system is under-determined and one cannot solve the equations of motion. Thus, all fluid cells require a thermodynamically consistent EoS, and one would prefer an EoS with the widest range possible in order to avoid contamination from backup EoSs that do not contain the relevant physics for the problem at hand. However, given the limited range of existing EoSs, as discussed in Fig.~\ref{fig:snBrange}, backup EoSs are still required, especially for lower $\sqrt{s_{NN}}$ simulations. Here we briefly summarize the three backup EoSs that were implemented in Ref.~\cite{Plumberg:2024leb}: the $\tanh$-conformal EoS, the conformal EoS, and the conformal-diagonal EoS.

The logic of the backup EoS is hierarchical. First, one attempts to invert an EoS table (e.g., lattice-QCD-based) for a given set of hydrodynamic variables (e.g., $(s,n_B)$) on a fixed grid. If no thermodynamically acceptable solution can be found in the original table EoS, the fluid cell is passed to the $\tanh$-conformal EoS. If that also fails, the code then tries the conformal EoS. Finally, if the conformal EoS does not yield a solution for that particular set of hydrodynamic variables, the code uses the conformal-diagonal EoS, which is designed to guarantee a solution for the remaining pathological cells as long as it is above some minimum energy density cut off. The purpose of this ordering is to stay as close as possible to the original table EoS whenever possible, while still ensuring that the hydrodynamic evolution can continue in a thermodynamically controlled way.

The first backup EoS is the $\tanh$-conformal EoS. It is designed to smoothly connect the table EoS near the edge of its range to a conformal EoS outside of it. Its pressure is
\begin{align}
p_\mathrm{tc}(T,\vec{\mu})
&= \frac{1}{2} A_0 T_0^4
\left[1+\tanh\left(\frac{T-T_c}{T_s}\right)\right]
\nonumber\\
&\times
\left[
\left(\frac{T}{T_0}\right)^2
+\sum_{X=B,S,Q}
\left(\frac{\mu_X}{\mu_{X,0}}\right)^2
\right]^2 ,
\end{align}
where the parameters are chosen to match the pressure scale of the lattice QCD EoS near the relevant boundary of the table. 
The hyperbolic tangent factor suppresses the conformal contribution at low $T$ and turns it on smoothly as $T$ increases. Physically, this choice is meant to mimic the expectation that the high-$T$ QCD EoS approaches a more nearly conformal form, while avoiding a sharp discontinuity when a fluid cell leaves the table EoS. 
This is the least intrusive backup EoS because it is still anchored to the table EoS pressure scale and is only meant to extend the table smoothly.

The second backup EoS is the conformal EoS,
\begin{equation}
p_\mathrm{c}(T,\vec{\mu})
=
A_0 T_0^4
\left[
\left(\frac{T}{T_0}\right)^2
+\sum_{X=B,S,Q}
\left(\frac{\mu_X}{\mu_{X,0}}\right)^2
\right]^2 .
\end{equation}
This EoS is genuinely conformal, meaning that it has the scaling structure expected for a system without an intrinsic dimensionful scale, apart from the normalization parameters used to match the table EoS. Unlike the $\tanh$-conformal EoS, there is no additional $T$-dependent switching function. Thus, this EoS is less tied to the crossover region and should be understood as a controlled mathematical extension rather than a realistic description of QCD at all $(s,n_X)$. It does, however, preserve the important feature that the pressure depends on both $(T,\mu_X)$ in a coupled way. In particular, expanding the pressure generates mixed terms such as $T^2\mu_X^2$ and $\mu_X^2\mu_Y^2$, which are qualitatively similar to the structure one expects in a multi-charge thermodynamic system.

The final backup EoS is the conformal-diagonal EoS,
\begin{equation}
p_\mathrm{cd}(T,\vec{\mu})
=
A_0 T_0^4
\left[
\left(\frac{T}{T_0}\right)^4
+\sum_{X=B,S,Q}
\left(\frac{\mu_X}{\mu_{X,0}}\right)^4
\right] .
\end{equation}
This EoS removes all mixed terms between $(T,\mu_X)$, as well as mixed terms between different conserved charges. For this reason, it is not intended to encode realistic QCD correlations between the $BSQ$ sectors. Instead, it is a final numerical safeguard. Because the pressure is diagonal in the thermodynamic variables, the inversion from hydrodynamic densities back to $\{T,\mu_B,\mu_S,\mu_Q\}$ becomes analytically much simpler. In Ref.~\cite{Plumberg:2024leb}, this property was used to derive a necessary and sufficient condition for the existence of a thermodynamic solution,
\begin{align}
\varepsilon
&\geq
\varepsilon_\mathrm{min}\left(\vec{\rho}\right)
\equiv
\frac{3}{4\cdot 2^{2/3}(A_0T_0^4)^{1/3}}
\sum_{X=B,S,Q}
\left(\mu_{X,0}|\rho_X|\right)^{4/3}.
\end{align}
Thus, the conformal-diagonal EoS should be viewed as a fail-safe that guarantees the hydrodynamic code can continue evolving a fluid cell even when that cell lies far outside the trustworthy region of the table EoS.

In this paper, we used 
\begin{align}
    A_0 &\equiv p_{T,0}/T_{\mathrm{scale}}^4 ,\nonumber \\
    T_0 &\equiv 1 \text{ fm}^{-1} ,\label{tanhConformalConstraints} \\
    \mu_{X,0} &\equiv \frac{A_0^{1/4} T_0 \mu_{X,{\max}}}{\sqrt{\sqrt{p_{X,\max}} - \sqrt{p_{T,0}}}} , \nonumber
\end{align}
where $X = B,S,Q$ and
\begin{align*}
    p_{T,0} &\equiv p_\mathrm{table}(T_{\mathrm{scale}}, \vec{0}) ,\\
    p_{B,\max} &\equiv p_\mathrm{table}(T_{\mathrm{scale}}, \mu_{B,{\max}}, 0, 0) ,\\
    p_{S,\max} &\equiv p_\mathrm{table}(T_{\mathrm{scale}}, 0, \mu_{S,{\max}}, 0) ,\label{EoS_constraints_definitions} \\
    p_{Q,\max} &\equiv p_\mathrm{table}(T_{\mathrm{scale}}, 0, 0, \mu_{Q,{\max}}) .
\end{align*}
Here, $T_{\mathrm{scale}} = 1.1 \, T_{\mathrm{FO}}$, $T_c = 220$ MeV and $T_s = 120$ MeV.

Users are free to adjust these coefficients and certain choices may be better motivated, depending on the given table EoS.

In practice, these backup EoSs are not meant to replace the table EoS. They are instead a controlled numerical extension that allows conservation laws to remain intact when a small number of fluid cells can no longer be represented within the table EoS. The physics interpretation of any simulation should therefore be based on the region where the table EoS dominates the evolution. The Backup EoSs thus provide a way to quantify the importance of the fluid cells which cannot be described by current MUSES EoSs, while also offering a thermodynamically consistent way to evolve those cells without introducing discontinuities or violating conservation laws.

\bibliography{reference,NOTinspire}

\end{document}